\documentclass[12pt,authoryear]{elsarticle}
\usepackage[utf8]{inputenc}
\usepackage{placeins}
\usepackage{graphicx}
\usepackage[a4paper,width=160mm,top=25mm,bottom=25mm,bindingoffset=6mm]{geometry}
\usepackage[compact]{titlesec}
\usepackage{hyperref}
\usepackage{tikz}
\usetikzlibrary{calc}
\usepackage{eso-pic}
\usepackage{setspace}
\usepackage{lipsum}
\usepackage[english]{babel}
\usepackage{fancyhdr}
\usepackage{array}
\usepackage{tabu}
\usepackage{multirow}
\usepackage{gensymb}
\usepackage{rotating}
 \usepackage{caption}
\usepackage{subcaption}
\usepackage{hanging}
\usepackage{todonotes}
\usepackage{changepage}
\usepackage{subcaption}
\usepackage{lscape}
\usepackage{chngcntr}
\usepackage{float}
\floatstyle{plaintop}
\restylefloat{table}
\usepackage{xcolor}
\usepackage{amsmath}
\usepackage{mathtools}
\usepackage{eufrak}
\usepackage[bbgreekl]{mathbbol}
\usepackage[makeroom]{cancel}
\usepackage{mathrsfs}
\DeclareSymbolFontAlphabet{\mathbbl}{bbold}
\usepackage{multirow,tabularx}
\usepackage{comment}
\usepackage{makecell}
\hypersetup{
    colorlinks,
    linkcolor={blue},
    citecolor={blue},
    urlcolor={blue}
}
\usepackage{longtable}
\usepackage{ragged2e}
\usepackage{tikz, blindtext}
\usepackage{fix-cm}
\usepackage{enumitem}
\usepackage{multicol}
\usepackage{longtable}
\usepackage{amsfonts}
\usepackage[font=footnotesize]{caption}
\usepackage[normalem]{ulem}

\usepackage[english]{babel}

\usepackage{amsthm}

\usepackage{algorithm}
\usepackage{algpseudocode}

\newif\ifthesis
\thesisfalse

\begin{document}
\begin{frontmatter}
    
\title{Homogenization of one-dimensional periodic rod networks as special Cosserat rods}

\author[a]{Vinayak}

\author[a]{Ajeet Kumar}


\affiliation[a]{organization={Department of Applied Mechanics},
addressline={Indian Institute of Technology Delhi}, 
city={New Delhi},
country={India}}




\begin{abstract}
One-dimensional periodic rod networks are the structures that are obtained by periodically assembling its microstructural unit, a network of rods itself, in just one direction. In this work, we present a scheme for obtaining the nonlinear constitutive response of such structures when homogenized macroscopically as a continuum rod. To accurately capture arbitrary and large deformations, the geometrically exact special Cosserat rod theory is used for modeling the rod at both micro- and macroscales. By assuming the periodic structure to be strained uniformly, at macroscale, along its arc length, the full structure problem is reduced to just that of its microstructural unit but subjected to helically periodic boundary condition. The microscale problem, consisting of a network of rods and formulated in a variational setting, is solved in the presence of rod-joint constraints and helically periodic boundary conditions. The expressions for the macroscale rod's stress resultants, i.e. internal contact force and moment, and stiffnesses are then obtained. Finally, several numerical examples having different microstructural units are presented to demonstrate our method. First, the results corresponding to simpler square and cross microstructural units are presented and validated with the existing literature. Square microstructural units having helical constituent rods are then taken up which have application as artificial muscle fiber material. Eventually, homogenization of auxetic tubular metamaterials is performed. It is shown how various design parameters of these microstructural units can be tuned to obtain the desired macroscopic response.  
\end{abstract}




\end{frontmatter}

\pagestyle{plain}
\section{Introduction}
\noindent In pursuit of exotic material behavior, several architected metamaterials have been proposed and developed. They have not only pushed the boundaries of stiffness and strength but have also opened the possibilities of tailored mechanical properties, e.g., negative Poisson's ratio, negative compressibility etc. \citep{jiao2023mechanicalbeyond}. Moreover, material properties beyond just mechanical such as thermal, electrical, electromechanical etc. are being envisioned \citep{jiao2023mechanical} that can be tuned as desired by tailoring their underlying microstructure. The advancement of additive manufacturing techniques has accelerated the development of such tailored microstructures resulting in the realization of exotic material properties. Predicting these properties beforehand forms an important cog in the design process of metamaterials. As conducting large scale experiments on such structures is not only difficult but also expensive, researchers have resorted to simulation driven design \citep{weeger2019digital}. Metamaterials are essentially heterogenous. Direct full-scale numerical simulation of such heterogeneous structures is again computationally expensive. One can thus make use of homogenization techniques in case of periodic metamaterials essentially replacing them as homogeneous continua. A metamaterial can be formed by periodically repeating its microstructural unit in all three, two or just one direction leading to three-dimensional bulk metamaterials, two-dimensional plate/shell-like metamaterials or one-dimensional rod-like metamaterials, respectively. The periodically repeating microstructural unit is called the representative volume element (RVE). Depending on the desired behaviour, an RVE can be constructed using network of rods \citep{fleck2010micro}, network of plates \citep{lee2016mechanical} or three-dimensional solids \citep{krishnan2022effective}. Rod-network based RVEs or lattices have conventionally been used for developing lightweight materials having high stiffnesses. However, their usage for high compliance applications such as for artificial tissues or as energy absorbing materials for high velocity impact applications \citep{chang2022mechanics,surjadi2025enabling} are being explored recently. The focus of the current work is on the homogenization of a subset of rod-like metamaterials whose microstructural unit is a network of rods. We call them one-dimensional periodic rod-networks. They have found application as cardiovascular stents \citep{han2018optimizing}, artificial muscle fibers \citep{sun2025recent}, robotic actuators \citep{parra2023modular}, metamaterial shafts \citep{wu2025metamaterial} etc.\\\\
Homogenization techniques require separation of micro and macro length scales \citep{saeb2016aspects}. For three-dimensional metamaterials, this length-scale separation should hold in all three periodically repeating directions. To this end, analytical \citep{deshpande2001effective}, asymptotic and computational models \citep{vigliotti2012linear,vigliotti2012stiffness} have been developed wherein a first-order homogenization technique has been used for small deformation of three-dimensional rod lattices. \citet{herrnbock2022homogenization} extended the computational homogenization for three-dimensional rod-lattices assuming finite strain at macroscale and geometrically exact elastoplastic beam elements at microscale. Similarly, there are two-dimensional micro-structured shells or shell-like metamaterials for which an appropriate homogenization technique has been developed recently by \citet{zhi2023multiscale, gruttmann2024fe2, geiger2025multiscale}. They first linearize the macroscale shell kinematics in shell coordinates and further add a fluctuation term in it to obtain the microscale kinematics for their RVE. Equating the RVE-averaged microscale strain with the macroscopic strain, they then derive appropriate periodic boundary condition that is consistent with the Hill-Mandel condition. Focusing now on rod-like continua or metamaterials, they typically undergo deformations such as bending and twisting which inherently involve significant level of strain gradient in cross-sectional plane. Therefore, first-order three-dimensional homogenization schemes are insufficient for such deformations. Another factor influencing the homogenization of such structures is that since it is periodic in only one direction, separation of length scale does not hold in other two directions and using three-dimensional homogenization schemes lead to size-effects \citep{yang2021size,alavi2022continualization,sarhil2023size}. The rod-like continua are often modeled using Cosserat rod theory \citep{antman2005problems}. In order to obtain constitutive response of continuum rods, both asymptotic \citep{yu2012variational,audoly2021asymptotic} and computational \citep{simo1991geometrically,chadha2019comprehensive,arora2019computational} homogenization models have been developed. These models are typically based on the assumption that the rod's strain parameters vary slowly along the arc-length of the rod; enabling the reduction of the full rod problem to just a cross-section. Thus, the two-dimensional cross-sectional plane becomes the RVE for homogenization. The cases of slow variation of bending and twisting which inherently involve large three-dimensional strain gradients in the cross-sectional plane are well covered in their homogenization schemes as the entire cross-section is homogenized simultaneously. Concerning the homogenization of one-dimensional periodic rod-networks which is the focus of this work, interest in space applications has driven the development of analytical models \citep{noor1978continuum,renton1984beam,burgardt1999continuum,kahla1995equivalent} that homogenized pin-jointed rod-like truss frames into continuum rods undergoing small deformations. More recent efforts have incorporated geometric nonlinearity at macroscale \citep{cao2020extended} and flexible joints at microscale \citep{liu2019equivalent}. Likewise, \citet{abdoul2018strain} and \citet{audoly:hal-04112136} developed asymptotic models for the homogenization of linear elastic beam networks with arbitrary geometry and periodicity, recovering the Euler-Bernoulli and Timoshenko beam models as the homogenized continua in cases of one-dimensional periodicity. Recently, \citet{saadat2023mixed} proposed a novel computational homogenization model for spiral strands using geometrically exact rod model at both macro and micro scales. Neglecting shear deformation at macroscale, they employed a linear macro-to-micro transition which, we note, is applicable only when the rod's strains are small enough. To homogenize discrete one-dimensional nanostructures such as carbon nanotubes as an effective Cosserat rod, \citet{kumar2016helical} introduced the \textit{helical Cauchy-Born (HCB) rule}. This approach leverages the helical periodicity in uniformly deformed rod-like structures to reduce the full structural problem to just its RVE. Their reduction leads to a fully non-linear macro-to-micro transition for one-dimensional structures. The HCB rule is geometrically exact and also supports general constitutive behavior at microscale. It has since been extended to model nonlinear constitutive responses of anisotropic elasto-plastic rods \citep{vinayak2023uniformly}, strips \citep{kumar2024computational} etc. \\\\
In this work, we further extend the HCB rule and propose a computational scheme for the homogenization of one-dimensional periodic rod networks wherein both the microscale rods and the macroscopic effective rod are modeled as special Cosserat rods. Unlike several existing homogenization schemes which use adapted forms of the classical Hill-Mandel condition, our scheme is based on the minimization of microscale energy of the RVE. The proposed scheme also uses fully nonlinear geometrically exact kinematics not just for the rod models at macro and micro scales but also for macro-to-micro transition map. It may be noted here that most of the existing homogenization techniques first linearize the macroscale kinematics in macroscale coordinates and further add a fluctuation while deriving microscale kinematics. The linearization here obviously leads to inaccuracy in macro-to-micro transition. We further assume arbitrary constitutive behavior of constituent rods in the metamaterial's microstructural unit. Unlike the original HCB rule for nanorods \citep{kumar2016helical} where the energy at microscale (inter-atomic energy) depended only on the positions of atoms, the microscale energy in the present work depends on both the position and orientation of directors at every nodal points of microscale rods. In summary, the following are the novel contributions of this work:
\begin{enumerate}
    \item Both microscale and macroscale rods are modeled as geometrically exact special Cosserat rods accounting for not just finite bending and twisting but also finite shear and stretch at both the scales.
    \item A geometrically exact nonlinear map is proposed for applying helically periodic boundary condition to both translational and rotational degrees of freedom at microscale. The map holds even at large strains of macroscale rod.
    \item The microscale rod's constitutive model can be arbitrary.
\end{enumerate}
The outline of the paper is as follows. In section \ref{geometrically exact rod theory}, we briefly describe the geometrically exact special Cosserat rod theory. Section \ref{sec:rve_problem_1d_1dh} is the main contribution of this work. Here we obtain the RVE problem for the homogenization of one-dimensional periodic rod networks. We first obtain the formula for the configuration variables of a periodic rod network strained uniformly (at macroscale) along its arclength. In particular, we derive how do microscale rod's centerline and directors in the metamaterial's RVE relate to the corresponding variables in the images of the RVE. This leads to what we call helically periodic boundary condition. We also derive mathematical form of various constraints such as joint constraints, helical boundary condition constraints and global constraints in the presence of which the RVE's energy needs to be minimized for homogenization. In Section \ref{stiffnesses}, we derive expressions of the homogenized rod's internal contact force, moment and stiffnesses at arbitrary state of the macroscale rod's strain. In Section \ref{discrete rod theory}, we present a procedure to discretize the rod's energy using what we call helical elements. In Section \ref{sec:results}, we present numerical examples involving several RVEs to demonstrate our scheme. First, we apply the proposed scheme to square- and cross-RVE based rod networks. Then, we consider more complex square RVEs formed by helical rods and auxetic RVEs. Section \ref{sec:conclusions} concludes our paper.\\\\
$\textbf{Notation}$:
The following notation is used unless otherwise specified. Vectors are denoted by lowercase bold letters whereas second-order tensors are denoted by uppercase bold letters. The symbol $(\cdot)^{\prime}$ represents derivative with respect to the rod’s undeformed arc-length. Finally, repeated Latin indices imply summation from 1 to 3 whereas repeated Greek indices imply summation from 1 to 2.

\pagestyle{plain}
\section{Brief description of the special Cosserat rod theory}\label{geometrically exact rod theory}
\begin{figure}[h!]
    \centering
    \includegraphics[width=.8\textwidth]{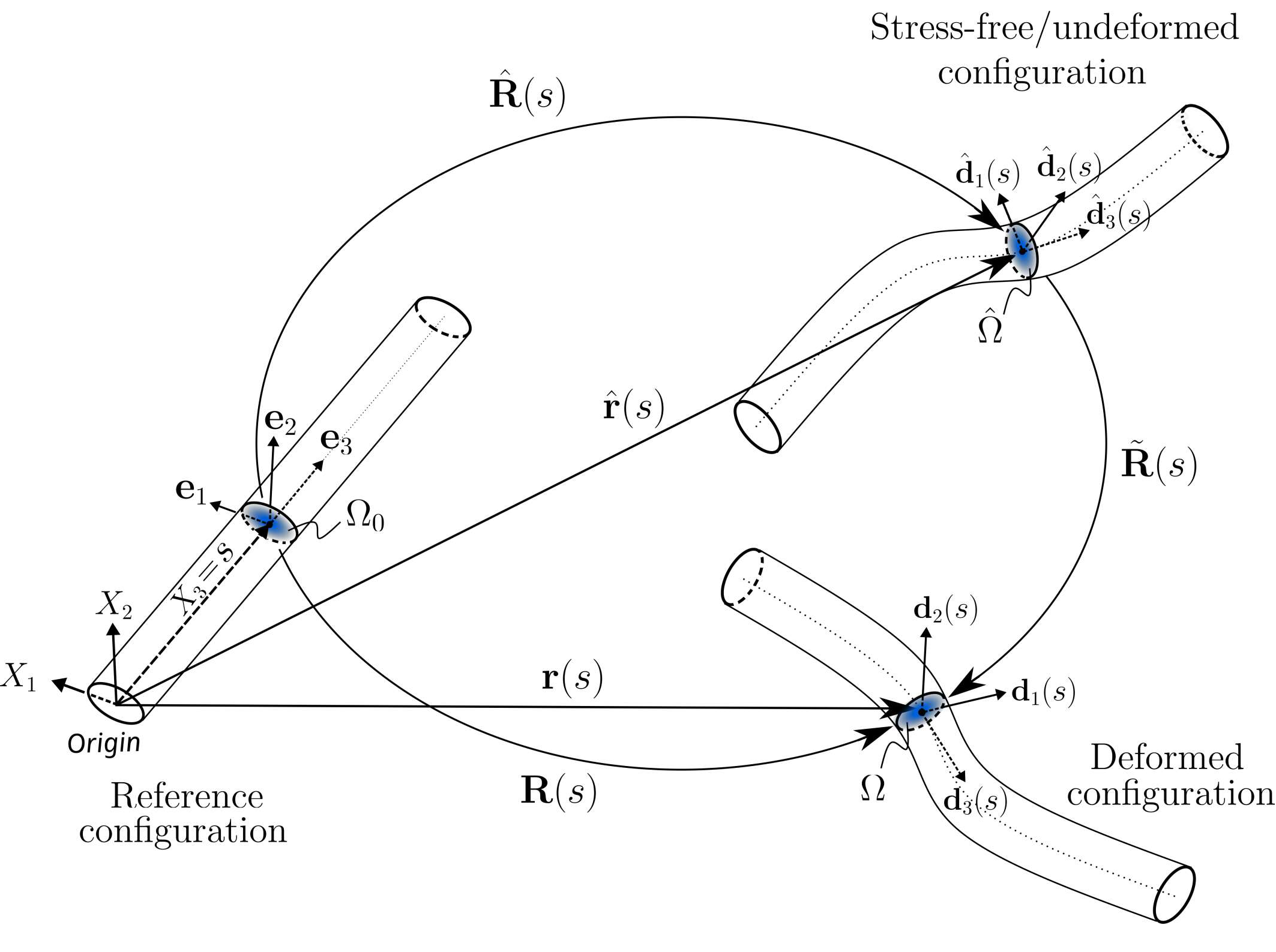}
    \caption{Kinematics of a special Cosserat rod}
    \label{fig:ge_rod_kinematics}
\end{figure}
A continuum rod can be described as a set of infinite cross-sections stacked together along a curve, known as the centreline of the rod. The orientation of the cross-section is given by a triad of orthonormal directors as shown in Figure \ref{fig:ge_rod_kinematics}. The reference centerline of the rod is represented by the straight line $s\textbf{e}_3$ where $s\in[0,l]$ and the orthonormal director triads coincide with the global basis in the reference state. In the stress-free configuration, the centreline is given by $\hat{\textbf{r}}(s)$ whereas the director triads are given by $(\hat{\textbf{d}}_{1}(s),\hat{\textbf{d}}_{2}(s),\hat{\textbf{d}}_{3}(s))$. 
The associated rotation tensor in the stress-free configuration is $\hat{\textbf{R}}(s)$ such that
\begin{align}\label{eq:stress_free_directors2}
    \hat{\textbf{d}}_i(s) = \hat{\textbf{R}}(s)\textbf{e}_i.
\end{align}
Finally, the deformed configuration is given by the centreline curve $\textbf{r}(s)$ and the director triads $({\textbf{d}}_{1}(s),{\textbf{d}}_{2}(s),{\textbf{d}}_{3}(s))$ such that
\begin{align}\label{eq:spatial_directors}
    \textbf{d}_i(s) = \textbf{R}(s)\textbf{e}_i.
\end{align}
In this work, we parameterize the rotation tensor using a vector $\boldsymbol{\theta}\in\mathbbl{R}^3$ whose direction coincides with the axis of rotation and whose magnitude is equal to the angle of rotation. The rotation tensor $\textbf{R}(\boldsymbol{\theta}(s))$ is then obtained using the Rodrigues' rotation formula (refer to equation \eqref{eq:rotation_formula}). Therefore, the variables $\textbf{r}(s)$ and $\boldsymbol{\theta}(s)$ are the kinematic variables of this theory. The strain measures in this theory are
\begin{align}
    &\textbf{v}_0 = \textbf{R}^T\textbf{r}^{\prime} = \text{v}_i\textbf{e}_i,\label{eq:v0}\\
    &\textbf{k}_0 = \text{axial}(\textbf{K}_0)=\kappa_i\textbf{e}_i~ \text{where} ~\textbf{K}_0 = \textbf{R}^T\textbf{R}^{\prime}\label{eq:k0}.
\end{align}
Here ($\text{v}_1$, $\text{v}_2$) are the shear strains, $\text{v}_3$ is the axial stretch, ($\kappa_1$, $\kappa_2$) are the bending curvatures and $\kappa_3$ is the twist. The spatial counterparts of the above strain measures are
\begin{align}
    &\textbf{v} = \textbf{R} \textbf{v}_0 = \textbf{r}^{\prime} = \text{v}_i\textbf{d}_i,\label{eq:v_spatial}\\
    &\textbf{k} = \text{axial}(\textbf{K}) = \kappa_i\textbf{d}_i\quad \text{where} ~\textbf{K} = \textbf{R}\textbf{K}_0\textbf{R}^T =\textbf{R}^{\prime}\textbf{R}^T.\label{eq:k_spatial}
\end{align}
For thin rods, the shear and extensional strains are negligible when compared to bending and twisting strains. The unshearability of the rod is incorporated by introducing the following constraints:
\begin{align}\label{eq:unshearability_constraint}
    \textbf{v}_0(s)\cdot\textbf{e}_1 &= \textbf{v}_0(s)\cdot\textbf{e}_2 = 0.
\end{align}
We call such a rod as an unshearable rod. For a Kirchhoff rod that is both unshearable and inextensible, the constraint is
\begin{align}\label{eq:kirchhoff_constraint}
    \textbf{v}_0(s) = \textbf{e}_3.
\end{align}
 
The internal contact force $\textbf{n}(s)$ and internal contact moment $\textbf{m}(s)$ acting on a cross-section of the rod in the deformed configuration are given by
\begin{align}\label{component}
\textbf{n}= \text{n}_i \textbf{d}_{i}, \quad\quad \textbf{m}=\text{m}_i\textbf{d}_{i}
\end{align}
while its rotational pull-backs are given by
\begin{align}\label{component1}
\textbf{n}_0= \text{n}_i \textbf{e}_{i},\quad\quad \textbf{m}_0= \text{m}_i \textbf{e}_{i},
\end{align}
respectively. Together, they are also called the rod's stress resultants. The linear and angular momentum balance equations of the rod assuming statics condition are given by
\begin{align}\label{global_eq}
\textbf{n}^{\prime} + \breve{\textbf{n}}&=\boldsymbol{0},\nonumber\\
\textbf{m}^{\prime}+\text{\textbf{v}}\times\textbf{n} +\breve{\textbf{m}}&=\boldsymbol{0}.
\end{align}
Here $\breve{\textbf{n}}$ and $\breve{\textbf{m}}$ are the distributed force and couple, respectively, that act on the rod. One further assumes the existence of a scalar-valued function $\phi(\textbf{v}_0,\textbf{k}_0)$ denoting stored energy per unit reference length such that
\begin{align}
    \textbf{n}_0 = \frac{\partial \phi}{\partial \textbf{v}_0},~~~~~~~~\textbf{m}_0 = \frac{\partial \phi}{\partial \textbf{k}_0},~~~~~~~~~\mathbbl{C}_0^{rod}=\frac{\partial^2\phi}{\partial[\textbf{v}_0,\textbf{k}_0]\partial[\textbf{v}_0,\textbf{k}_0]}.
\end{align}
Here $\mathbb{C}_0^{{rod}}$ is the rod's elasticity tensor whose diagonal components are the shearing, stretching, bending and twisting stiffnesses whereas the off-diagonal components are the various coupling stiffnesses. It assumes a diagonal form for isotropic rods. Often, the following quadratic energy model is assumed with $\mathbb{C}_0^{{rod}}$ being a constant tensor: 
\begin{align}\label{eq:rod_energy_function_quadratic}
    \phi(\textbf{v}_0,\textbf{k}_0) = \frac{1}{2}\begin{bmatrix}
        \textbf{v}_0 - \hat{\textbf{v}}_0 \\
        \textbf{k}_0 - \hat{\textbf{k}}_0
    \end{bmatrix}^T\mathbbl{C}_0^{rod}
    \begin{bmatrix}
        \textbf{v}_0 - \hat{\textbf{v}}_0 \\
        \textbf{k}_0 - \hat{\textbf{k}}_0
    \end{bmatrix}.
\end{align}
Here $\hat{\textbf{v}}_0$ and $\hat{\textbf{k}}_0$ are the strain measures in the rod's stress-free/natural configuration. The quadratic model works for small departure of strain values from their values in the natural state. For large enough departure from natural strain values, a nonlinear energy model would be required. It should include the effect of the underlying nonlinear three-dimensional material behavior of the rod material as well as nonlinear cross-sectional warping. \cite{arora2019computational} proposed a cross-sectional warping problem to obtain this nonlinear energy numerically for hyperelastic rods. They employ helical Cauchy-Born rule to subject the rod to a strain field which is uniform in the rod's arc-length. This enables the reduction of the equations of elasticity from the entire rod body to just its cross-section.\footnote{The cross-section can be thought of as the two-dimensional RVE here which is helically repeated continuously to form the entire helical rod.} The rod's energy is then obtained by integrating the three-dimensional stored energy density $W$ in the rod's cross-section as follows:
\begin{align}\label{eq:rod_energy_hyperelastic}
    \phi(\textbf{v}_0,\textbf{k}_0) = \int_{\Omega_0}W(\textbf{F}(\textbf{v}_0,\textbf{k}_0))d\text{X}_1d\text{X}_2.
\end{align}
Here $\textbf{F}$ is the three-dimensional deformation gradient tensor which is obtained through the map
\begin{align}\label{eq:warping_map}
    \textbf{x}(\text{X}_1,\text{X}_2,s) = \textbf{r}(s) + \textbf{R}(s)\textbf{x}_\text{w}(\text{X}_1,\text{X}_2).
\end{align}
Here
\begin{align}\label{eq:warping_function}
    \textbf{x}_\text{w}(\text{X}_1,\text{X}_2) = \text{X}_{\alpha}\textbf{e}_{\alpha} + \text{u}_i(\text{X}_1,\text{X}_2)\textbf{e}_i
\end{align}
is the warped cross-section's map and $\text{u}_i$'$s$ denote the unknown in-plane and out-of-plane warping displacement components which are obtained by solving the above mentioned cross-sectional warping problem. In this work, we use a similar approach to obtain $\phi$ for a one-dimensional periodic rod-network. As we show next, instead of solving the cross-sectional warping problem, one has to solve the metamaterial RVE's warping problem. 
\section{The RVE problem for one-dimensional periodic rod networks}\label{sec:rve_problem_1d_1dh}
\begin{figure}[h!]
    \centering
    \includegraphics[width=0.4\textwidth]{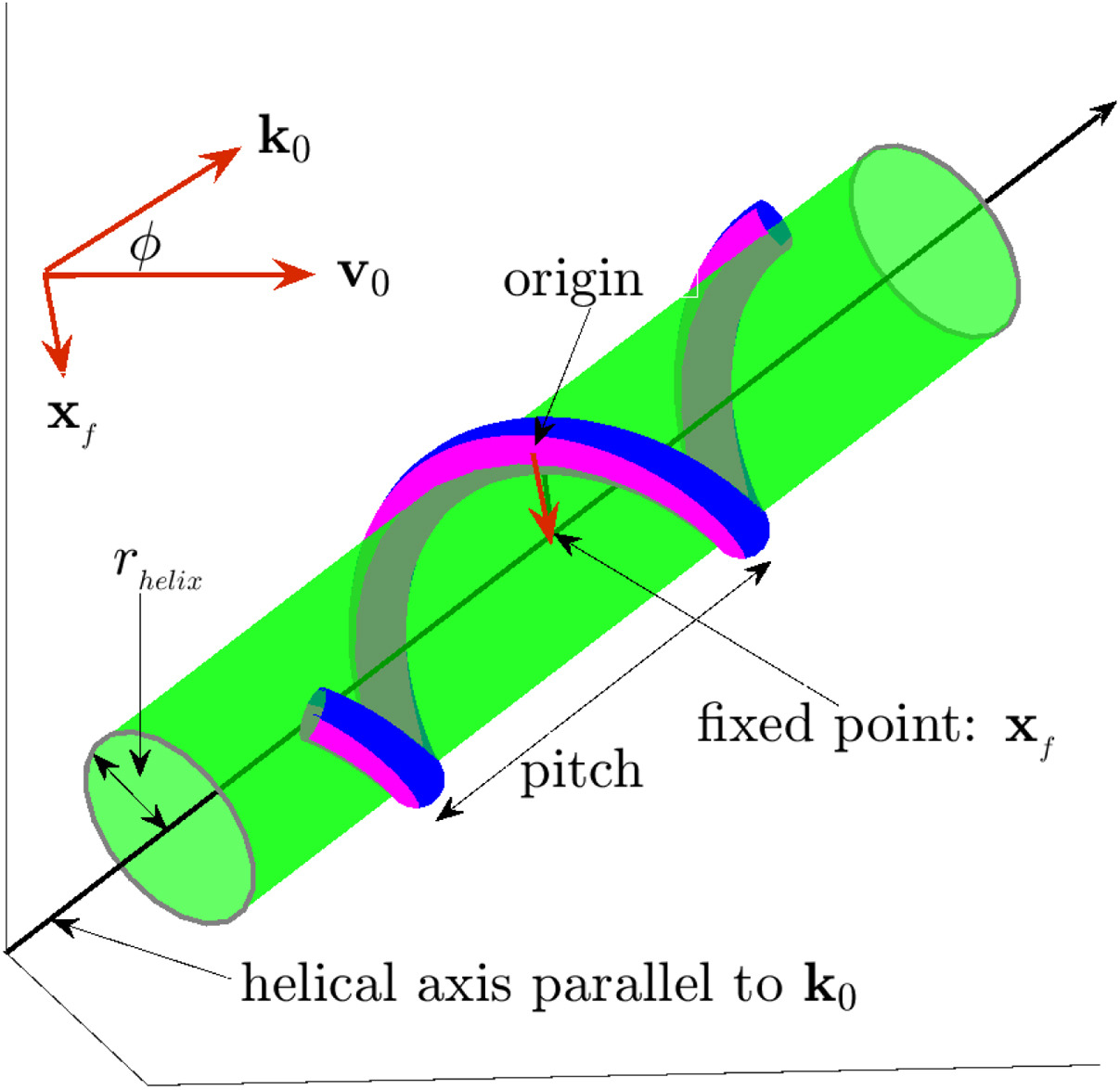}
    \includegraphics[width=0.56\textwidth]{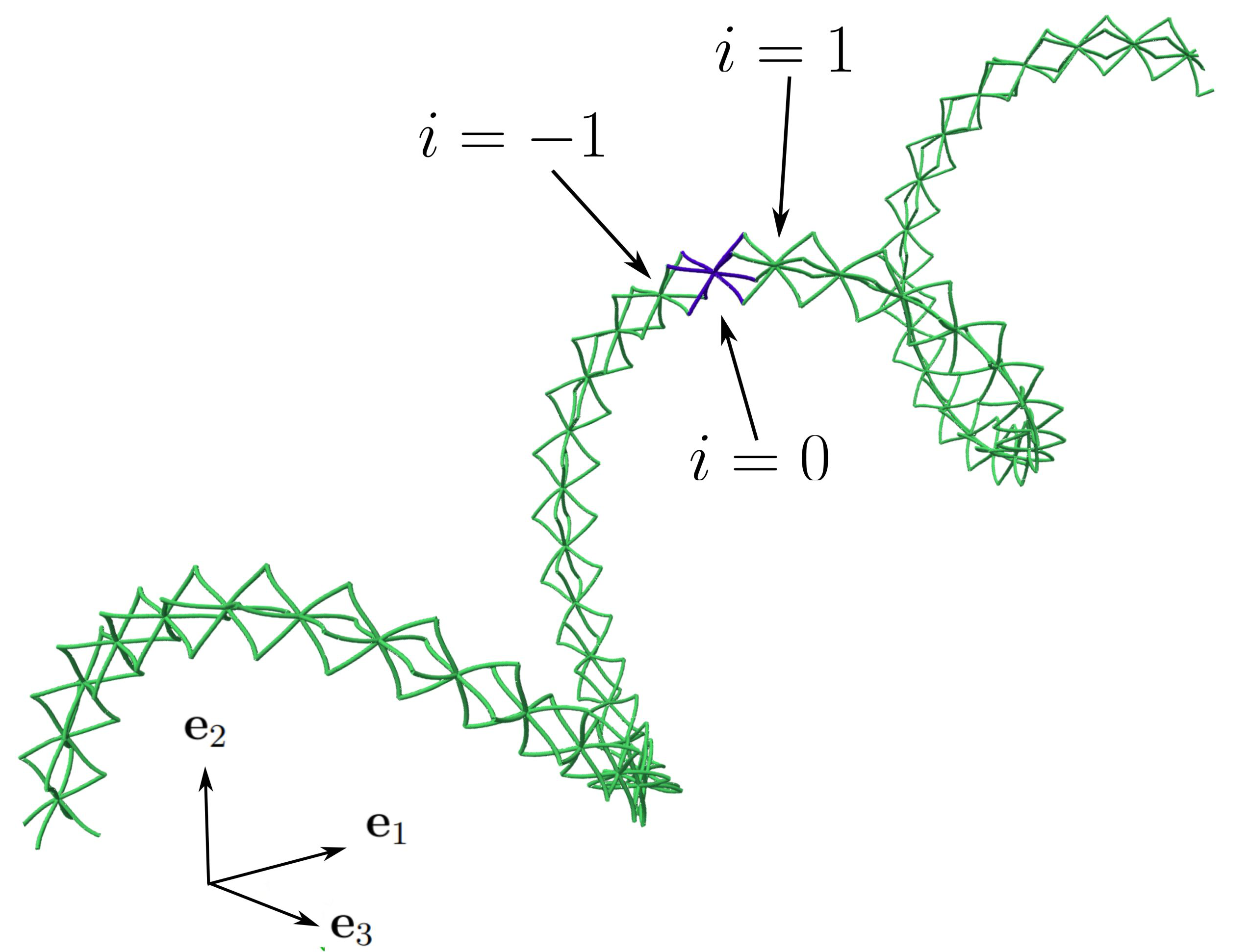}
    \caption{A continuum rod (left) and a one-dimensional periodic rod network (right) both deformed uniformly along their arclength}
    \label{fig:helicalrod}
\end{figure}
In order to obtain $\phi$ for rod-like metamaterials, we think of macroscopically subjecting it to a strain field which is uniform in its arc-length at macroscale, i.e., $\textbf{v}_0(s^M) = \textbf{v}_0^M$ and $\textbf{k}_0(s^M)=\textbf{k}_0^M$. Here $s^M$ denotes the macroscale arc-length parameter. The macroscopic configuration ($\textbf{r}^M,\textbf{R}^M$) of such a homogenized rod is obtained by integrating \eqref{eq:v0} and \eqref{eq:k0} which results in a uniform helical rod given by\footnote{The superscript $M$ here denotes that they are macroscopic variables.} (also see Figure \ref{fig:helicalrod}a):
\begin{align}\label{eq:macro_centreline}
    \textbf{r}^{M}(s^M) = \bigg(\int_0^{s^M}\textbf{R}^M(l) dl\bigg)\textbf{v}_0^M,\quad
        \textbf{R}^M(s^M) = e^{s^M\textbf{K}_0^M}.
\end{align}
Substituting \eqref{eq:macro_centreline} in \eqref{eq:warping_map} and further generalizing it for rod-like metamaterials, we get
\begin{align}\label{eq:macro_centreline2 hcb rule}
        &\textbf{x}^i(X_1^m,X_2^m,s^m) = \bigg(\int_0^{iL}\textbf{R}^M(l) dl\bigg)\textbf{v}_0^M + \textbf{R}^M(iL)~\textbf{x}^{RVE}(X_1^m,X_2^m,s^m)
\end{align}
where $\textbf{x}^i$ is the $i^{th}$ image of the metamaterial's RVE, corresponding to $i=0$ in Figure \ref{fig:helicalrod}b, whose deformed position is denoted by $\textbf{x}^{RVE}$ in the above equation. The macroscale arc-length parameter $s^M$ is replaced here by $iL$ where $L$ is the extent of the RVE along the axis of the homogenized rod in its reference state. Likewise, $(X_1^m,X_2^m,s^m)$ denote the referential microscopic coordinates of material points in the RVE. Note that $s^m\ne s^M$: the variable $s^m$ typically runs within an RVE. Also note that the role of the unknown warped cross-section $\textbf{x}_w$ in \eqref{eq:warping_map} is overtaken by the unknown warped/deformed RVE, i.e., $\textbf{x}^{RVE}$ in equation \eqref{eq:macro_centreline2 hcb rule}. In case, the RVE comprises of a network of rods (as in Figure \ref{fig:helicalrod}b) which is the focus in this work, the unknowns in the RVE would simply be the centerline and rotation of all the rods in the RVE, i.e., $\{\textbf{r}^{\alpha}(s^{\alpha}),\textbf{R}^{\alpha}(s^{\alpha})\}_{\alpha=1}^{N_{rods}}$ where $s^{\alpha}$ denotes the arc-length parameter for the $\alpha^{th}$ rod in the RVE and $N_{rods}$ is the total number of rods in the RVE. Equation \eqref{eq:macro_centreline2 hcb rule} then gets modified as follows:
\begin{align}\label{eq:hcb_rule_rod_continuum}
            &\textbf{r}^{i,\alpha}(s^{\alpha}) = \bigg(\int_0^{iL}\textbf{R}^M(l) dl\bigg)\textbf{v}_0^M + \textbf{R}^M(iL)~\textbf{r}^{\alpha}(s^{\alpha}),\nonumber\\
            &\textbf{R}^{i,\alpha}(s^{\alpha}) = \textbf{R}^M(iL)~\textbf{R}^{\alpha}(s^{\alpha}).
\end{align}
Here $(\textbf{r}^{i,\alpha}(s^{\alpha}),\textbf{R}^{i,\alpha}(s^{\alpha}))$ denotes the centerline and rotation tensor of the $\alpha^{th}$ rod in the $i^{th}$ image of the RVE. We now propose the RVE problem to determine the unknowns $\{\textbf{r}^{\alpha},\textbf{R}^{\alpha}\}_{\alpha=1}^{N_{rods}}$. 
\begin{figure}[h!]
    \centering
    \includegraphics[width = .9\textwidth]{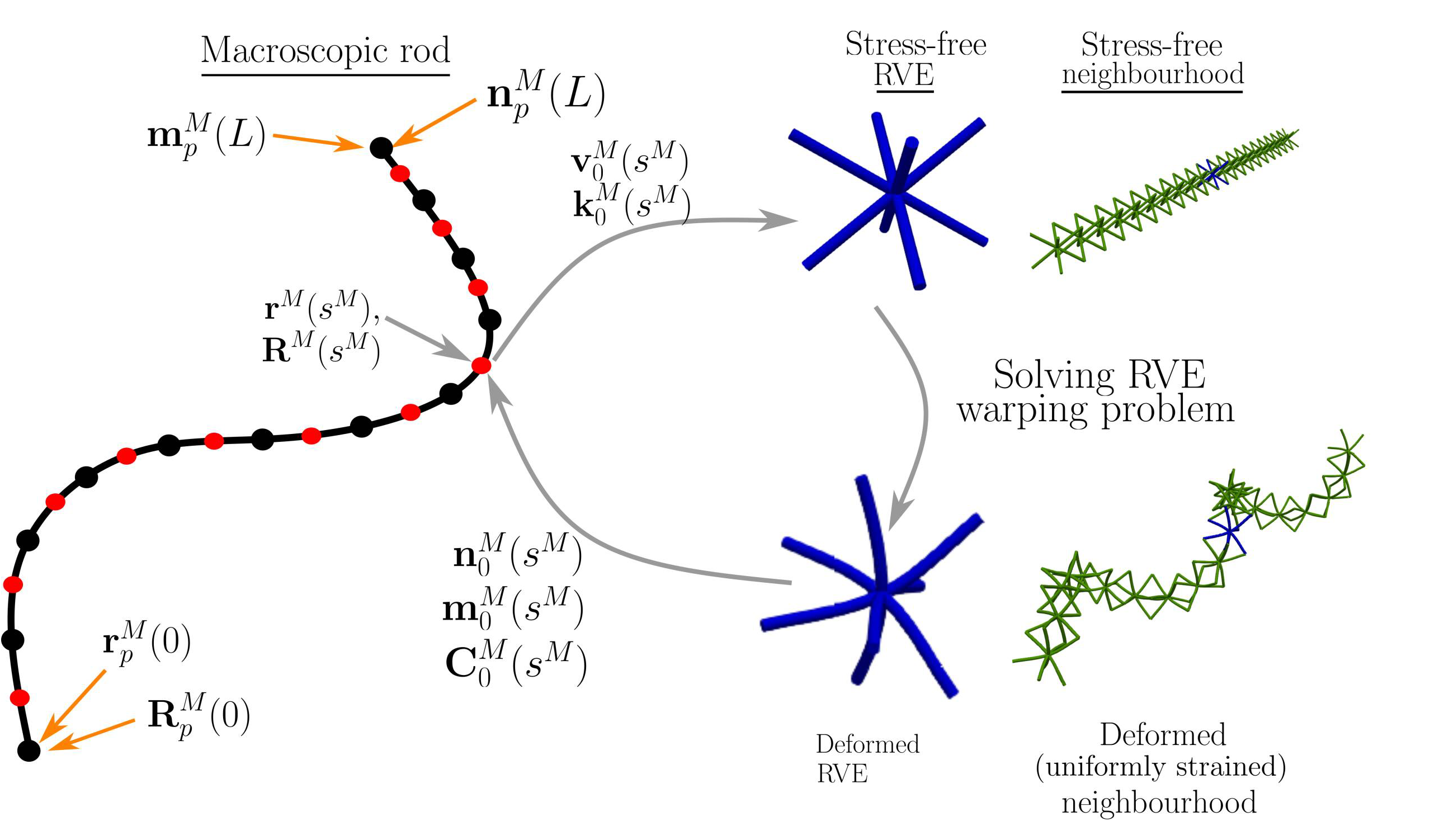}
    \caption{A multiscale framework for one-dimensional periodic rod networks modeled as special Cosserat rods. On the macroscopic rod: black dots denote nodal points, red dots denote quadrature points and $(\cdot)_p$ denote prescribed quantities at the two ends.}
    \label{fig:slender_mm}
\end{figure}
Figure \ref{fig:slender_mm} shows the centerline of a long homogenized macroscopic rod whose cross-section is formed by the metamaterial's RVE.\footnote{A rod's cross-section has zero thickness in a continuum sense. However, the metamaterial's RVE has finite extent along the rod's arclength. As the length of the RVE is typically much smaller compared to the metamaterial's total length (only then the separation of length scale holds), we can therefore call the RVE here to be the metamaterial's ``cross-section". } A non-uniform strain field $(\textbf{v}_0^M(s^M),\textbf{k}_0^M(s^M))$ typically gets induced in the macroscopic rod. If this strain field varies slowly along the arc-length, one can think of the strain field as being locally uniform. One can then construct a fictitious metamaterial that is uniformly strained with the macroscopic rod's local value of strain $(\textbf{v}_0^M,\textbf{k}_0^M)$. This fictitious metamaterial will deform helically obeying equation \eqref{eq:hcb_rule_rod_continuum} as also shown in Figure \ref{fig:slender_mm}. In this section, we use this helical symmetry to formulate the microscale RVE warping problem to obtain the energy density $\Phi^M$ for one-dimensional periodic rod networks. We write:
\begin{align}\label{eq:rve_warping_problem}
 \Phi^M(\textbf{v}_0^M,\textbf{k}_0^M) = \frac{1}{L}\min_{\{\textbf{r}^{\alpha},\textbf{R}^{\alpha}\}}  \mathcal{E}^{RVE} = \frac{1}{L}\min_{\{\textbf{r}^{\alpha},\textbf{R}^{\alpha}\}}\left(\sum_{\alpha=1}^{N_{rods}}\int^{L^{\alpha}}_0 \phi^{\alpha}(\textbf{v}^{\alpha}_0,\textbf{k}^{\alpha}_0)ds^{\alpha}\right)
\end{align}
where $\mathcal{E}^{RVE}$ is the elastic energy stored in the RVE of the fictitious uniformly deformed metamaterial. Note that we rely on energy minimization to achieve homogenization, whereas several other approaches rely on the classical Hill-Mandel condition which equates virtual works at macro- and microscales \citep{saeb2016aspects,herrnbock2022homogenization,zhi2023multiscale,gruttmann2024fe2}. The minimization problem \eqref{eq:rve_warping_problem} needs to be solved in the presence of following three constraints: (i) internal joint constraints to enforce connectivity of rods within the RVE (ii) helical boundary condition constraint to apply helical periodicity and (iii) global constraints to restrict rigid translation and rotation of the RVE as a whole and also to keep the macroscopic strain parameters fixed during minimization.
\begin{figure}
    \centering
    \includegraphics[width=.9\textwidth]{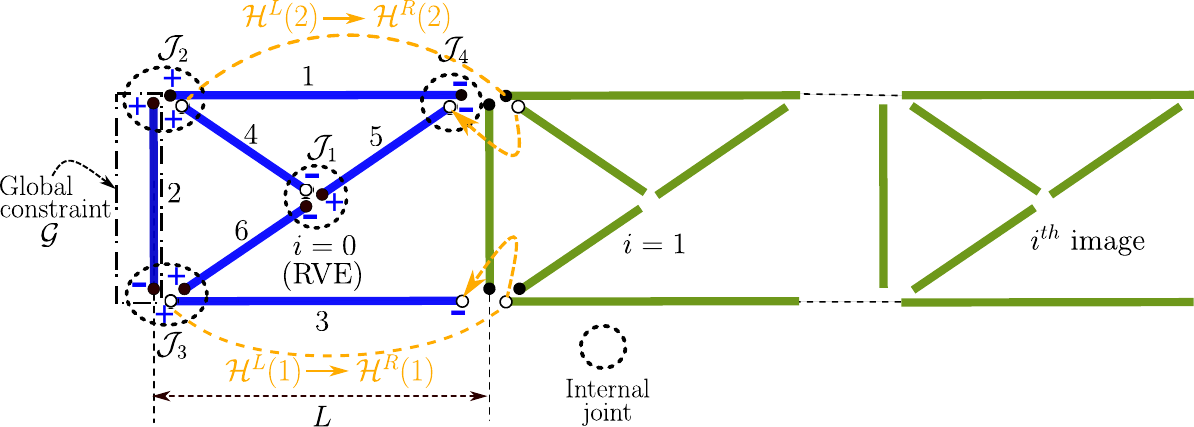}
    \caption{Various constraints needed for the homogenization of one-dimensional periodic rod networks}
    \label{fig:slender_mm_fd}
\end{figure}
A pictorial illustration of these constraints is shown in Figure \ref{fig:slender_mm_fd} for a representative one-dimensional periodic rod network. This rod network has six rods in its RVE all shown in blue. The green vertical rod at the right boundary is not a part of the RVE but its right image. We also use the following notation for the position of different nodes of a rod in the RVE:
\begin{align}
    \textbf{r}^{\alpha}_{k} &= \textbf{r}^{\alpha}(s^{\alpha}_k)\quad \text{for the $k^{th}$ internal node located at $s_k\in(0,L^{\alpha})$}.
\end{align}
For the rod's two end-points however, we use the following special notation:
\begin{align}
    \textbf{r}^{\alpha}_{\beta} =     \begin{cases}
        \textbf{r}^{\alpha}_{+} & \text{for } \textbf{r}^{\alpha}(L^{\alpha})\\
        \textbf{r}^{\alpha}_{-} & \text{for } \textbf{r}^{\alpha}(0)\\
    \end{cases}
\end{align} 
These notations will be used heavily while discussing constraints. We begin with internal joint constraint.
\subsection{Internal joints in the RVE}
\begin{figure}
    \centering
    \includegraphics[width=.7\linewidth]{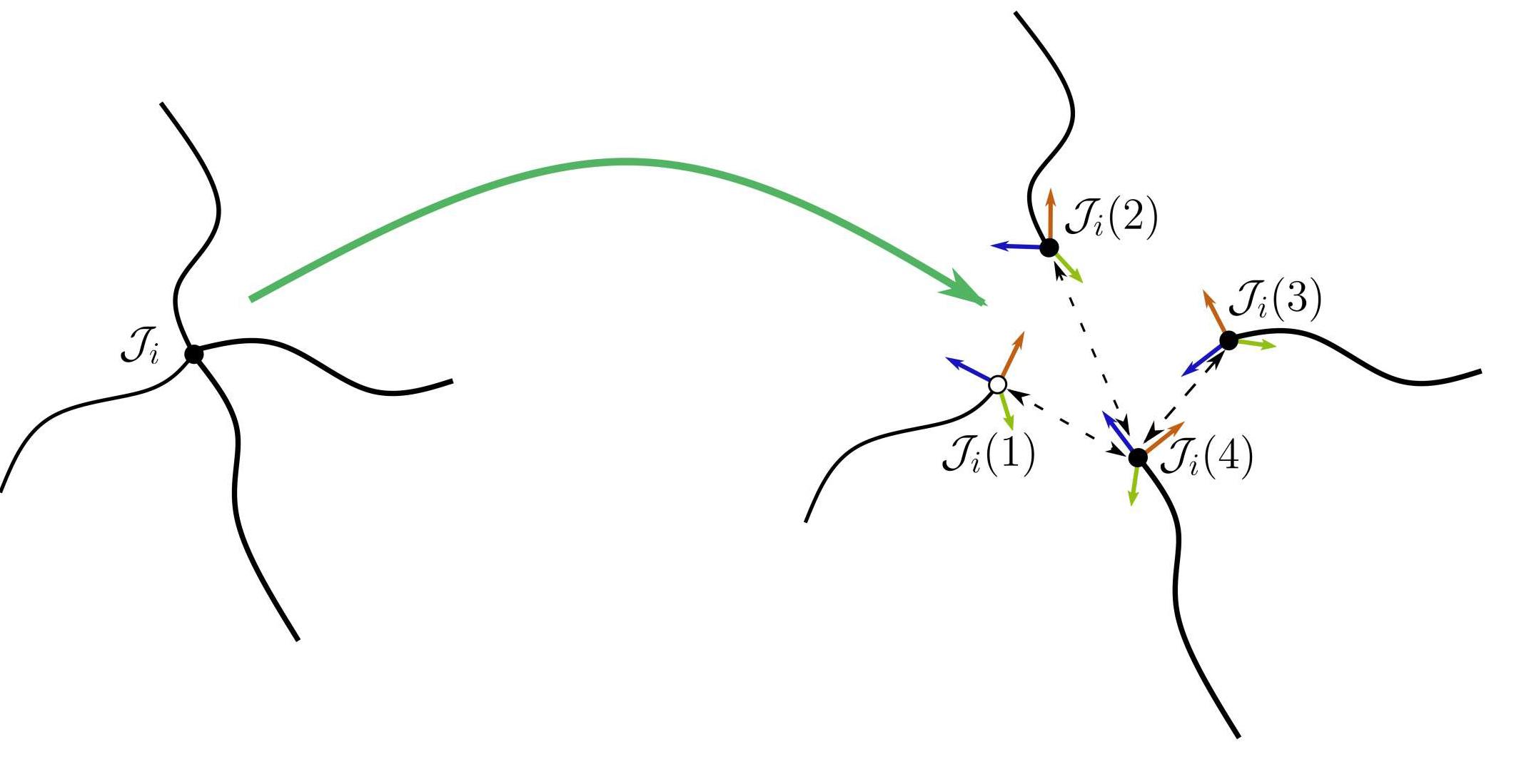}
    \caption{A schematic of multiple rods connecting at a joint. Here, the joint $\mathcal{J}_i$ contains $J_i=4$ nodes. The master node given by $\mathcal{J}_i(1)$ is shown as a white circle. All the other nodes, shown as black circles, are slaves to the master node at this joint.}
    \label{fig:internal joint}
\end{figure}
We assume that the RVE is a network of rods consisting of $N_J$ internal joints. By \textit{internal}, we imply that all the rods participating in a joint are internal to the RVE. Let us assume the connection between the rods to be rigid and use the master-slave approach to formulate the internal joint constraint \citep{jelenic1996non,ibrahimbegovic2000rigid}.\footnote{Other kinds of joints such as pinned joint or ball-and-socket joint can also be easily incorporated.} For the $i^{th}$ joint in the RVE, we assume it to be formed by $J_i$ number of nodes each belonging to different rods that join there and denote this set of nodes by $\mathcal{J}_i$ - see Figure \ref{fig:internal joint}. It is defined as follows:
 \begin{align}
     \mathcal{J}_i = \{(\alpha,\beta):\textbf{r}_{\beta}^{\alpha}\text{ lies at the $i^{th}$ joint\}},~ \vert\mathcal{J}_i\vert = J_i \quad\forall~ i=1: N_J
 \end{align}
Here $\beta$ takes either `+' or `-' value. Each set $\mathcal{J}_i$ is an ordered set such that its first element $\mathcal{J}_i(1)=(\alpha_1,\beta_1)$ is the master node and all the other elements $\mathcal{J}_i(k)=(\alpha_k,\beta_k)$ where $k\in[2,J_i]$ are slave nodes. In the RVE of Figure \ref{fig:slender_mm_fd}, there are four internal joints whose sets are given by
\begin{align}
    &\mathcal{J}_1 = \{(4,-), (6,-), (5,+)\},\nonumber\\
    &\mathcal{J}_2 = \{(4,+), (2,+),(1,+)\},\nonumber\\
    &\mathcal{J}_3 = \{(3,+),(2,-),(6,+)\},\nonumber\\
    &\mathcal{J}_4 = \{(5,-),(1,-)\}.
\end{align}
Once internal joints are defined, for an $i^{th}$ joint, we have the following translational constraint relating the positions of master and slave nodes:
 \begin{align}\label{eq:internal_jt_trans_constraint}
     \boldsymbol{\mathcal{J}}^{trans}_{i,k} \equiv \textbf{r}_{\beta_1}^{{\alpha}_1} - \textbf{r}_{\beta_k}^{\alpha_k} = \textbf{0} \quad \forall ~ k\in [2,J_i] ~\text{where}~(\alpha_k,\beta_k) = \mathcal{J}_i(k).
 \end{align}
Likewise, the relative angle between the directors of master and slave nodes is constrained as follows:
\begin{align}
    \textbf{d}^{\alpha_1}_{\beta_1,p}\cdot\textbf{d}^{\alpha_k}_{\beta_k,q} = \hat{\textbf{d}}^{\alpha_1}_{\beta_1,p}\cdot\hat{\textbf{d}}^{\alpha_k}_{\beta_k,q}  \quad \forall~ k\in [2,J_i] ~ \text{and} ~ p,q\in{1,2,3}.
\end{align}
Here $\textbf{d}^{\alpha_k}_{\beta_k,q}$ refers to the $q^{th}$ director of $k^{th}$ node at $i^{th}$ internal joint in the RVE. The above equation leads to
\begin{align}\label{eq:internl_joint_rotation_constraint1}
    (\textbf{R}^{\alpha_{k}}_{\beta_{k}})^T~\textbf{R}^{\alpha_1}_{\beta_1} &= (\hat{\textbf{R}}^{\alpha_k}_{\beta_k})^T~\hat{\textbf{R}}^{\alpha_1}_{\beta_1}
\end{align}
implying that the relative three-dimensional rotation tensor between the director frames of the master node and all slave nodes is preserved during deformation. This will also automatically ensure that the relative rotation between director frames of every pair of slave nodes is also preserved. Thus, the internal joint rotational constraint at $i^{th}$ joint is given by 
\begin{equation}\label{eq:internal_jt_rotation_constraint2}
     \boldsymbol{\mathcal{J}}^{rot}_{i,k} \equiv axial\left(\log\left(\left((\textbf{R}^{\alpha_{k}}_{\beta_{k}})^T~\textbf{R}^{\alpha_1}_{\beta_1}\right)\left((\hat{\textbf{R}}^{\alpha_k}_{\beta_k})^T~\hat{\textbf{R}}^{\alpha_1}_{\beta_1}\right)^T\right)\right)=\textbf{0}\quad \forall ~ k\in [2,J_i].
\end{equation}
Here the $axial$ function returns the axial vector of the skew-symmetric tensor. We mention here that there are no dangling rods in the rod network, i.e., both the ends of every rod belong to some joint.
\subsection{Helical boundary condition}
The idea here is to express the position and directors associated with the nodes located at the RVE's right boundary in terms of their corresponding values from the RVE's left boundary. This enforces what we call \textit{helical boundary condition}. Let us construct two sets $\mathcal{H}^R$ and $\mathcal{H}^L$ of all such nodes at the right boundary  and their counterpart nodes at the left boundary, respectively.\footnote{In case a joint lies at the right boundary, only the master node from the joint is taken in the set $\mathcal{H}^R$.} These sets are also assumed to be ordered such that the $k^{th}$ element of the sets $\mathcal{H}^L$ and $\mathcal{H}^R$ are related to each other. They are denoted as follows: $\mathcal{H}^L(k)=(\alpha_{k_L},\beta_{k_L})$, $\mathcal{H}^R(k)=(\alpha_{k_R},\beta_{k_R})$ for all $k\in\{1,2,3,.., |\mathcal{H}^L|=|\mathcal{H}^R|=N_H\}$. For the RVE in Figure \ref{fig:slender_mm_fd}, $N_H=2$ and the helical sets $\mathcal{H}^L$ and $\mathcal{H}^R$ are given by
\begin{align}
 \mathcal{H}^L = \{(3,+),(4,+)\} \quad \text{and} \quad  \mathcal{H}^R = \{(3,-),(5,-)\}.
\end{align}
As mentioned earlier, since joint $\mathcal{J}_4$ lies at the right boundary, only its master node, i.e., $(5,-)$ is considered in the helical set $\mathcal{H}^R$. We now obtain the mathematical relations which enforce \textit{helical boundary condition}. Using HCB rule \eqref{eq:hcb_rule_rod_continuum} on the left boundary nodes from the set $\mathcal{H}^L$, we have
\begin{align}\label{eq:position helical constraint}
	\textbf{r}^{1,\alpha_{k_L}}_{\beta_{k_L}} = \bigg(\int_0^{L}\textbf{R}^M(l)dl\bigg)\textbf{v}_0^M
+\textbf{R}^M(L)\textbf{r}^{\alpha_{k_L}}_{\beta_{k_L}} \quad \forall \quad k\in[1,N_H]~ \text{such that}~(\alpha_{k_L},\beta_{k_L}) = \mathcal{H}^L(k).
\end{align}
Here, the node $((1,\alpha_{k_L}),\beta_{k_L})$ is the image of the node $(\alpha_{k_L},\beta_{k_L})$ in the neighboring image $i=1$. It overlaps with the right boundary node of the RVE and forms a rigid joint with it such that
\begin{align}\label{eq:position_helical_rigid_constraint}
    \textbf{r}_{\beta_{k_R}}^{\alpha_{k_R}} = \textbf{r}^{1,\alpha_{k_L}}_{\beta_{k_L}}. 
\end{align}
Using equations \eqref{eq:position helical constraint} and \eqref{eq:position_helical_rigid_constraint}, we then get the following translational helical constraint relating the left and right boundary nodes in the two helical sets:
\begin{align}\label{eq:position_constraint_hcb}
  \boldsymbol{\mathcal{H}}_k^{trans} :=  \textbf{r}_{\beta_{k_R}}^{\alpha_{k_R}} - \bigg(\int_0^{L}\textbf{R}^M(l)dl\bigg)\textbf{v}_0^M
	&-\textbf{R}^M(L)\textbf{r}^{\alpha_{k_L}}_{\beta_{k_L}} = \textbf{0} \quad \forall~ k\in[1,N_H].
\end{align}
Likewise, the director triads of the left and right boundary nodes also need to be related. Essentially, the angle between the macroscopic directors at the RVE's left boundary and the microscopic directors of the RVE's left boundary nodes should be the same as the angle between the macroscopic directors at the left boundary of the image RVE and the microscopic directors of the image RVE's left boundary nodes, i.e.,
\begin{align}\label{director_constraint1}
	\textbf{d}_p^{M}(0) \cdot \textbf{d}^{\alpha_{k_L}}_{\beta_{k_L},q}
	= \textbf{d}_p^{M}(iL) \cdot \textbf{d}^{i,\alpha_{k_L}}_{\beta_{k_L},q} \quad \forall p,q\in1,2,3 
\end{align}
where
\begin{align}\label{macroscopic_directors}
	\textbf{d}_p^{M}(0) = \textbf{R}^{M}(0)\textbf{e}_p=\textbf{e}_p,\quad	\textbf{d}_p^{M}(iL) = \textbf{R}^{M}(iL)\textbf{e}_p
\end{align}
are the director vectors associated with the macroscopic rod's cross-section. We have taken $\textbf{R}^M(0)=\textbf{I}$ here without any loss of generality. The above two equations then yield
\begin{align}\label{rotation_constraint1}
	\textbf{R}^{i,\alpha_{k_L}}_{\beta_{k_L}}
	  = \textbf{R}^{M}(iL)~\textbf{R}^{\alpha_{k_L}}_{\beta_{k_L}}.
\end{align}
We have obtained a relation for rotation between the nodes lying on the left boundary of the RVE and the corresponding nodes lying in the $i^{th}$ image of the RVE. Let us now use the joint constraint equation \eqref{eq:internl_joint_rotation_constraint1} to relate rotation between right boundary nodes of the RVE and the left boundary nodes of the neighboring image, i.e,
\begin{align}\label{eq:rotation_helical_rigid_constraint}
\bigg(\textbf{R}^{{1,\alpha_{k_L}}}_{\beta_{k_L}}\bigg)^T	\textbf{R}_{\beta_{k_{R}}}^{\alpha_{k_{R}}} = \bigg(\hat{\textbf{R}}^{{1,\alpha_{k_L}}}_{\beta_{k_L}}\bigg)^T\hat{\textbf{R}}^{\alpha_{k_R}}_{\beta_{k_R}} \quad \forall \quad k\in[1,N_H].
\end{align}
Let us further assume without loss of generality that the metamaterial or the macroscopic rod is straight and untwisted in its reference configuration, i.e, $\hat{\textbf{R}}^M(iL)=\textbf{I}$. Hence, using \eqref{rotation_constraint1}, we can write
\begin{align}\label{eq:reference_rotation_hcb}
    \hat{\textbf{R}}^{i,\alpha_{k_L}}_{\beta_{k_L}}
	  = \hat{\textbf{R}}^{\alpha_{k_L}}_{\beta_{k_L}}.
\end{align}
Substituting equations \eqref{rotation_constraint1} and \eqref{eq:reference_rotation_hcb} into \eqref{eq:rotation_helical_rigid_constraint}, we then obtain
\begin{align}\label{eq:helical_constraint_rotmat_eq}
	&\textbf{R}^{\alpha_{k_R}}_{\beta_{k_R}} 
 = \textbf{R}^M(L)~\textbf{R}^{\alpha_{k_L}}_{\beta_{k_L}}~
\bigg(\hat{\textbf{R}}^{{\alpha_{k_L}}}_{\beta_{k_L}}\bigg)^T\hat{\textbf{R}}^{\alpha_{k_R}}_{\beta_{k_R}}\nonumber\\
&\Rightarrow \textbf{R}^{\alpha_{k_R}}_{\beta_{k_R}}\bigg(\hat{\textbf{R}}^{\alpha_{k_R}}_{\beta_{k_R}}\bigg)^T = \textbf{R}^M(L)~\textbf{R}^{\alpha_{k_L}}_{\beta_{k_L}}~
\bigg(\hat{\textbf{R}}^{{\alpha_{k_L}}}_{\beta_{k_L}}\bigg)^T.
 \end{align}
 The above equation can be interpreted as follows: the relative rotation between current and reference states of microscopic directors at right boundary nodes differs from those of microscopic directors at left boundary nodes by macroscopic rotation $\textbf{R}^M(L)$. The rotational helical boundary constraint then becomes
 \begin{equation}\label{eq:rotation_constraint_hcb}
    \boldsymbol{\mathcal{H}}_{k}^{rot} \equiv axial\left(\log\left(\left(\textbf{R}^{\alpha_{k_R}}_{\beta_{k_R}}\left(\hat{\textbf{R}}^{\alpha_{k_R}}_{\beta_{k_R}}\right)^T\right)\left(\textbf{R}^M(L)~\textbf{R}^{\alpha_{k_L}}_{\beta_{k_L}}~
\left(\hat{\textbf{R}}^{{\alpha_{k_L}}}_{\beta_{k_L}}\right)^T\right)^T\right)\right)=0.
\end{equation}
Equations \eqref{eq:position_constraint_hcb} and \eqref{eq:rotation_constraint_hcb} together form the helical boundary condition constraints for the RVE problem. It is through these constraints that the cognizance of macroscopic rod's strain parameters $(\textbf{v}_0,\textbf{k}_0)$ is taken while solving the microscopic RVE problem. Hence, they can also be called the macro-to-micro transition maps in this theory. These maps are nonlinear, geometrically exact and hold even for large values of strain parameters $(\textbf{v}_0,\textbf{k}_0)$. In contrast, the macro-to-micro transition map proposed in \citet{saadat2023mixed} holds only when the rod's strain parameters are small enough. Also note that here we have determined the helical boundary condition constraints purely based on kinematics - similar to the approach in \cite{vigliotti2012stiffness} for three-dimensional homogenization of rod networks. This is in contrast with other approaches where matching of three-dimensional macroscopic strain with RVE-averaged three-dimensional microscopic strain together with the Hill-Mandel condition is used to derive periodic boundary condition \citep{saeb2016aspects,saadat2023mixed, zhi2023multiscale,gruttmann2024fe2}. In another approach, \citet{herrnbock2022homogenization} use matching of macroscopic stress tensor with RVE-averaged microscopic stress tensor together with the Hill-Mandel condition to derive periodic boundary condition.
\subsection{Global constraints}\label{sec:global constraint}
\noindent In order to restrict rigid translation and rigid rotation of the RVE about the helical axis as well as to keep the imposed macroscopic strain parameters ($\text{v}_0^M,\text{k}_0^M$) unchanged during energy minimization, we also need to apply certain global constraints \citep{kumar2016helical}. Basically, one has to ensure that the centerline position of the macroscopic rod at $s^M=0$ coincides with the origin and the rotation of the macroscopic rod's cross-section at $s^M=0$ is identity in order to satisfy the macroscopic helical rod equation \eqref{eq:macro_centreline}.  Microscopically, at the RVE level, this is enforced by ensuring that the centroid of the material points on the left boundary of the RVE remains at origin and the left face of the RVE remains oriented in ($\textbf{e}_1-\textbf{e}_2$) plane in an average sense.\footnote{In our numerical experiments, we note that applying these global constraints on the left face alone is enough. If one also includes internal nodes of the RVE in these global constraints, the global constraint is found to be too restrictive leading to less accurate response.} Let $\mathcal{G}$ denote the set of all nodes lying on the left boundary of the RVE. In case, an internal joint lies on the left boundary, only the master node from the joint is considered in the set $\mathcal{G}$. It is also possible that a full rod lies on the left boundary in which case even the internal nodes of this rod would lie in the set $\mathcal{G}$. For the RVE in Figure \ref{fig:slender_mm_fd}, this set can be written as follows:
\begin{equation}
    \mathcal{G}=\{(4,+),(3,+)\}+~\text{all internal nodes of rod \#2}
\end{equation}
The number of elements in this set will be denoted by $N_G$. In order to fix the center of all the nodes in $\mathcal{G}$ to be at the origin, we have the following constraint:
\begin{align}\label{eq:constraint_mass_centre_continuum}
    \boldsymbol{\mathcal{G}}^{trans} \equiv \sum_{\substack{k=1\\(\alpha_k,\beta_k)=\mathcal{G}(k)}}^{N_G} \textbf{r}^{\alpha_{k}}_{\beta_{k}} = \textbf{0}.
 \end{align}
 To set the orientation of the left face of the RVE, we simply set all the three mixed moments of area corresponding to the nodes in $\mathcal{G}$ to zero, i.e., 
\begin{align}\label{eq:mixed_moment_gc}
    \boldsymbol{\mathcal{G}}^{rot} =\sum_{\substack{k=1\\(\alpha_k,\beta_k)=\mathcal{G}(k)}}^{N_G} \boldsymbol{\mathfrak{m}}^{\alpha_k}_{\beta_k}=\textbf{0}
\end{align}
where
\begin{align}
	\boldsymbol{\mathfrak{m}}^{\alpha_k}_{\beta_k} =  r^{\alpha_k}_{\beta_k,2}~r^{\alpha_k}_{\beta_k,3}\textbf{e}_1 + r^{\alpha_k}_{\beta_k,1}~r^{\alpha_k}_{\beta_k,3}\textbf{e}_2 + r^{\alpha_k}_{\beta_k,1}~r^{\alpha_k}_{\beta_k,2}\textbf{e}_3. 
\end{align}
\cite{yeoh2022multiscale} imposed a similar constraint wherein the average moment of the microscopic and macroscopic displacements on the left boundary of a three dimensional continuum RVE were equaled. \cite{zhi2023multiscale}, in the context of homogenization of laminated thin-shell structures, extended the constraint of \cite{yeoh2022multiscale} by applying it to the volume of the RVE. This was shown to restrict the rigid body rotation arising due to shear. We call the constraint in equation \eqref{eq:mixed_moment_gc} as the ``mmm" constraint - `m' stands here for mixed moment of area. For RVEs with symmetric left face (such as in the case of a BCC lattice), one needs to modify the third constraint in \eqref{eq:mixed_moment_gc} as follows:
\begin{align}
   \mathfrak{m}^{\alpha_k}_{\beta_k,3} =
	\bigg[\arctan\bigg(\frac{r^{\alpha_k}_{\beta_k,2}}{r^{\alpha_k}_{\beta_k,1}}\bigg)-\arctan\bigg(\frac{\hat{r}^{\alpha_k}_{\beta_k,2}}{\hat{r}^{\alpha_k}_{\beta_k,1}}\bigg)\bigg].
\end{align}
This new constraint is called ``mmt" constraint which is more versatile but at the same time also more complex than ``mmm" constraint. Note that we have used positional coordinates of nodes in $\mathcal{G}$ to preserve orientation of the RVE. However, we also have directors associated with every node which represent microscopic rotations within the RVE. One could also think of tying the average of these micro-rotations with macroscopic orientation. This could be done by ensuring that the average change in rotation of the directors of nodes in set $\mathcal{G}$ must be zero, i.e.,
\begin{align}\label{eq:delta_theta_constraint}
        \boldsymbol{\mathcal{G}}^{rot} = \sum_{k=1}^{N_G}\Delta\boldsymbol{\theta}^{\alpha_k}_{\beta_k}=\textbf{0}.
\end{align}
Here the rotation vector $\Delta\boldsymbol{\theta}^{\alpha_k}_{\beta_k}$ is defined as follows:
\begin{align}\label{delta_theta_spatial}
    \textbf{R}(\boldsymbol{\theta}^{\alpha_k}_{\beta_k}) = \exp^{\Delta\boldsymbol{\Theta}^{\alpha_k}_{\beta_k}}~{\textbf{R}}(\hat{\boldsymbol{\theta}}^{\alpha_k}_{\beta_k}).
\end{align}
We call this constraint as ``delta-theta" constraint. One could also define $\Delta\boldsymbol{\theta}^{\alpha_k}_{\beta_k}= \boldsymbol{\theta}^{\alpha_k}_{\beta_k}-\hat{\boldsymbol{\theta}}^{\alpha_k}_{\beta_k}$ as in \citet{abdoul2018strain} and \citet{vcanic2022optimal}.\footnote{Although the change in rotation is being measured here with respect to a fixed reference state, the same could also be measured with respect to a converged solution, say the solution in the previous load step.} These two definitions become the same for planar rotations as planar rotations are additive. In a departure from kinematics-based constraints, \cite{klarmann2020homogenization,gruttmann2024fe2} have used a stress-based approach to suppress the rigid rotation of the RVE. In this work, we use the ``delta-theta" constraint given by \eqref{eq:delta_theta_constraint}. 
\subsection{Constrained minimization of RVE energy}\label{sec:rve_problem}
To solve the minimization problem \eqref{eq:rve_warping_problem} in the presence of constraints discussed above, we define the following constrained energy functional:
\begin{align}\label{eq:energy_functional_constraind}
    \mathcal{E}^{cons} = \mathcal{E}^{RVE} + \mathcal{C}^G + \mathcal{C}^H + \mathcal{C}^J
\end{align}
wherein the constraint terms are as follows:
\begin{align}\label{eq:constraints_terms}
    \mathcal{C}^G =
    \begin{bmatrix}
        \boldsymbol{\lambda} \\
        \boldsymbol{\mu}
    \end{bmatrix}\cdot
    \begin{bmatrix}
        \boldsymbol{\mathcal{G}}^{trans} \\
        \boldsymbol{\mathcal{G}}^{rot}
    \end{bmatrix},~
    \mathcal{C}^H = \sum_{k=1}^{N_H}
    \begin{bmatrix}
        \textbf{h}^n_k \\
        \textbf{h}^m_k
    \end{bmatrix}\cdot
    \begin{bmatrix}
        \boldsymbol{\mathcal{H}}_{k}^{trans} \\
        \boldsymbol{\mathcal{H}}_{k}^{rot}
    \end{bmatrix},~
    \mathcal{C}^J =\sum_{i=1}^{N_{J}}\sum_{k=2}^{J_i} \begin{bmatrix}
        \textbf{J}^n_{i,k} \\
        \textbf{J}^m_{i,k}
    \end{bmatrix}\cdot
    \begin{bmatrix}
        \boldsymbol{\mathcal{J}}_{i,k}^{trans} \\
        \boldsymbol{\mathcal{J}}_{i,k}^{rot}
    \end{bmatrix}.
\end{align}
Here $\left((\boldsymbol{\lambda},\boldsymbol{\mu}), (\textbf{h}^n_k,\textbf{h}^m_k), (\textbf{J}^n_{i,k},~\textbf{J}^m_{i,k})\right)$ are the three sets of unknown Lagrange multipliers to enforce global constraint, helical boundary constraint and internal joint constraints, respectively. The first variation of the above functional is obtained by varying the unknowns as follows:
\begin{align}\label{eq:perturb_config}
    \textbf{r}^{\alpha}_{\epsilon}(s) &= \textbf{r}^{\alpha}(s^{\alpha}) + \epsilon \delta \textbf{r}^{\alpha}(s^{\alpha}),\quad \quad \textbf{R}^{\alpha}_{\epsilon}(s^{\alpha}) = e^{\epsilon\delta\boldsymbol{\Theta}^{\alpha}(s^{\alpha})}\textbf{R}^{\alpha}(s^{\alpha}),\nonumber\\
    \boldsymbol{\lambda}_{\epsilon} &= \boldsymbol{\lambda} + \epsilon \delta \boldsymbol{\lambda},\quad \quad \quad \quad \quad \quad\boldsymbol{\mu}_{\epsilon} = \boldsymbol{\mu} + \epsilon \delta \boldsymbol{\mu},\nonumber\\
    \textbf{h}_{k}^n &= \textbf{h}_k^n + \epsilon \delta \textbf{h}_k^n, \quad \quad \quad \quad \quad{\textbf{h}}_{k}^m = \textbf{h}_k^m + \epsilon \delta \textbf{h}_k^m,\nonumber\\
    \textbf{J}_{i,k}^n &= \textbf{J}_{i,k}^n + \epsilon \delta \textbf{J}_{i,k}^n, \quad \quad\quad\quad~ \textbf{J}_{i,k}^m = \textbf{J}_{i,k}^m + \epsilon \delta \textbf{J}_{i,k}^m.
\end{align}
For the variation of constraint terms in \eqref{eq:constraints_terms}, we refer the readers to \ref{appendix:contraint_linearization}. The first variation of the constrained energy functional for prescribed macroscopic strains $(\textbf{v}^M_0,\textbf{k}^M_0)$ yields the following weak form:
\begingroup
\allowdisplaybreaks
\begin{align}\label{eq:weak_form_continuum}
    G \equiv 
    \frac{d}{d\epsilon}\mathcal{E}^{cons}\bigg |_{\epsilon=0} &= \frac{d\mathcal{E}^{RVE}}{d\epsilon}\bigg |_{\epsilon=0} + \frac{d\mathcal{C}^{G}}{d\epsilon}\bigg |_{\epsilon=0} + \frac{d\mathcal{C}^{H}}{d\epsilon}\bigg |_{\epsilon=0} + \frac{d\mathcal{C}^{J}}{d\epsilon}\bigg |_{\epsilon=0}=0 \nonumber\\
    &= \sum_{\alpha=1}^{N_{rods}}\int_0^{L^{\alpha}}(\textbf{n}^{\alpha}\cdot(\delta\textbf{r}^{\alpha})^{\prime} + 
                (\textbf{n}^{\alpha}\times(\textbf{r}^{\alpha})^{\prime})\cdot\delta\boldsymbol{\theta}^{\alpha} +\textbf{m}^{\alpha}\cdot(\delta\boldsymbol{\theta}^{\alpha})^{\prime})ds^{\alpha}\nonumber\\
                 & \quad+\sum_{k=1}^{N_{B}}\begin{bmatrix}
        \boldsymbol{\lambda} \\
        \boldsymbol{\Xi}^{\boldsymbol{\theta},\alpha_k}_{\beta_k}\boldsymbol{\mu}
    \end{bmatrix}\begin{bmatrix}
        \delta\textbf{r}^{\alpha_{k}}_{\beta_{k}} \\
        \delta\boldsymbol{\theta}^{\alpha_k}_{\beta_k}
    \end{bmatrix} + \begin{bmatrix}
        \boldsymbol{\mathcal{G}}^{trans} \\
        \boldsymbol{\mathcal{G}}^{rot}
    \end{bmatrix}\cdot
    \begin{bmatrix}
        \delta\boldsymbol{\lambda} \\
        \delta\boldsymbol{\mu}
    \end{bmatrix} \nonumber\\
    &\quad+ \sum_{k=1}^{N_H}\bigg\{\begin{bmatrix}
        \boldsymbol{\mathcal{H}}^{trans}_{k} \\
                \boldsymbol{\mathcal{H}}^{rot}_{k}
    \end{bmatrix}\cdot\begin{bmatrix}
        \delta \textbf{h}^n_{k} \\
        \delta \textbf{h}^m_{k}
    \end{bmatrix} + \begin{bmatrix}
        \textbf{h}^n_{k} \\
        -\textbf{R}^{\alpha_{k_R}}_{\beta_{k_R}}\textbf{T}^T(\boldsymbol{\mathcal{H}}^{rot}_k)\textbf{h}^m_{k}
    \end{bmatrix}\cdot\begin{bmatrix}
        \delta\textbf{r}^{\alpha_{k_R}}_{\beta_{k_R}}\\ 
        \delta\boldsymbol{\theta}^{\alpha_{k_R}}_{\beta_{k_R}}
    \end{bmatrix}\nonumber\\
    &\qquad+ \begin{bmatrix}
        -(\textbf{R}^M)^T\textbf{h}^n_{k} \\
        (\textbf{R}^M)^T\textbf{R}^{\alpha_{k_R}}_{\beta_{k_R}}\textbf{T}^T(\boldsymbol{\mathcal{H}}^{rot}_k)\textbf{h}^m_{k}
    \end{bmatrix}\cdot\begin{bmatrix}
        \delta\textbf{r}^{\alpha_{k_L}}_{\beta_{k_L}} \\ 
        \delta\boldsymbol{\theta}^{\alpha_{k_L}}_{\beta_{k_L}}
    \end{bmatrix}\bigg\}\nonumber\\
   &\quad+ \sum_{i=1}^{N_J}\sum_{k=2}^{J_i}\bigg\{\begin{bmatrix}
        \boldsymbol{\mathcal{J}}^{trans}_{i,k} \\
                \boldsymbol{\mathcal{J}}^{rot}_{i,k}
    \end{bmatrix}\cdot\begin{bmatrix}
        \delta \textbf{J}^n_{i,k} \\
        \delta \textbf{J}^m_{i,k}
    \end{bmatrix} + \begin{bmatrix}
        \textbf{J}^n_{i,k} \\
        [\textbf{R}^{\alpha_k}_{\beta_k}\textbf{T}^T(\boldsymbol{\mathcal{J}}^{rot}_{i,k})]\textbf{J}^m_{i,k}
    \end{bmatrix}\cdot\begin{bmatrix}
        \delta\textbf{r}^{\alpha_1}_{\beta_1} \\
        \delta\boldsymbol{\theta}^{\alpha_1}_{\beta_1}
    \end{bmatrix} \nonumber\\
    &\qquad+ \begin{bmatrix}
        -\textbf{J}^n_{i,k} \\
        -[\textbf{R}^{\alpha_k}_{\beta_k}\textbf{T}^T(\boldsymbol{\mathcal{J}}^{rot}_{i,k})]\textbf{J}^m_{i,k}
    \end{bmatrix}\cdot\begin{bmatrix}
        \delta\textbf{r}^{\alpha_k}_{\beta_k} \\
        \delta\boldsymbol{\theta}^{\alpha_k}_{\beta_k}
    \end{bmatrix}\bigg\} = 0.
\end{align}
\endgroup
In order to obtain the Euler-Lagrange equations, one simply needs to do integration by parts in the integral terms above. This will yield the usual balance equations (see equation \eqref{global_eq}) for every rod in the RVE along with boundary conditions. These boundary conditions will also contain the unknown Lagrange multipliers for which the earlier derived joint constraint, helical boundary constraint and global constraint equations will be required. The weak form \eqref{eq:weak_form_continuum} takes care of all of these by the way which could be suitably discretized and solved using standard numerical solvers. In this work, the energy of each constituent rod within the RVE is approximated using a discrete rod formulation as described in section \ref{discrete rod theory}. After solving the weak form \eqref{eq:weak_form_continuum}, we obtain the solution of the RVE warping problem in \eqref{eq:rve_warping_problem} and the rod's energy density can then be written as
 \begin{align}\label{eq:macro_energy}
     \Phi^M(\textbf{v}_0^M,\textbf{k}_0^M) = \frac{1}{L} \mathcal{E}^{RVE}\left(\tilde{\textbf{r}}^{\alpha}(\textbf{v}_0^M,\textbf{k}_0^M),\tilde{\boldsymbol{\theta}}^{\alpha}(\textbf{v}_0^M,\textbf{k}_0^M)\right)
 \end{align}
where $\{\tilde{\textbf{r}}^{\alpha}(\textbf{v}_0^M,\textbf{k}_0^M),\tilde{\boldsymbol{\theta}}^{\alpha}(\textbf{v}_0^M,\textbf{k}_0^M)\}_{\alpha=1}^{N_{rods}}$ denotes the configuration of the RVE at equilibrium. As the constraints get automatically satisfied at equilibrium, one can also write 
\begin{align}\label{eq:macro_energy_constrained}
    \Phi^M(\textbf{v}_0^M,\textbf{k}_0^M) = \frac{1}{L}\tilde{\mathcal{E}}^{cons}
\end{align}
which we will use to derive expressions of macroscopic internal contact force, moment and stiffnesses in the next section. Henceforth, we omit the superscript $\tilde{(\cdot)}$ for the sake of brevity.
\section{Expressions of internal contact force, moment and stiffnesses of the macroscopic rod}\label{stiffnesses}
As depicted in Figure \ref{fig:slender_mm}, once the RVE problem is solved at microscale, one needs to go back to macroscale with the information of the rod's stress resultants and stiffnesses whose expressions we derive now. The macroscopic internal force and moment are given by
\begin{align}\label{eq:macroscopic_force_moment}
    \begin{bmatrix}
        \textbf{n}_0^M \\
        \textbf{m}_0^M
    \end{bmatrix} &= \frac{\partial \Phi^M}{\partial [\textbf{v}_0^M,\textbf{k}_0^M]^T}.
    \end{align}
We can substitute \eqref{eq:macro_energy_constrained} in the above equation to obtain the partial derivatives which we denote succinctly as $\delta \equiv \frac{\partial}{\partial p^M}$ where $p^M \in \{\text{v}^M_1,\text{v}^M_2, \text{v}^M_3,\kappa^M_1,\kappa^M_2, \kappa^M_3\}$. First, using \eqref{eq:energy_functional_constraind}, we obtain
\begingroup
\allowdisplaybreaks
\begin{align}
    \delta \mathcal{E}^{cons} &=  \sum_{\alpha=1}^{N_{rods}}\int_0^{L^{\alpha}}(\textbf{n}^{\alpha}\cdot(\delta\textbf{r}^{\alpha})^{\prime} + 
                (\textbf{n}^{\alpha}\times(\textbf{r}^{\alpha})^{\prime})\cdot\delta\boldsymbol{\theta}^{\alpha} +\textbf{m}^{\alpha}\cdot(\delta\boldsymbol{\theta}^{\alpha})^{\prime})ds^{\alpha}\nonumber\\
                 &\quad+ \sum_{k=1}^{N_{B}}\begin{bmatrix}
        \boldsymbol{\lambda} \\
        \boldsymbol{\Xi}^{\boldsymbol{\theta},\alpha_k}_{\beta_k}\boldsymbol{\mu}
    \end{bmatrix}\begin{bmatrix}
        \delta\textbf{r}^{\alpha_{k}}_{\beta_{k}} \\
        \delta\boldsymbol{\theta}^{\alpha_k}_{\beta_k}
    \end{bmatrix} + \begin{bmatrix}
        \boldsymbol{\mathcal{G}}^{trans} \\
        \boldsymbol{\mathcal{G}}^{rot}
    \end{bmatrix}\cdot
    \begin{bmatrix}
        \delta\boldsymbol{\lambda} \\
        \delta\boldsymbol{\mu}
    \end{bmatrix} \nonumber\\
    &\quad+ \sum_{k=1}^{N_H}\bigg\{\begin{bmatrix}
        \boldsymbol{\mathcal{H}}^{trans}_{k} \\
                \boldsymbol{\mathcal{H}}^{rot}_{k}
    \end{bmatrix}\cdot\begin{bmatrix}
        \delta \textbf{h}^n_{k} \\
        \delta \textbf{h}^m_{k}
    \end{bmatrix} + \begin{bmatrix}
        \textbf{h}^n_{k} \\
        -\textbf{R}^{\alpha_{k_R}}_{\beta_{k_R}}\textbf{T}^T(\boldsymbol{\mathcal{H}}^{rot}_k)\textbf{h}^m_{k}
    \end{bmatrix}\cdot\begin{bmatrix}
        \delta\textbf{r}^{\alpha_{k_R}}_{\beta_{k_R}}\\ 
        \delta\boldsymbol{\theta}^{\alpha_{k_R}}_{\beta_{k_R}}
    \end{bmatrix}\nonumber\\
    &\qquad+ \begin{bmatrix}
        -(\textbf{R}^M)^T\textbf{h}^n_{k} \\
        (\textbf{R}^M)^T\textbf{R}^{\alpha_{k_R}}_{\beta_{k_R}}\textbf{T}^T(\boldsymbol{\mathcal{H}}^{rot}_k)\textbf{h}^m_{k}
    \end{bmatrix}\cdot\begin{bmatrix}
        \delta\textbf{r}^{\alpha_{k_L}}_{\beta_{k_L}} \\ 
        \delta\boldsymbol{\theta}^{\alpha_{k_L}}_{\beta_{k_L}}
    \end{bmatrix}\nonumber\\
    &\qquad +\textcolor{blue}{\begin{bmatrix}
        -\textbf{h}_k^n \\
        \textbf{h}_k^m    \end{bmatrix}
        \cdot
        \begin{bmatrix}
        \left(\int_0^L\delta\textbf{R}^M(l)dl\right)~\textbf{v}^M_0 + \int_0^L\textbf{R}^M(l)dl~\delta\textbf{v}^M_0+ \delta \textbf{R}^M(L)\textbf{r}_{{\beta}_{k_L}}^{{\alpha}_{k_L}}\\
        \textbf{T}(\boldsymbol{\mathcal{H}}^{rot}_k)(\textbf{R}^{\alpha_{k_R}}_{\beta_{k_R}})^T\text{axial}(\delta \textbf{R}^M(\textbf{R}^{M})^T)
    \end{bmatrix}}\bigg\}\nonumber\\
   &\quad+ \sum_{i=1}^{N_J}\sum_{k=2}^{J_i}\bigg\{\begin{bmatrix}
        \boldsymbol{\mathcal{J}}^{trans}_k \\
                \boldsymbol{\mathcal{J}}^{rot}_k
    \end{bmatrix}\cdot\begin{bmatrix}
        \delta \textbf{J}^n_{k} \\
        \delta \textbf{J}^m_{k}
    \end{bmatrix} + \begin{bmatrix}
        \textbf{J}^n_{i,k} \\
        \textbf{R}^{\alpha_k}_{\beta_k}\textbf{T}^T(\boldsymbol{\mathcal{J}}^{rot}_{i,k})
    \end{bmatrix}\cdot \begin{bmatrix}
        \delta \textbf{r}^{\alpha_1}_{\beta_1} \\
        \delta \boldsymbol{\theta}^{\alpha_1}_{\beta_1}
    \end{bmatrix}\nonumber\\
    &\quad-\begin{bmatrix}
        \textbf{J}^n_{i,k} \\
        [\textbf{R}^{\alpha_k}_{\beta_k}\textbf{T}^T(\boldsymbol{\mathcal{J}}^{rot}_{i,k})]\textbf{J}^m_{i,k}
    \end{bmatrix}\cdot \begin{bmatrix}
        \delta \textbf{r}^{\alpha_{k}}_{\beta_{k}} \\
        \delta \boldsymbol{\theta}^{\alpha_k}_{\beta_k}
    \end{bmatrix}\bigg\}.
\end{align}
\endgroup
Note that the above equation consists of both explicit (blue terms) and implicit (black terms) derivatives of constrained energy with respect to macroscopic strain $p^M$. The explicit derivative terms arise only from helical boundary constraints. The closed form expressions for the terms  $\int \textbf{R}^Mdl$, $\delta \textbf{R}^M$ and $\int \delta\textbf{R}^Mdl$, wherein $\delta=\frac{\partial}{\partial \kappa_i^M}$, can be seen in Appendix B of \citet{kumar2016helical}. At equilibrium, using \eqref{eq:weak_form_continuum}, the implicit derivative terms vanish as a whole. Therefore, we get
\begin{align}\label{eq:del_e_cons_del_macro_strains}
    \delta \mathcal{E}^{cons} =   \textcolor{blue}{\sum_{k=1}^{N_H}\begin{bmatrix}
        -\textbf{h}_k^n \\
        \textbf{h}_k^m    \end{bmatrix}
        \cdot
        \begin{bmatrix}
        \left(\int_0^L\delta\textbf{R}^M(l)dl\right)~\textbf{v}^M_0 + \int_0^L\textbf{R}^M(l)dl~\delta\textbf{v}^M_0 + \delta \textbf{R}^M(L)\textbf{r}_{{\beta}_{k_L}}^{{\alpha}_{k_L}}\\
        (\textbf{R}^{\alpha_{k_R}}_{\beta_{k_R}})^T\text{axial}(\delta \textbf{R}^M(\textbf{R}^{M})^T)
    \end{bmatrix}.}
\end{align}
 Using equation \eqref{eq:macro_energy_constrained} and \eqref{eq:del_e_cons_del_macro_strains}, for $\delta = \frac{\partial}{\partial \text{v}^M_i}$, we get the macroscopic internal contact force, i.e.,
 \begin{align}
    \text{n}^M_i =\frac{\partial \Phi^M}{\partial \text{v}^M_i} = -\frac{1}{L}\sum_{k=1}^{N_H}\textbf{h}_k^n\cdot \left(\int_0^L\textbf{R}^M(l)dl\right)~\textbf{e}_i
 \end{align}
while for $\delta = \frac{\partial}{\partial \kappa^M_i}$, we get the macroscopic internal contact moment, i.e.,
 \begin{align}
     \text{m}^M_i = \frac{\partial \Phi^M}{\partial \kappa_i^M} &= -\frac{1}{L}\sum_{k=1}^{N_H}\textbf{h}_k^n\cdot\left[\left(\int_0^L\frac{\partial \textbf{R}^M(l)}{\partial \kappa^M_i}dl\right)\textbf{v}_0^M\right]\nonumber\\
     &+ \frac{1}{L}\sum_{k=1}^{N_H}\left(\textbf{R}^{\alpha_{k_R}}_{\beta_{k_R}}\textbf{h}_k^m+\textbf{R}^M\left({(\textbf{R}^M})^T\textbf{h}_k^n\times\textbf{r}^{{\alpha}_{k_L}}_{{\beta}_{k_L}}\right)\right)\cdot\text{axial}\left(\frac{\partial \textbf{R}^M(L)}{\partial \kappa^M_i}{(\textbf{R}^{M}})^T\right). 
 \end{align}
Next, in order to obtain the macroscopic stiffnesses, we simply need to further differentiate the above expressions of macroscopic force and moment with respect to macroscopic strain parameters, i.e.,
\begin{align}\label{eq:macro_rod_elasticity_tensor}
    \mathbb{C}_0^M = \frac{\partial^2 \Phi^M}{\partial[\textbf{v}^M_0,\textbf{k}^M_0]^T~\partial[\textbf{v}^M_0,\textbf{k}^M_0]^T}.
\end{align}
First, we differentiate the macroscopic internal force equation as follows:
\begin{align}\label{eq:C_macros_vivq}
    \frac{\partial^2 \Phi^M}{\partial \text{v}^M_i \partial \text{v}^M_q} =  -\frac{1}{L}\sum_{k=1}^{N_H}\bigg[\frac{\partial \textbf{h}^n_k}{\partial \text{v}^M_q}\bigg]\cdot\bigg(\int_0^{L}\textbf{R}^M(l)dl\bigg)\textbf{e}_i,
\end{align}
\begin{align}\label{eq:C_macros_vikq}
     \frac{\partial^2 \Phi^M}{\partial \text{v}^M_i \partial \kappa^M_q} &= -\frac{1}{L}\Bigg\{\sum_{k=1}^{N_H}\bigg[\frac{\partial \textbf{h}^n_k}{\partial \kappa^M_q}\bigg]\cdot\bigg(\int_0^{L}\textbf{R}^M(l)dl\bigg)\textbf{e}_i+\sum_{k=1}^{N_H}\textbf{h}^n_k\cdot\bigg(\int_0^{L}\frac{\partial \textbf{R}^M(l)}{\partial \kappa^M_q}dl\bigg)\textbf{e}_i\Bigg\}.
\end{align}
Similarly, we obtain the derivative of macroscopic internal moment as follows:
\begin{align}\label{eq:C_macros_kivq}
     \frac{\partial^2 \Phi^M}{\partial \kappa^M_q \partial \text{v}^M_i} &= \frac{1}{L}\Bigg\{-\sum_{k=1}^{N_H}\frac{\partial \textbf{h}^n_k}{\partial \text{v}^M_i}\cdot\bigg(\int_0^{L}\frac{\partial \textbf{R}^M(l)}{\partial \kappa^M_q}dl\bigg)\textbf{v}_0^M-\sum_{k=1}^{N_H}\textbf{h}^n_{k}\cdot \bigg(\int_0^{L}\frac{\partial \textbf{R}^M(l)}{\partial \kappa^M_q}dl\bigg)\textbf{e}_i\nonumber\\
     &+  \sum_{k=1}^{N_H}\bigg[\frac{\partial \boldsymbol{\theta}^{\alpha_{k_R}}_{\beta_{k_R}}}{\partial \text{v}^M_i}\times(\textbf{R}^{\alpha_{k_R}}_{\beta_{k_R}}\textbf{h}^m_k) + \textbf{R}^{\alpha_{k_R}}_{\beta_{k_R}}\frac{\partial \textbf{h}^m_k}{\partial \text{v}^M_i}\nonumber\\
     &+\textbf{R}^M\bigg((\textbf{R}^M)^T\frac{\partial \textbf{h}^n_k}{\partial \text{v}^M_i}\times\textbf{r}^{\alpha_{k_L}}_{\beta_{k_L}}+(\textbf{R}^M)^T\textbf{h}^n_{k}\times\frac{\partial \textbf{r}^{\alpha_{k_L}}_{\beta_{k_L}}}{\partial \text{v}^M_i}\bigg)\bigg]\cdot\text{axial}\left(\frac{\partial \textbf{R}^M(L)}{\partial \kappa^M_q}{\textbf{R}^{M}}^T\right)\Bigg\},
\end{align}
\resizebox{1.0\textwidth}{!}{
\begin{minipage}{\linewidth}
\begin{align}\label{eq:C_macros_kikq}
    \frac{\partial^2 \Phi^M}{\partial \kappa^M_i \partial \kappa^M_q} &= \frac{1}{L}\Bigg\{-\sum_{k=1}^{N_H}\frac{\partial \textbf{h}^n_k}{\partial \kappa^M_q}\cdot\bigg(\int_0^{L}\frac{\partial \textbf{R}^M(l)}{\partial k^M_i}dl\bigg)\textbf{v}_0^M-\sum_{k=1}^{N_H}\textbf{h}^n_{k}\cdot \bigg(\int_0^{L}\frac{\partial \textbf{R}^M(l)}{\partial \kappa^M_i\partial \kappa^M_q}dl\bigg)\textbf{v}_0^M\nonumber\\
    &+\sum_{k=1}^{N_H}\bigg[\frac{\partial \boldsymbol{\theta}^{\alpha_{k_R}}_{\beta_{k_R}}}{\partial \kappa^M_q}\times(\textbf{R}^{\alpha_{k_R}}_{\beta_{k_R}}\textbf{h}^m_k)+\textbf{R}^{\alpha_{k_R}}_{\beta_{k_R}}\frac{\partial \textbf{h}^m_k}{\partial \kappa^M_q}+\frac{\partial \textbf{R}^M}{\partial \kappa^M_q}[(\textbf{R}^M)^T\textbf{h}^n_{k}\times\textbf{r}^{\alpha_{k_L}}_{\beta_{k_L}}]\nonumber\\
    &+\textbf{R}^M\bigg(\frac{\partial (\textbf{R}^M)^T}{\partial \kappa^M_q}\textbf{h}^n_k\times\textbf{r}^{\alpha_{k_L}}_{\beta_{k_L}} + (\textbf{R}^M)^T\frac{\partial \textbf{h}^n_{k}}{\partial \kappa^M_q}\times\textbf{r}^{\alpha_{k_L}}_{\beta_{k_L}}+(\textbf{R}^M)^T\textbf{h}^n_{k}\times\frac{\partial \textbf{r}^{\alpha_{k_L}}_{\beta_{k_L}}}{\partial \kappa^M_q}\bigg)\bigg]\cdot \text{axial}\left(\frac{\partial \textbf{R}^M(L)}{\partial \kappa^M_i}{\textbf{R}^{M}}^T\right)\nonumber\\
    &+ \sum_{k=1}^{N_H}\bigg[\textbf{R}^{\alpha_{k_R}}_{\beta_{k_R}}\textbf{h}^m_{k}+\textbf{R}^M((\textbf{R}^M)^T\textbf{h}^n_{k}\times\textbf{r}^{\alpha_{k_L}}_{\beta_{k_L}})\bigg]\cdot\bigg[\text{axial}\bigg(\frac{\partial^2 \textbf{R}^M}{\partial \kappa^M_i\partial \kappa^M_q}(\textbf{R}^M)^T+\frac{\partial \textbf{R}^M}{\partial \kappa^M_i}\frac{\partial (\textbf{R}^M)^T}{\partial \kappa^M_q}\bigg)\bigg]\Bigg\}.
\end{align}
\end{minipage}%
}
As these expressions are obtained by taking the derivative of the same scalar energy function $\Phi^M$, the mixed derivatives in \eqref{eq:C_macros_vikq} and \eqref{eq:C_macros_kivq}, although appearing to have different mathematical expressions, actually yield the same value. The above stiffness expressions also contain the derivatives of the variables $(\textbf{r}^{\alpha_{k_L}}_{\beta_{k_L}},\boldsymbol{\theta}^{\alpha_{k_L}}_{\beta_{k_L}},\textbf{r}^{\alpha_{k_R}}_{\beta_{k_R}},\boldsymbol{\theta}^{\alpha_{k_R}}_{\beta_{k_R}},\textbf{h}^n_k,\textbf{h}^m_k)$ with respect to strain parameters $\text{v}^M_q$ and $\kappa^M_q$. These unknown derivatives are obtained by following a standard procedure of taking the derivative of the weak form \eqref{eq:weak_form_continuum} with respect to each of the macroscopic strain parameters $q^M$ $\in\{\text{v}^M_1,\text{v}^M_2,\text{v}^M_3,\kappa^M_1,\kappa^M_2,\kappa^M_3\}$ one-by-one (see section 5.2 in \citet{arora2019computational} for details of this procedure). This makes the nonlinear weak form \eqref{eq:weak_form_continuum} linear in those unknown derivatives. An appropriate discretization then yields the following linear matrix equation for the unknowns:
\begin{align}
    \underbrace{\mathbb{K}~\Delta_{q^M}}_{\text{implicit}} + \underbrace{\mathbb{R}_{q^M}}_{\text{explicit}} = \textbf{0}.
\end{align}
The first term in the above equation generates from the implicit derivative of the weak form with respect to strain parameters while the second term generates from the explicit derivative. The unknown vector $\Delta_{q^M}$ in the above equation is the collection of the derivative of the RVE unknowns $
(\textbf{r},\boldsymbol{\theta}),(\boldsymbol{\lambda},\boldsymbol{\mu}), (\textbf{h}^n_k,\textbf{h}^m_k), (\textbf{J}^n_{i,k},~\textbf{J}^m_{i,k})$ with respect to the strain parameter $q^M$ at each nodal point. In the next section, we discuss a scheme to discretize the rod's energy.

\pagestyle{plain}
\section{Discrete rod formulation}\label{discrete rod theory}
\begin{figure}[h!]
\begin{center}
\includegraphics[width = 0.9\textwidth]{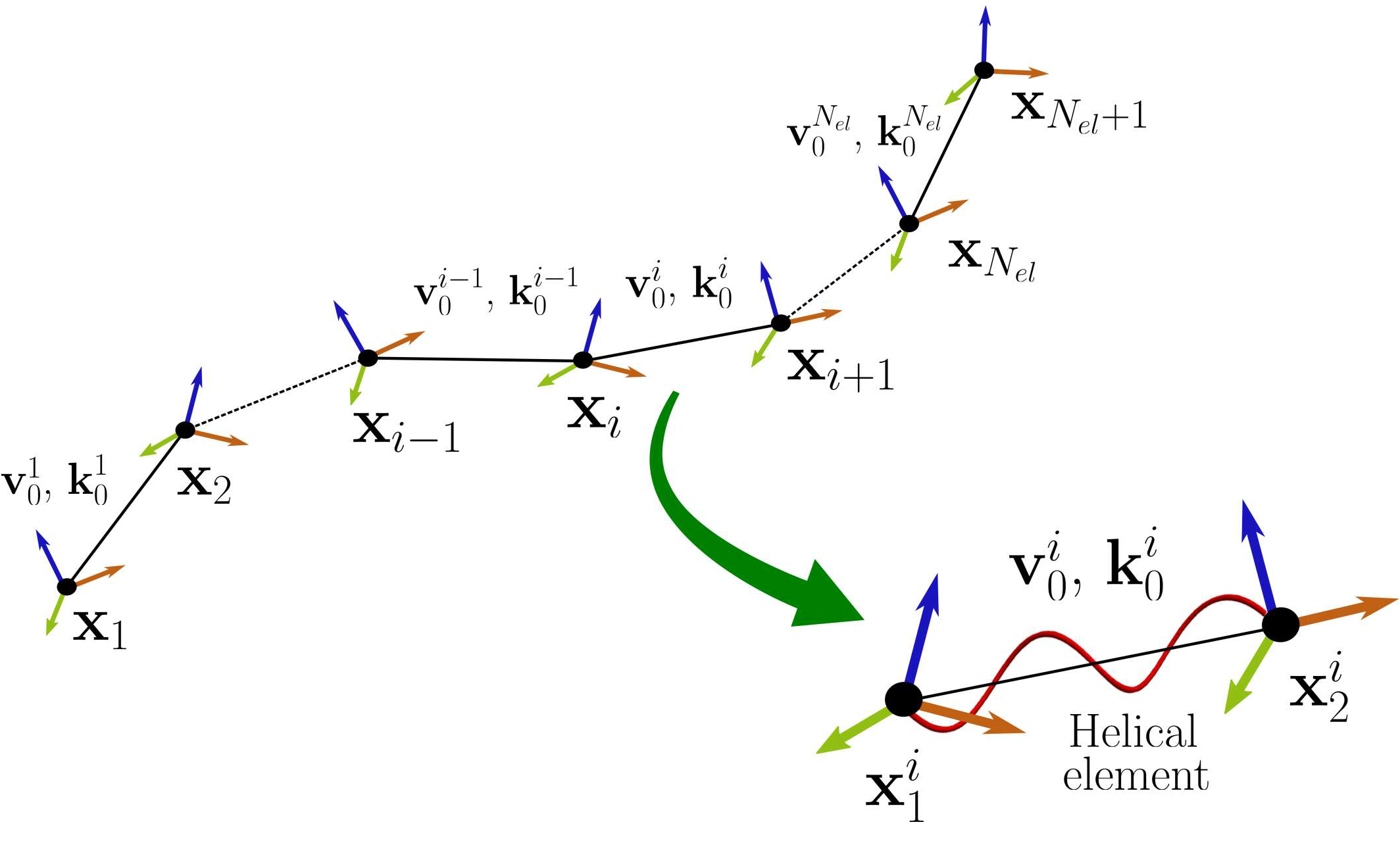}
\end{center}
\caption{A schematic of a discrete helical rod: each segment is assumed to be subjected to different strain set (but uniform within the segment) making the segment helical and the entire rod as a set of distinct helices joined together.}
\label{discrete helical schematic}
\end{figure}

In this section, we present a scheme for discretizing the rod's energy. It results in a locking-free element. The approach is similar to the $SE(3)$ framework presented in \cite{sonneville2014geometrically}. A rod can be thought of as a set of discrete points with their positions $\textbf{r} = \{\textbf{r}_1$, $\textbf{r}_2$,.., $\textbf{r}_i$,.., $\textbf{r}_{{N_{el}}+1}\}$ denoting the deformed centerline of a rod. Here $N_{el}$ is the total number of elements in which the rod is discretized and $\textbf{r}_i$ denotes the position of the rod's centerline at arclength coordinate $s_i$, i.e., $\textbf{r}_i = \textbf{r}(s_i)$. The arc-length coordinates need not be equispaced. A set of orthonormal director frames $\{\textbf{d}_{I,1}$, $\textbf{d}_{I,2}$,...,$\textbf{d}_{I,i}$,...,$\textbf{d}_{I,N_{el}+1}\}_{I=1}^3$ is further attached to those points to describe the orientation of deformed cross-sections (see Figure \ref{discrete helical schematic}). As mentioned earlier, the director frames can also be described by rotation vectors, i.e.,  $\{\boldsymbol{\theta}_1$, $\boldsymbol{\theta}_2$,..., $\boldsymbol{\theta}_{N_{el}+1}\}$. Thus, a discrete rod can be represented by the generalized coordinate $\textbf{x}=\{\textbf{x}_1,\textbf{x}_2,...,\textbf{x}_{N_{el}+1}\}$ where $\textbf{x}_i=\{\textbf{r}_i,\boldsymbol{\theta}_i\}$. One could also view a discrete rod to be a set of $N_{el}$ segments in which the $i^{th}$ segment is prescribed by its two end points given by its local positions $\textbf{r}_1^i$ (globally $\textbf{r}_{i}$) and $\textbf{r}_2^i$ (globally $\textbf{r}_{i+1}$), and local cross-section orientations $\boldsymbol{\theta}_1^i$ (globally $\boldsymbol{\theta}_{i}$) and $\boldsymbol{\theta}_2^i$ (globally $\boldsymbol{\theta}_{i+1}$). 
Considering uniform strain field $(\textbf{v}_0^i,\textbf{k}_0^i)$ in the $i^{th}$ segment, each segment becomes helical as per formula \eqref{eq:macro_centreline} due to which we also call them helical elements. The two end points of a segment then get related as follows: 
\begin{align}\label{r2-r1-1}
    \textbf{r}_2^i 
    &= \textbf{r}_1^i +\textbf{R}_1^i\left[\frac{\textbf{v}_0^i\cdot\textbf{k}_0^i}{||\textbf{k}_0^i||^2}~h^i\textbf{k}_0^i + (\textbf{I}-e^{h^i\textbf{K}_0^i})\frac{\textbf{K}_0^i\textbf{v}_0^i}{||\textbf{k}_0^i||^2}\right]\quad\text{(see \citet{kumar2016helical})}\nonumber\\
    \Rightarrow\frac{\textbf{r}_2^i-\textbf{r}_1^i}{h^i} 
    &=  \textbf{R}^i_1\left[\textbf{I}+\left(\frac{\textbf{K}^{i}_0}{||\textbf{k}_0||}\right)^2-\left(\alpha(h^i\textbf{k}^i_0)\textbf{K}^i_0+\frac{\beta(h^i\textbf{k}_0^i)}{2} h^i\textbf{K}^{i^2}_0\right)\frac{\textbf{K}^i_0}{||\textbf{k}_0||^2}\right]\textbf{v}_0^i\nonumber\\
    &=  \textbf{R}^i_1\left[\textbf{I}+\frac{\beta(h^i\textbf{k}_0^i)}{2}h^i\textbf{K}^i_0+\left(1-\alpha(h^i\textbf{k}_0^i)\right)\left(\frac{\textbf{K}^{i}_0}{||\textbf{k}_0||}\right)^2\right]\textbf{v}_0^i\nonumber\\
    &= \textbf{R}^i_1({\textbf{T}}^{-1})^i~\textbf{v}_0^i \quad (\text{using }\eqref{eq:tangent_operator}).
\end{align}
where $(\textbf{T}^{-1})^i:=\textbf{T}^{-1}(h^i\textbf{k}^i_0)$ and $h^i=s_{i+1}-s_i$ is the length of the $i^{th}$ segment. The scalar functions $\left(\alpha(\cdot),\beta(\cdot)\right)$ used in the above expression are defined in equation \eqref{eq:rotation_formula2} while the tensor function $\textbf{T}^{-1}(\cdot)$ is defined in equation \eqref{eq:tangent_operator} in \ref{appendix:rotation_formulas}. We thus have
\begin{align}\label{v01}
    \textbf{v}_0^i =\textbf{T}^i~\textbf{R}_1^{i^T}\bigg(\frac{\textbf{r}_2^i-\textbf{r}_1^i}{h^i}\bigg).
\end{align}  
where $\textbf{T}^i := \textbf{T}(h^i\textbf{k}_0^i)$. Likewise, we have
\begin{align}\label{eq:k0_discrete}
     \textbf{R}_2^i= \textbf{R}_1^{i}e^{h^i\textbf{K}_0^i}\Rightarrow     \textbf{k}_0^i = \frac{1}{h^i}rv(\textbf{R}_1^{i^T}\textbf{R}_2^i)
\end{align}
where $rv$ denotes the rotation vector of the corresponding rotation tensor. Having obtained the strains $\textbf{v}_0^i$ and $\textbf{k}_0^i$ in each of the rod's segments in terms of the segment's end positions and rotations using \eqref{v01} and \eqref{eq:k0_discrete}, we can finally write the discrete rod's energy as
\begin{align}\label{eq:energy_discrete_sc_rod}
    \mathcal{I}(\textbf{x}) = \sum_{i=1}^{N_{el}}\phi(\textbf{v}_0^i,\textbf{k}_0^i)h^i - W^{\text{ext}} 
\end{align}
where $\phi(\textbf{v}_0^i,\textbf{k}_0^i)$ is the energy density of the discrete rod segment with strains $(\textbf{v}_0^i,\textbf{k}_0^i)$. Likewise, $W^{ext}$ is the work done by external loads. For the Kirchhoff rod model, the constrained discrete energy can be written as
\begin{align}\label{eq:energy_discrete_kirchhoff_rod}
 \mathcal{I}(\textbf{x},\boldsymbol{\lambda}_0) = \sum_{i=1}^{N_{el}}\phi(\textbf{k}_0^i)h^i+\sum_{i=1}^{N_{el}}\boldsymbol{\lambda}_0^i\cdot(\textbf{v}^i_0-\textbf{e}_3) -W^{ext}
\end{align}
where $\boldsymbol{\lambda}_0^i$ is the Lagrange multiplier vector that enforces both the unshearability and inextensibility constraints in \eqref{eq:kirchhoff_constraint}. For the unshearable (but extensible) rod type, the constrained energy function would be
\begin{align}\label{eq:energy_discrete_unshearable_rod}
 \mathcal{I}(\textbf{x}, \boldsymbol{\lambda}_1,\boldsymbol{\lambda}_2) = \sum_{i=1}^{N_{el}}\phi(\text{v}^i_3,\textbf{k}_0^i)h^i+\sum_{i=1}^{N_{el}}[\lambda_1^i(\textbf{v}_0^i\cdot\textbf{e}_1)+\lambda_2^i(\textbf{v}_0^i\cdot\textbf{e}_2)]-W^{ext}
\end{align}
where the scalar Lagrange multipliers $\lambda_1^i$ and $\lambda_2^i$ enforce only the unshearability constraints (see \eqref{eq:unshearability_constraint}) in the discrete element. Analogous to the continuous setting, the governing equations of the three rod models discussed above can be obtained by minimizing the corresponding discrete energy  with respect to unknowns, i.e., by setting
\begin{align}\label{deltaI}
    &\frac{d}{d\textbf{x}}\mathcal{I}(\textbf{x})=0\quad (\text{special Cosserat rod}),\quad \frac{d}{d[\textbf{x},
\boldsymbol{\lambda}_0]}\mathcal{I}(\textbf{x},\boldsymbol{\lambda}_0)=0\quad (\text{Kirchhoff rod}),\nonumber\\
\quad &\frac{d}{d[\textbf{x},\boldsymbol{\lambda}_1,\boldsymbol{\lambda}_2]}\mathcal{I}(\textbf{x},\boldsymbol{\lambda}_1,\boldsymbol{\lambda}_2)=0\quad (\text{unshearable rod}).
\end{align}
The readers are referred to \ref{appendix:euler_lagrange_discrete element} for the derivation of the governing equations.
\section{Numerical examples} \label{sec:results}
In this section, we solve the RVE warping problem proposed in Section \ref{sec:rve_problem} and further obtain various macroscopic stiffnesses as detailed in Section \ref{stiffnesses}. First, we take up simple planar RVEs - cross and square ones to validate the macroscopic stiffnesses obtained using our scheme with those from the existing literature. Later, we take up more complex three-dimensional RVEs. To this end, an in-house code was developed on the deal.II platform \citep{2024:africa.arndt.ea:deal}. The microscale rods are discretized using the discrete helical rod model presented in Section \ref{discrete rod theory}. The discretized RVE energy along with the relevant constraints are then substituted in the constrained RVE energy \eqref{eq:energy_functional_constraind} whose first derivative with respect to all the unknowns (both configuration variables and Lagrange multipliers) when set to zero yields the governing equations for the RVE problem. The equations turn out to be nonlinear and are solved using the Newton-Raphson scheme. Table \ref{table:common_choice} summarizes the common choice of simulation parameters for all the numerical results to be discussed in this section.
\begin{table}[H]
\centering
\begin{tabular}{c|c} 
\hline
global constraint type & \textit{delta-theta} \\ 
\hline
global constraint set& 
left boundary nodes of the RVE\\
\hline
joint constraint type& rigid\\
\hline
\makecell[c]{number of elements for \\discretizing constituent rods}& 40 (straight rod), 100 (helical rod)\\
\hline
rod model used& \makecell[c]{special Cosserat rod\\(unless otherwise specified)}\\
\hline
\end{tabular}
\caption{Common choice of simulation parameters for all the numerical results in this section}
\label{table:common_choice}
\end{table} 
The constituent rods are assumed to have circular cross-section and obey the strain energy function given in equation \eqref{eq:rod_energy_function_quadratic} wherein $\mathbbl{C}^{rod}_0$ is taken to be a diagonal tensor with its components $\mathbbl{C}^{rod}_{11}=\mathbbl{C}^{rod}_{22}=kGA$, $\mathbbl{C}^{rod}_{33} = EA$, $\mathbbl{C}^{rod}_{44}=\mathbbl{C}^{rod}_{55} = EI$, $\mathbbl{C}^{rod}_{66}=GJ$. Here $E$ is the Young's modulus, $G$ is the shear modulus and $k$ is the shear correction factor. The symbols $A$, $I$ and $J$ denote cross-sectional area, second moment of area and polar moment of area, respectively. In the schematics shown henceforth, we use blue color to denote the RVE and green color to denote its repetitions.
\subsection{Cross RVE}\label{section:cross_rve}
    \begin{figure}[h!]
        \centering
        \includegraphics[width=\textwidth]{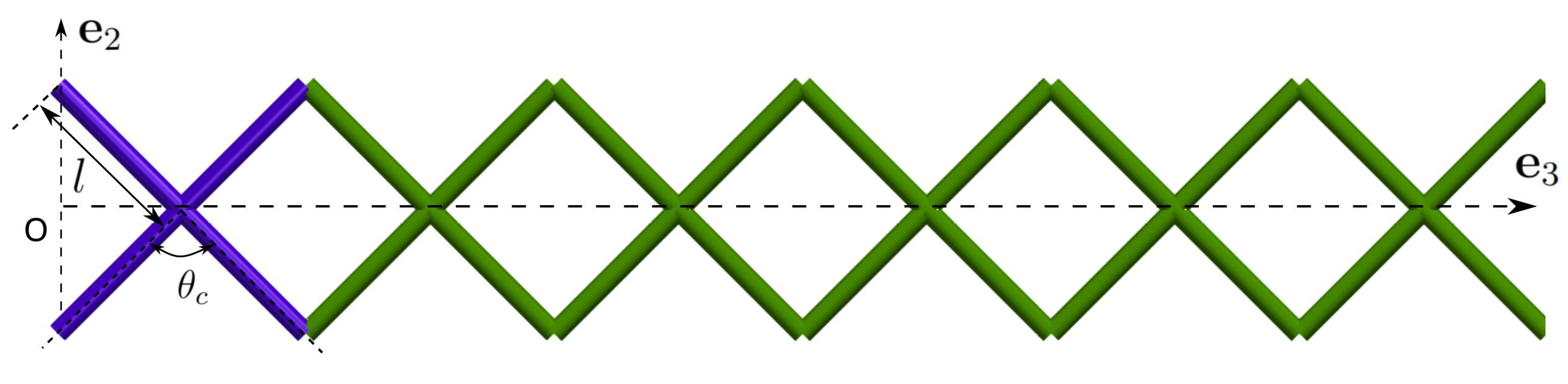}
        \caption{A rod-like metamaterial having cross-shaped RVE.}
        \label{fig:cross_rve_problem}
    \end{figure}
Let us first consider a rod-like pantographic metamaterial \citep{dell2020discrete} as shown in Figure \ref{fig:cross_rve_problem}. Such a metamaterial could be constructed by periodically repeating a cross-shaped RVE in one direction. The cross RVE consists of four rods of equal length $l$ rigidly connected to each other at the center of the RVE. The geometric design parameters for this RVE are: slenderness parameter $\frac{r}{l}$ of each rod and the cross angle $\theta_c$. Their values are taken to be the following in numerical simulation unless specified otherwise:
\begin{equation}
    \frac{r}{l}=0.05,~\theta_c=\pi/2.
\end{equation}
\begin{figure}[h!]
    \centering
    \begin{subfigure}{0.49\textwidth}
        \includegraphics[width=\textwidth]{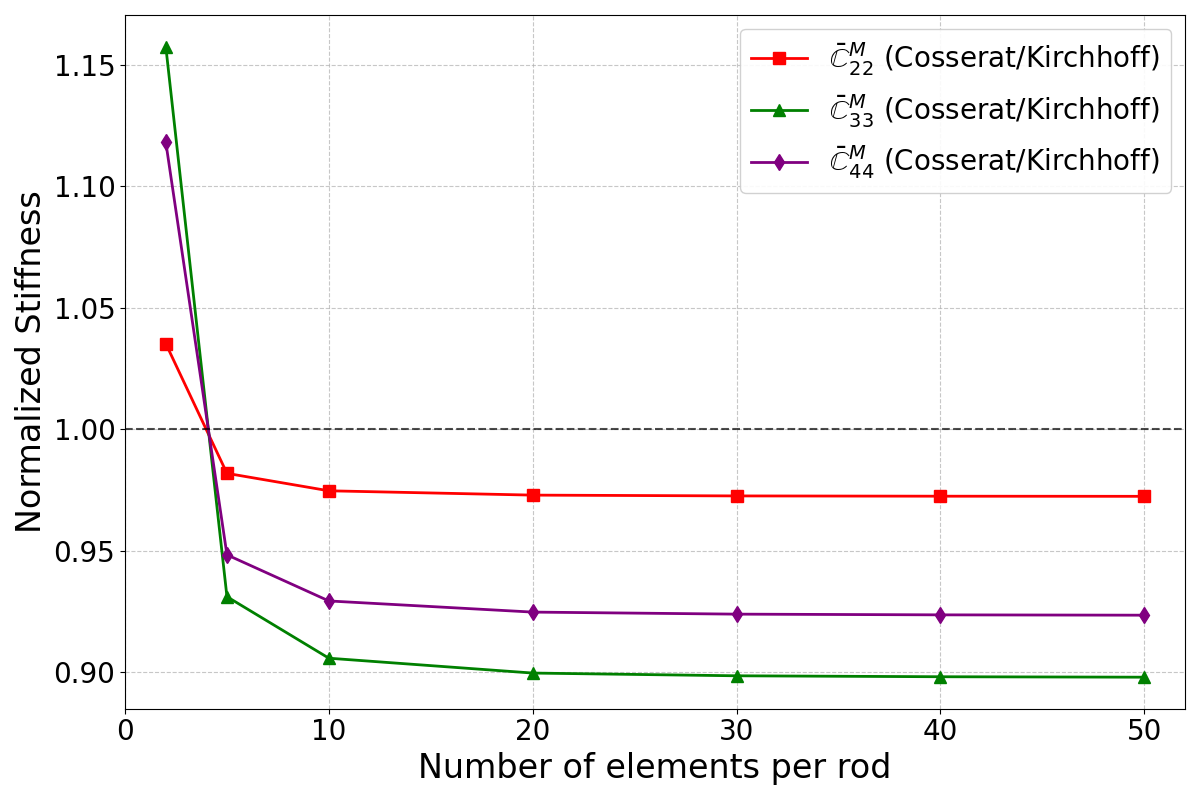}
        \caption{Mesh convergence study at $\frac{r}{l}=0.1$}
    \label{subfig:cross_zero_strain_stiffness1}
    \end{subfigure}
    \begin{subfigure}{0.49\textwidth}
         \includegraphics[width=\textwidth]{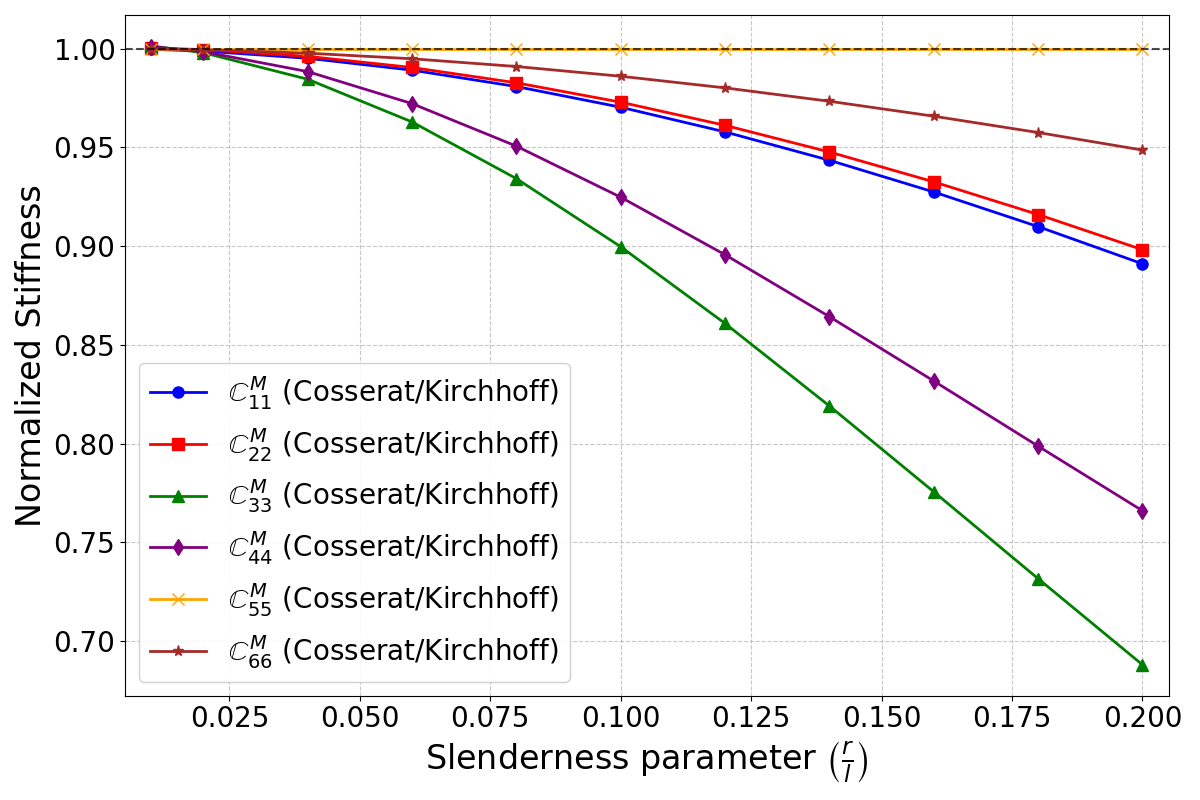}
         \caption{Effect of slenderness parameter $\left(\frac{r}{l}\right)$}
    \label{subfig:cross_zero_strain_stiffness2} 
    \end{subfigure}
    \caption{Cross RVE normalized stiffnesses at zero macroscopic strain: Cosserat vs Kirchhoff constituent rods.}
    \label{fig:cross_zero_strain_stiffness}
\end{figure}
We first examine the effect of including shear and extension in microscale rods on the metamaterial's macroscopic stiffnesses. Figure \ref{fig:cross_zero_strain_stiffness} shows the stiffnesses at zero strain obtained using special Cosserat microscale rods. They have been normalized with their corresponding values when microscale rods are assumed to be Kirchhoff type. The formulas of the stiffnesses $\mathbb{C}_{22}^M$, $\mathbb{C}_{33}^M$ and $\mathbb{C}_{44}^M$ for the Kirchhoff case are also obtained analytically in \ref{cross_rve_stiffness} and further summarized in Table \ref{table:cross_rve_parameters}. In Figure \ref{subfig:cross_zero_strain_stiffness1}, we first do a mesh convergence study. We note that the normalized stiffnesses are converging to a value different from unity since $r/l=0.1$ here which isn't small enough for Cosserat rod based stiffnesses to become equal to those of Kirchhoff rod based stiffnesses. Notice that the stiffnesses reach their converged limit at $N_{el}=30$. For all other examples done in this section, the result showed convergence to within 1\% of 
accuracy at maximum 40 elements. We then see from Figure \ref{subfig:cross_zero_strain_stiffness2} that the effect of including shear and extension in microscale rods is significant at large slenderness parameter $(r/l)$. Only the out-of-plane bending stiffness is independent of the slenderness ratio since microscale rods do not stretch or shear during infinitesimal out-of-plane bending . Further note that all the normalized stiffnesses converge to unity as $r/l\rightarrow0$.
\begin{table}[h!]
\centering
\begin{tabular}{c|c|c} 
\hline
$\mathbb{C}^M_{22}~\text{(in-plane shear)}$ & $\mathbb{C}^M_{33}~\text{(extension)}$ & $\mathbb{C}^M_{44}~\text{(in-plane bending)}$ \\ 
\hline
$\frac{6EI}{l^2} \frac{1}{\sin \frac{\theta_c}{2}}$& 
$\frac{24EI}{l^2}\frac{\tan \frac{\theta_c}{2}}{\cos \frac{\theta_c}{2}}$ & 
$8EI \sin \frac{\theta_c}{2}$
\rule[-3mm]{0pt}{0pt}
\\
\hline
\end{tabular}
\caption{Analytical formulas of stiffnesses for cross RVEs considering its constituent rods to be Kirchhoff type}
\label{table:cross_rve_parameters}
\end{table}
\begin{figure}[h!]
    \centering
    \includegraphics[width=0.49\linewidth]{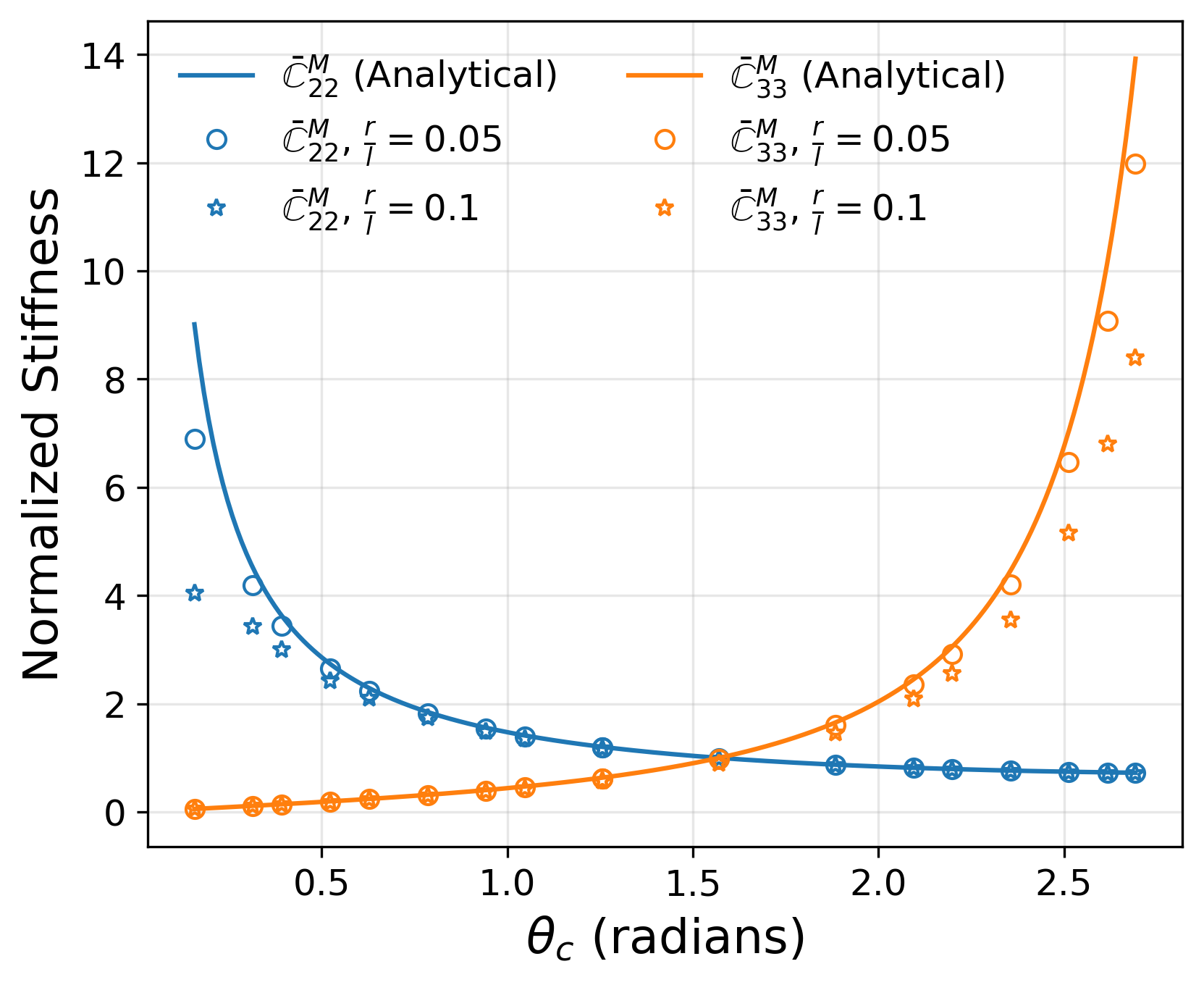}
    \includegraphics[width=0.49\linewidth]{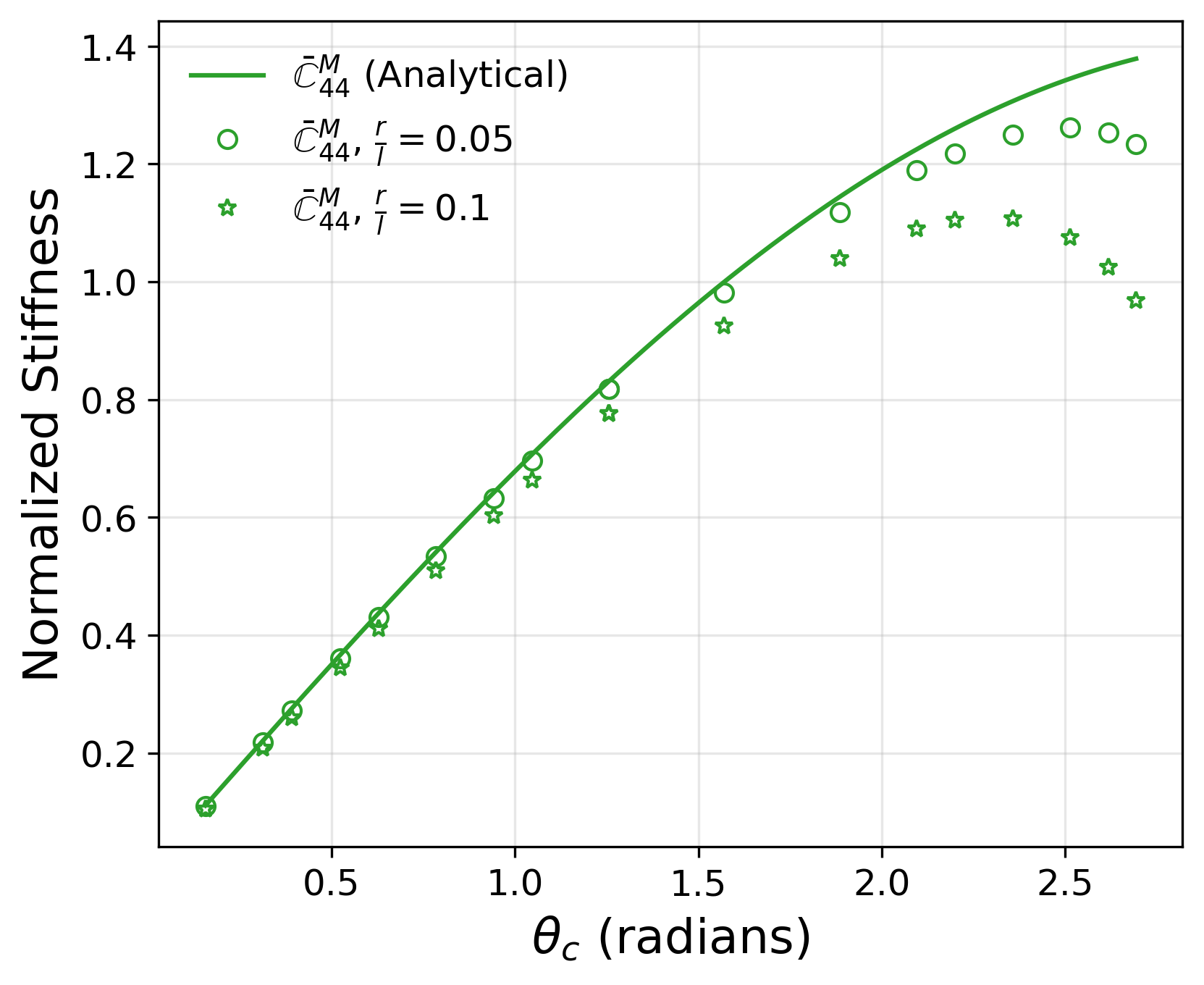}
    \caption{Variation of normalized stiffnesses at zero strain vs. the cross-RVE parameter $\theta_c$: The analytical results plotted are obtained assuming the constituent rods to be of Kirchhoff type (see Table \ref{table:cross_rve_parameters}). For the numerical stiffnesses obtained using the present approach, each constituent rod is modeled as a special Cosserat rod. All the stiffnesses are normalized by the corresponding analytical stiffness value at $\theta_c=\pi/2$.}
    \label{fig:cross_angle_study}
\end{figure}
\begin{figure}[h!]
    \centering
    \includegraphics[width=0.5\linewidth]{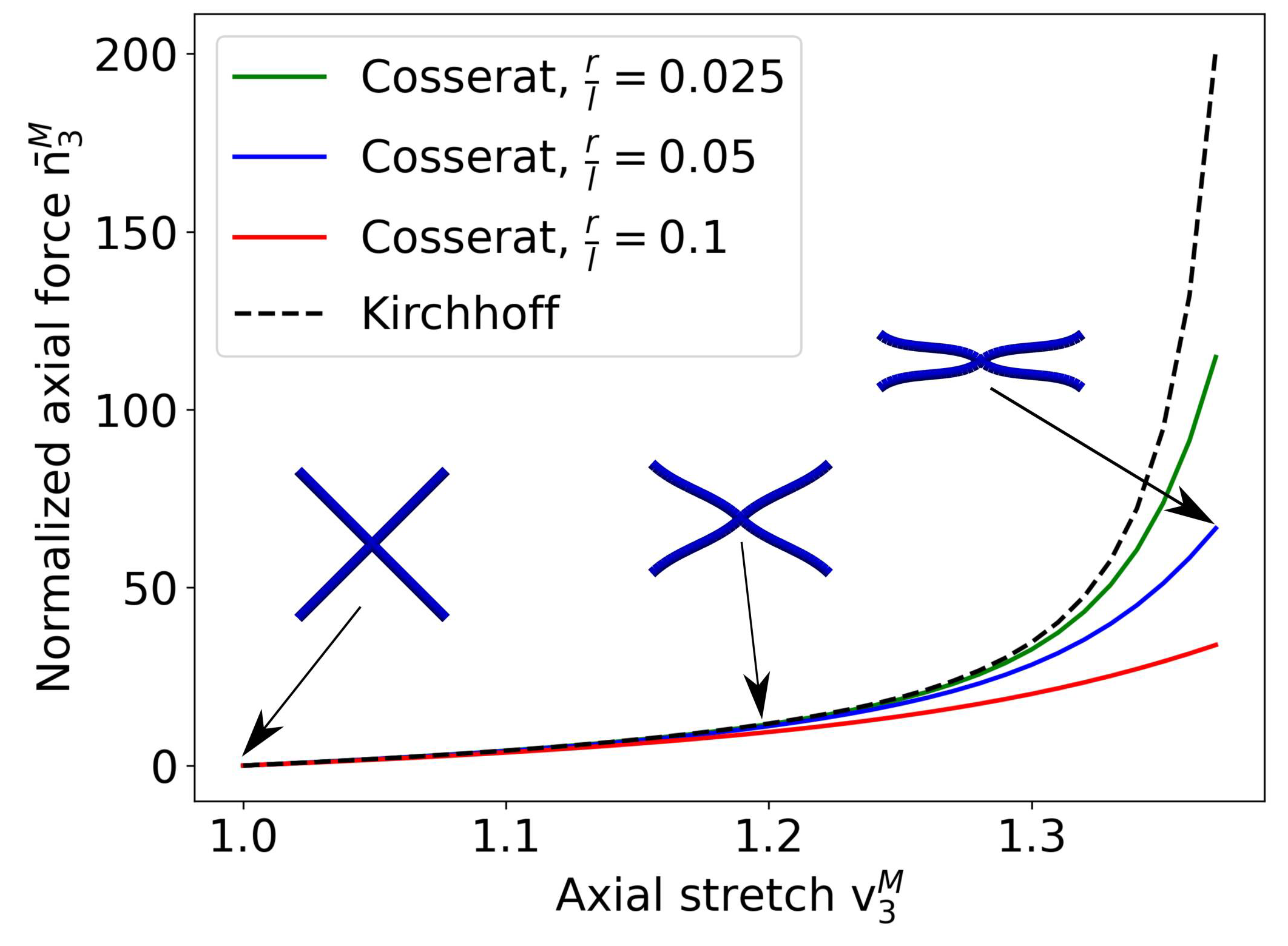}
    \caption{Finite stretching of cross-RVEs: deformation within the RVE transitions from being bending dominated to stretching dominated as seen from the insets of the RVE at different macroscopic stretch. The normalized axial force $\bar{\text{n}}^M_3=\text{n}_3^M/(EI/l^2)$.}
    \label{fig:cross_rve_finite_stretch}
\end{figure}
Next, we study the effect of geometric parameter $\theta_c$ on macroscopic stiffnesses at zero strain (see Figure \ref{fig:cross_angle_study}). The stiffnesses have been numerically obtained for two different slenderness ratios and compared with their Kirchhoff-limit analytical formulas from Table \ref{table:cross_rve_parameters}. All the stiffnesses have been normalized with their corresponding analytical formulas at $\theta_c=\pi/2$. We note that the in-plane shearing stiffness approaches infinity as $\theta_c\to0$. This is because the microscopic deformation due to macroscopic shearing is stretching-dominated in this regime. For the same reason, the macroscopic extensional stiffness approaches infinity as $\theta_c\to\pi$. Finally, as expected, the three macroscopic stiffnesses match with the analytical formulas more for smaller slenderness ratio of constituent microscale rods. We then study finite stretching of this RVE. Figure \ref{fig:cross_rve_finite_stretch} shows the axial force versus axial stretch response for different $\frac{r}{l}$ ratios. We see that as the RVE is stretched, initially the deformation mode of microscale rods is bending dominated. Later, as the microscale rods get aligned with the macroscopic stretching direction, the deformation mode of microscale rods transitions to being stretching-dominated. This leads to a J-shaped response as depicted in the figure. In addition, it can be seen that as the initial deformation is primarily bending-dominated, the slenderness parameter $\frac{r}{l}$ does not influence the initial macroscopic response. However, once the behavior enters the stretching-dominated regime, the impact of $\frac{r}{l}$ becomes pronounced.
\begin{figure}[h!]
    \centering
    \includegraphics[width=0.8\linewidth]{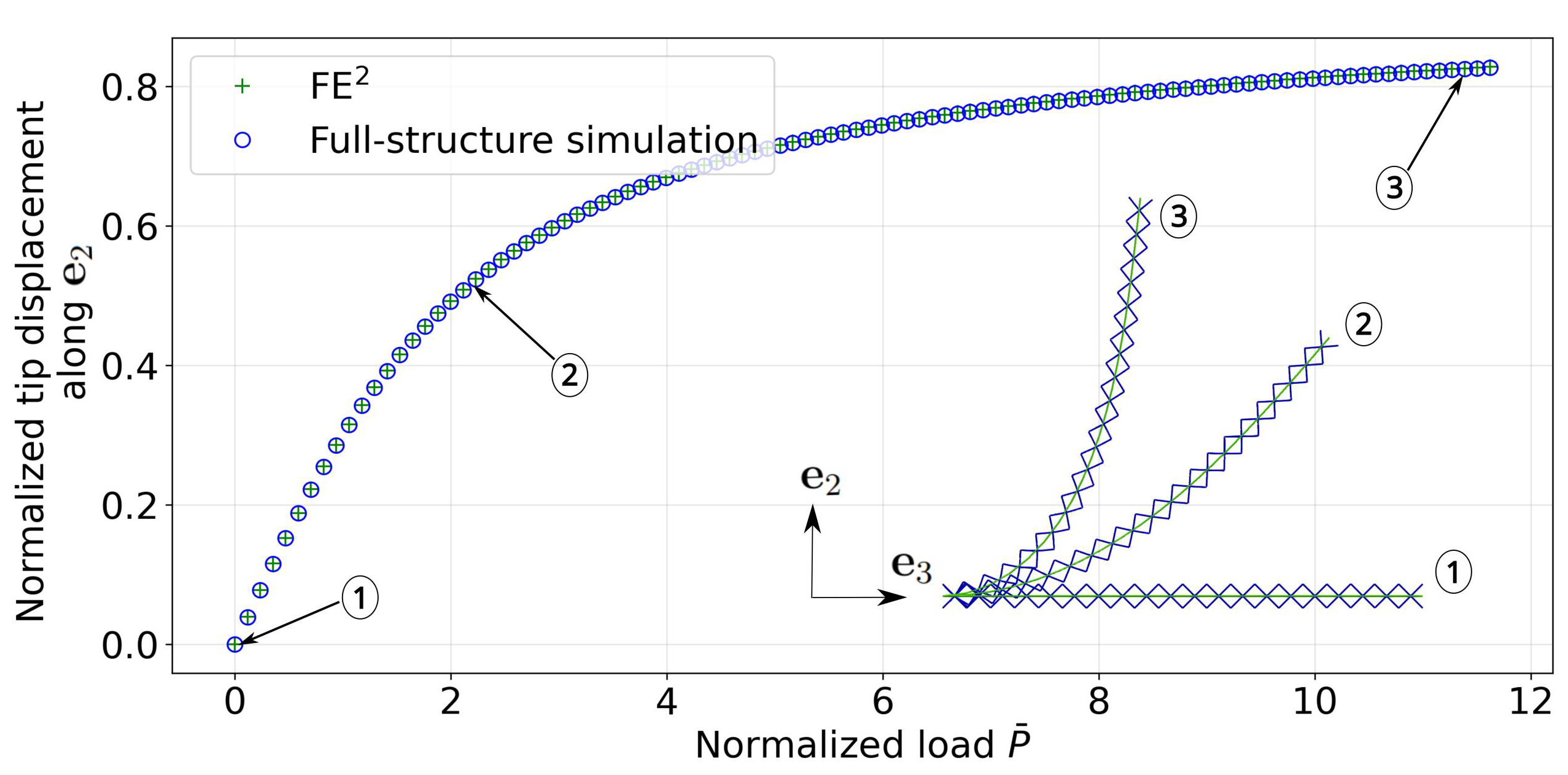}
    \caption{A comparison between $\text{FE}^2$ versus full-scale simulation for a cross-RVE based metamaterial. The inset shows the deformed configurations of both the full-structure (blue) and the homogenized rod depicted by its centerline (green). The transverse load on the horizontal axis is normalized as follows: $ \bar{P}=P/(\mathbb{C}_{44}^M/L_{full}^2)$.}
    \label{fig:cross_transverse_load}
\end{figure}
 Next, we compare the full-scale simulation of a cross-RVE metamaterial with the simulation of a homogenized single rod in an $\text{FE}^2$ framework wherein the rod's finite-strain constitutive relationship is obtained using the presented homogenization scheme - see Figure \ref{fig:cross_transverse_load}. The metamaterial is formed by assembling 20 RVEs along $\textbf{e}_3$ direction, resulting in a structure of length $L_{full} = 20L$. The boundary condition of the problem is as follows: the left end ($s=0$) is clamped, while at the right end ($s=L_{full}$), a concentrated normalized transverse force ${\textbf{F}} =P\textbf{e}_2$ is applied where $P$ is the magnitude of load. In the full structure simulation, this load is divided equally among the two nodes at the end of the metamaterial. We see a good agreement between the full-scale simulation and the $\text{FE}^2$ approach, validating our homogenization scheme. \\
\begin{figure}[h!]
    \centering
    \includegraphics[width=0.49\linewidth]{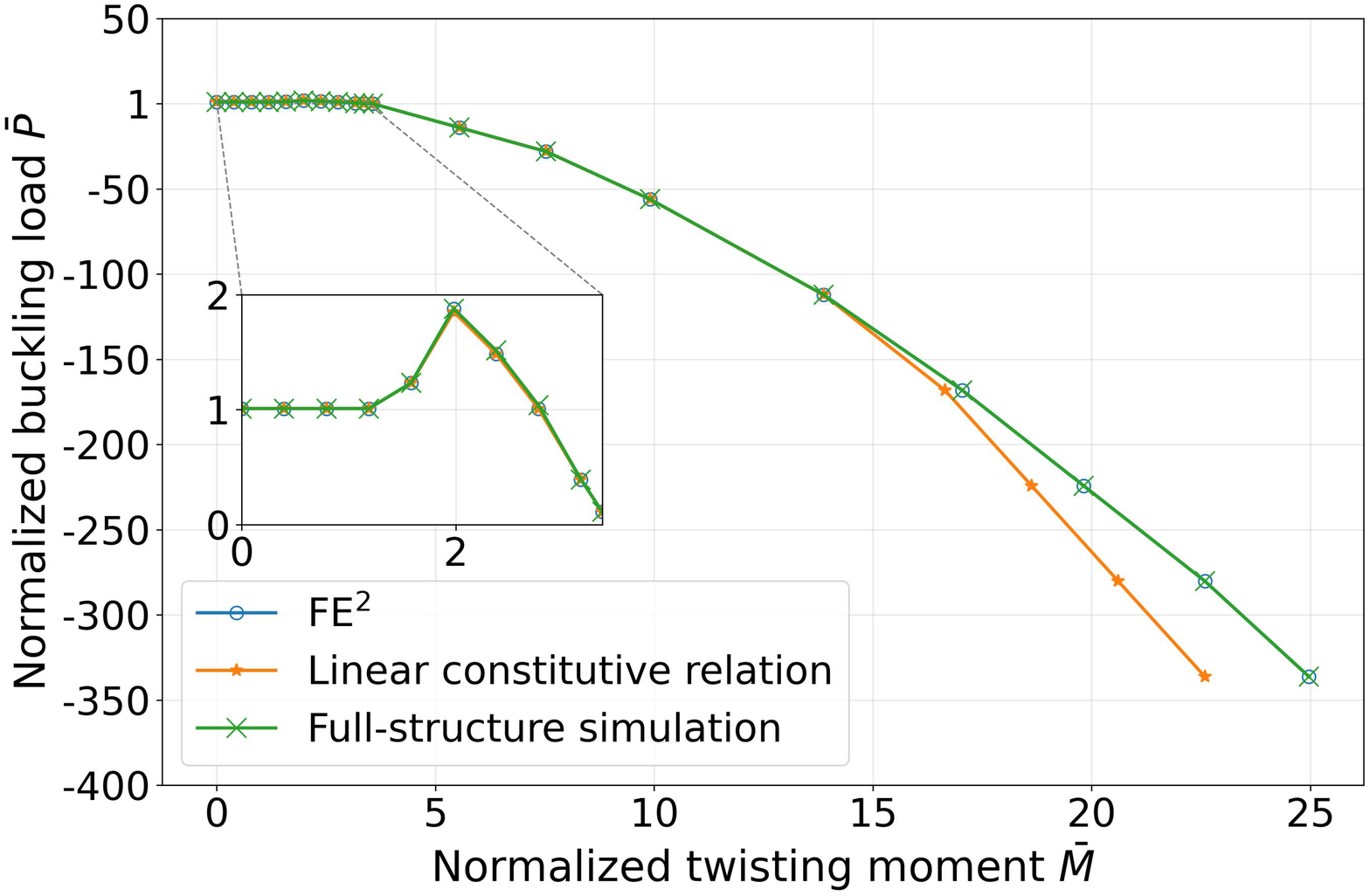}
    \raisebox{2em}{\includegraphics[width=0.495\linewidth]{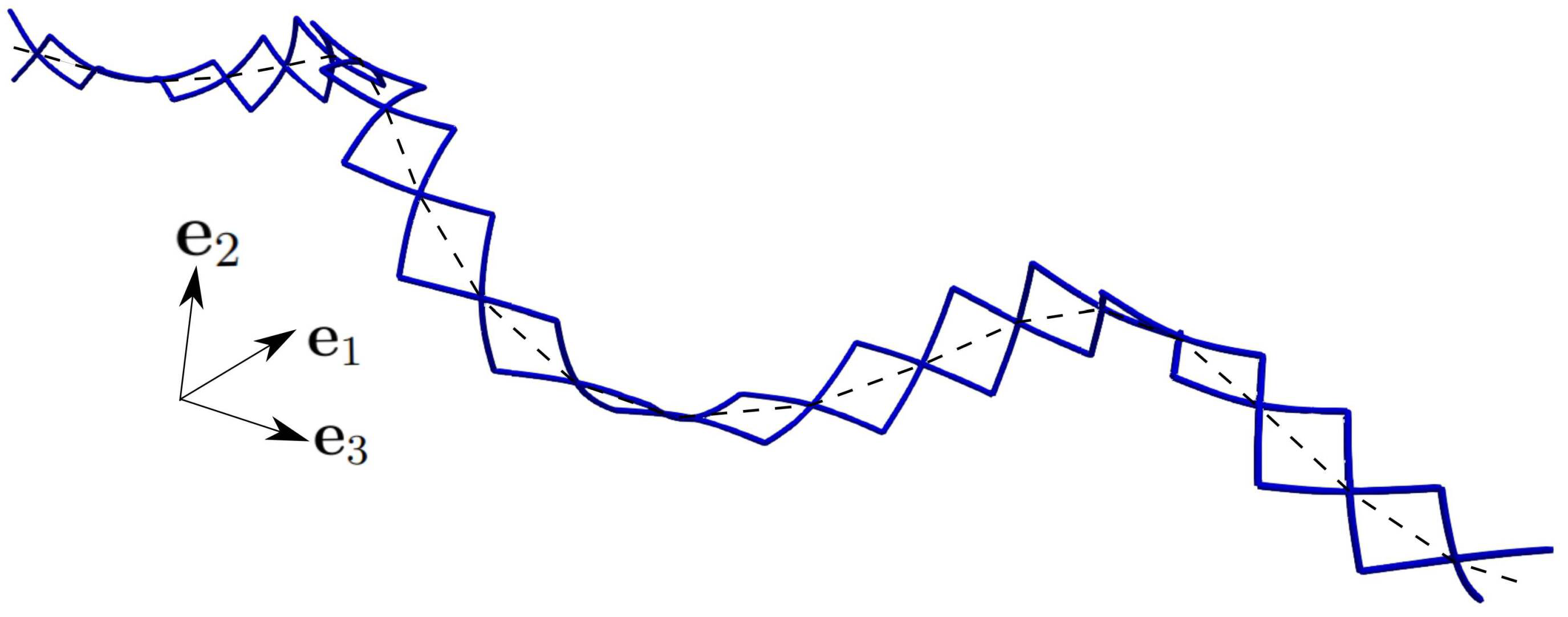}}
    \caption{Variation in buckling load of a cross-RVE metamaterial in the presence of twisting moment. On the right, the buckled mode shape is also shown whose centerline is non-planar due to the presence of twisting moment. The inset shows initial increase in buckling load due to bending anisotropy in cross-RVE metamaterials. The normalization is done as follows: $\bar{P}=P/\frac{\pi^2 \mathbb{C}^M_{44}}{4{L^2_{full}}},~\bar{M}=M/\frac{\pi^2 \mathbb{C}^M_{44}}{4{L_{full}}}$.}
    \label{fig:cross_twist_buckling}
\end{figure}
Finally, we illustrate the effect of including non-linearity in the homogenized constitutive relationship. To this end, we consider the problem of Euler buckling of this metamaterial in the presence of twisting moment - see Figure \ref{fig:cross_twist_buckling}. The Euler buckling load for a clamped-free untwisted Kirchhoff rod is given by $P_{cr}=\frac{\pi^2 \mathbb{C}^M_{44}}{4{L^2_{full}}}$ which is used for normalization here. The boundary condition of this problem is as follows: left end ($s=0$) is clamped while at the right end ($s= L_{full}$), a concentrated axial compressive force $P$ and a twisting moment $M$ is applied. We notice that at zero twisting moment, the normalized buckling load is unity due to our choice of normalization. The buckling load then reduces as the twisting moment is increased and, in fact, becomes tensile when the applied twisting moment is large enough. We again notice that the buckling load obtained using full-structure simulation matches pretty well with $\text{FE}^2$ simulation even at large twisting moment. However, when we use a linear constitutive relation, i.e, the homogenized rod's stiffnesses are not changed as the rod deforms, the curve shows deviation from full-structure simulation at large twisting moment. It is also interesting to note the initial increase in buckling load (see the inset of Figure \ref{fig:cross_twist_buckling}). This is a classical result which is attributed to anisotropy in bending deformation. At zero twisting moment, buckling happens in the easy-bending plane. However, as the rod is twisted, no such easy bending plane exists globally and hence the buckling load increases initially.
\begin{figure}[h!]
    \centering
\includegraphics[width=\textwidth]{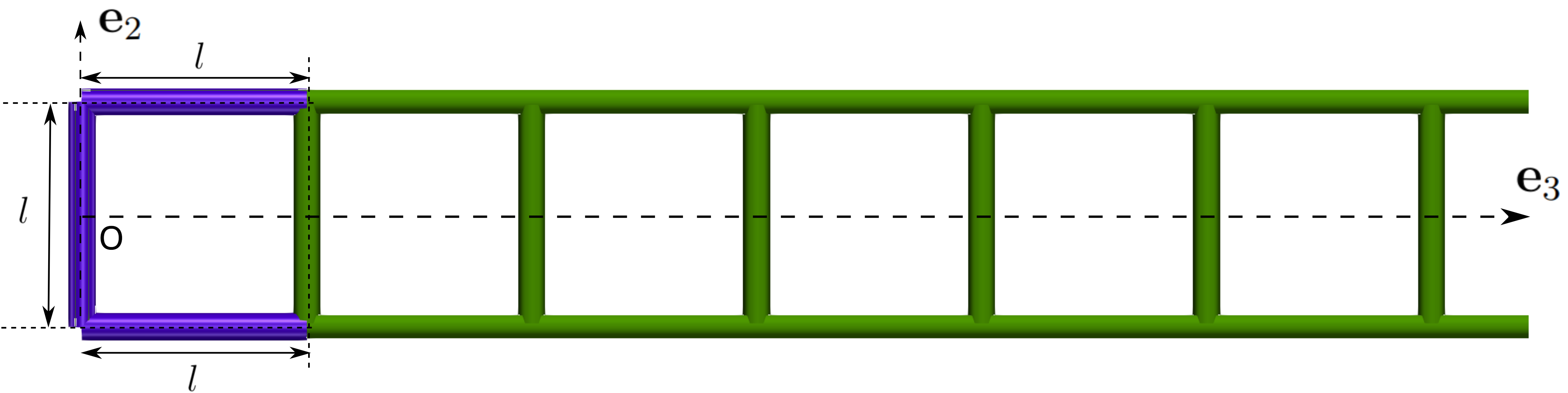}
    \caption{A square-periodic rod-like metamaterial}
    \label{fig:square_rve_problem}
\end{figure}
\subsection{Square RVEs}\label{section:square_rve}
\begin{figure}[h!]
    \centering
    \includegraphics[width=0.7\textwidth]{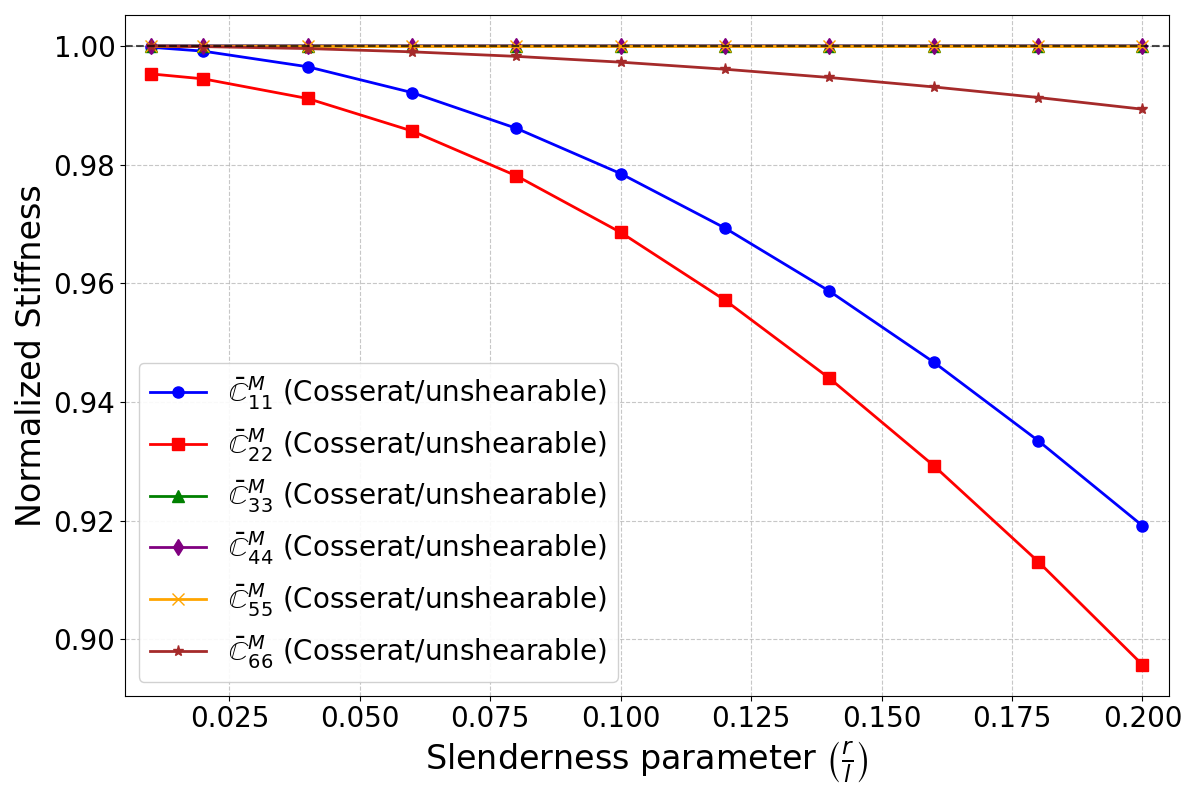}
    \caption{Square RVE normalized stiffnesses at zero macroscopic strain: Cosserat vs unshearable constituent
rods.}
    \label{fig:square_rve_stiffnesses}
\end{figure}

In this section, we consider a square-periodic beam. The RVE of this beam consists of three straight rods of equal length $l$ as shown in blue in Figure \ref{fig:square_rve_problem}. Let us first study the effect of including shear deformation in microscale rods on the metamaterial's macroscopic stiffnesses.\footnote{Unlike the cross RVE case where the microscale rods could also be made Kirchhoff-type, we can, at best, simplify the microscale rods in the square RVE case to be unshearable type only because a Kirchhoff microscale rod will lead to locking behavior when we impose macroscopic stretching or bending.} Figure \ref{fig:square_rve_stiffnesses} shows the macroscopic stiffnesses normalized by their corresponding values when the microscale rods are assumed unshearable. The derived analytical formulas for the latter case are shown in Table \ref{table:square_rve_parameters}. Note from Figure \ref{fig:square_rve_stiffnesses} that the macroscopic stretching and bending stiffnesses remain unaffected by shearing of microscale rods. However, the in-plane shearing stiffness is affected significantly. \cite{audoly:hal-04112136,abdoul2018strain} also obtained the expression for macroscopic shearing stiffness assuming the microscale rods to be unshearable (see Table \ref{table:square_rve_parameters}).\\  
\begin{table}[h!]
\centering
\caption{Analytical formulas of stiffnesses for square RVEs assuming the constituent rods to be unshearable: the formula of $\mathbb{C}_{22}^M$ was taken from \cite{audoly:hal-04112136,abdoul2018strain} whereas other formulas have been derived.}
\label{table:square_rve_parameters}
\begin{tabular}{c|c|c|c} 
\hline
\rule{0pt}{4mm}
 \shortstack{ $\mathbb{C}^M_{22}$ \\ (in-plane shear) } & \shortstack{ $\mathbb{C}^M_{33}$ \\ (extension) } & \shortstack{ $\mathbb{C}^M_{44}$ \\ (in-plane bending) } & \shortstack{ $\mathbb{C}^M_{55}$ \\ (out-of-plane bending) } \\ 
\hline
\rule{0pt}{6mm}
$\frac{8EI}{l^2}$ & 
2EA & 
$\frac{EAl^2}{2}+2EI$ & $2EI$
\rule[-3mm]{0pt}{0pt}
\\
\hline
\end{tabular}
\end{table}
\begin{figure}[h!]
    \centering
    \includegraphics[width=\textwidth]{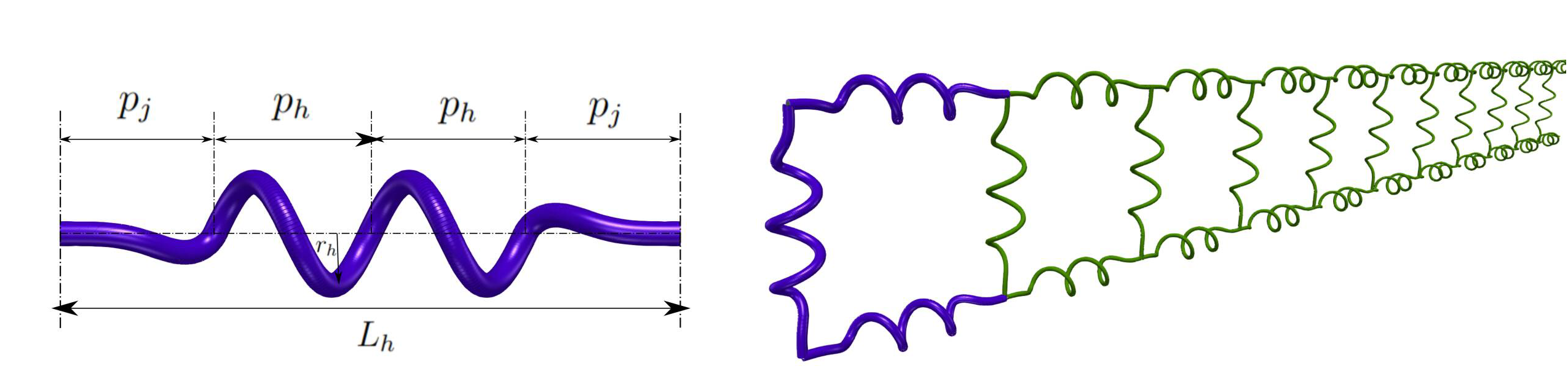}
    \caption{A rod-like metamaterial formed using square RVEs wherein the constituent rods are helical in their stress-free state \citep{chang2022mechanics}.}
    \label{fig:helical_rve_geometry}
\end{figure}
Let us again consider the square-shaped RVE but now having helical constituent rods (see Figure \ref{fig:helical_rve_geometry}). The inspiration for such rod networks comes from random helical and folded molecular chains making up biological networks such as cytoskeletons \citep{bathe2008cytoskeletal,purohit2011protein}. The idealized periodic network in  Figure \ref{fig:helical_rve_geometry} also serves as a useful model for artificial tissues and muscle fibers \citep{chang2022mechanics}. The rods in this RVE are not strictly helical in order to enable them to connect smoothly at each joint. They are made up of a uniform helix in the middle and curved rod segments at the two ends as proposed in \cite{chang2022mechanics}. The formula for the stress-free configuration of this helical rod is reproduced in \ref{appendix:helical_config}. It has the following geometric design parameters: rod diameter $d$, number of turns $N_h$, helix pitch $p_h$, helix radius $r_h$, and curved section parameter $p_j$ (see Figure \ref{fig:helical_rve_geometry}). As in \cite{chang2022mechanics}, we consider the following parameter values to model each helical rod unless otherwise specified: 
\begin{equation}
    d/r_h=0.3, ~p_h/r_h = 3.0,~p_j/r_h = 3.0.
\end{equation}
The end-to-end length of the helical rod is $L_h=2p_j + N_h~p_h$.
\begin{figure}[h!]
    \centering
    \begin{subfigure}{0.49\textwidth}
\includegraphics[width=\linewidth]{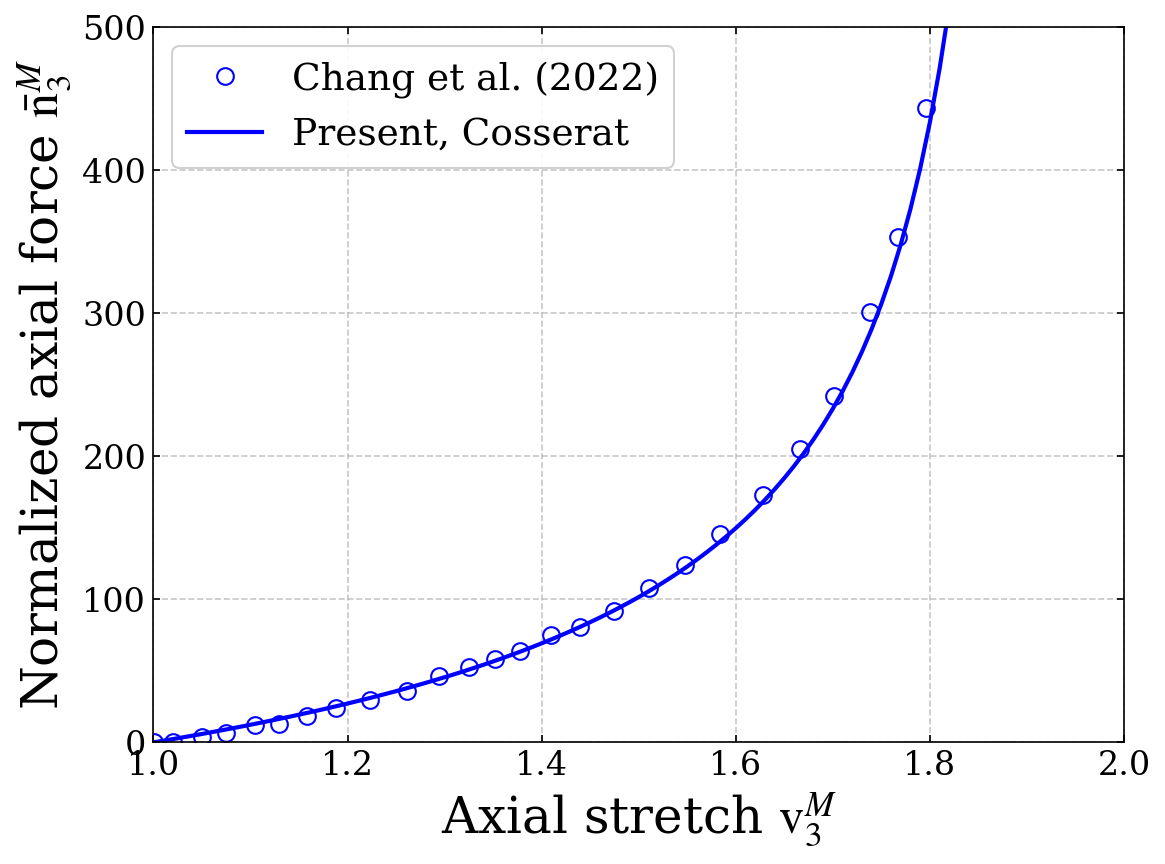}
    \caption{Stretching of a single helical rod ($N_h=3,d=0.3$).}
\label{subfig:single_helical_rod_stretching}
    \end{subfigure}
    \begin{subfigure}{0.49\textwidth}
\includegraphics[width=\linewidth]{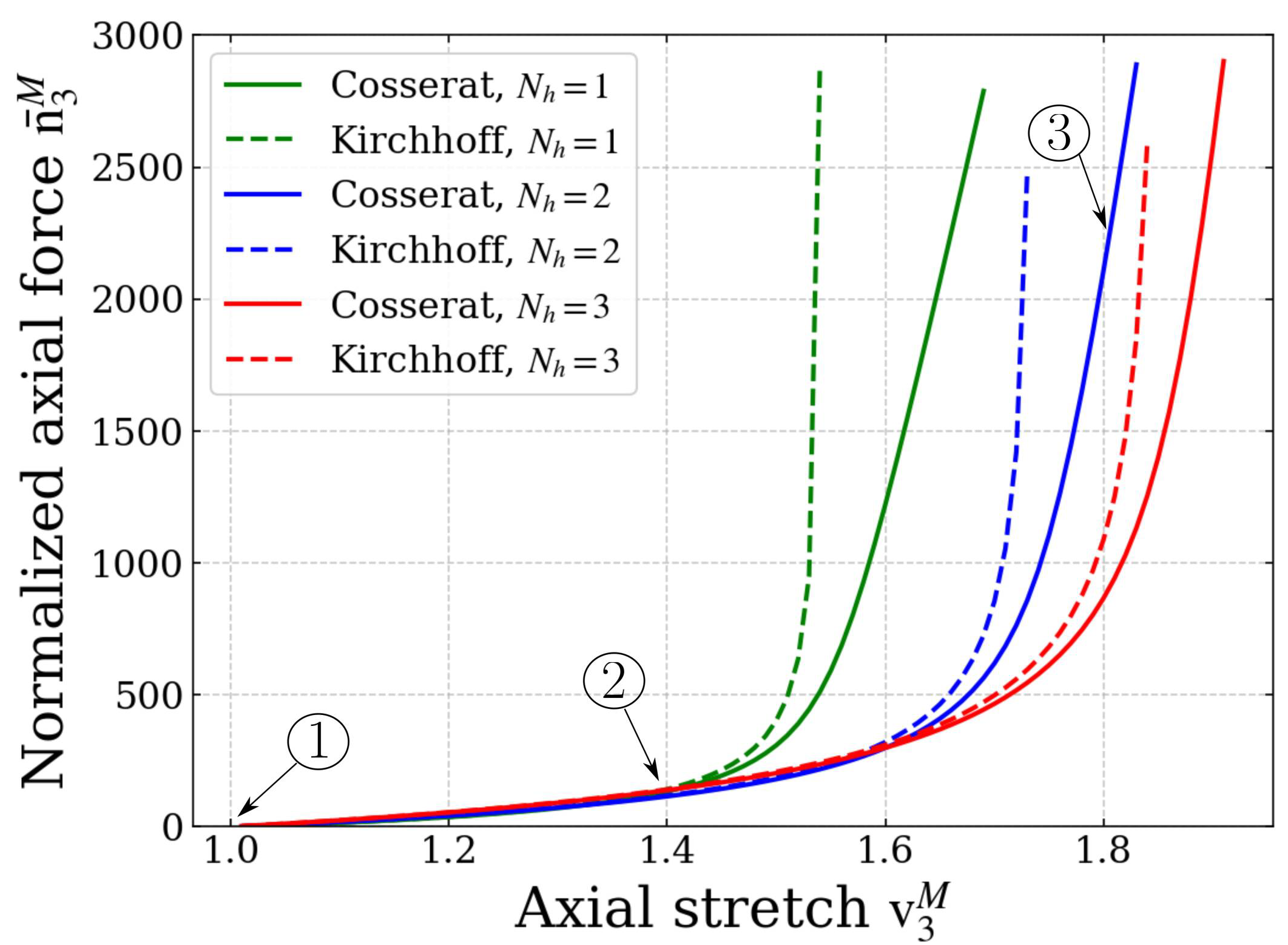}
    \caption{Stretching of a square helical RVE with varying number of helical turns in its constituent rods ($d=0.1$)}
    \label{subfig:square_helical_stretching}
    \end{subfigure}
    \begin{subfigure}{\textwidth}
        \includegraphics[width=\textwidth]{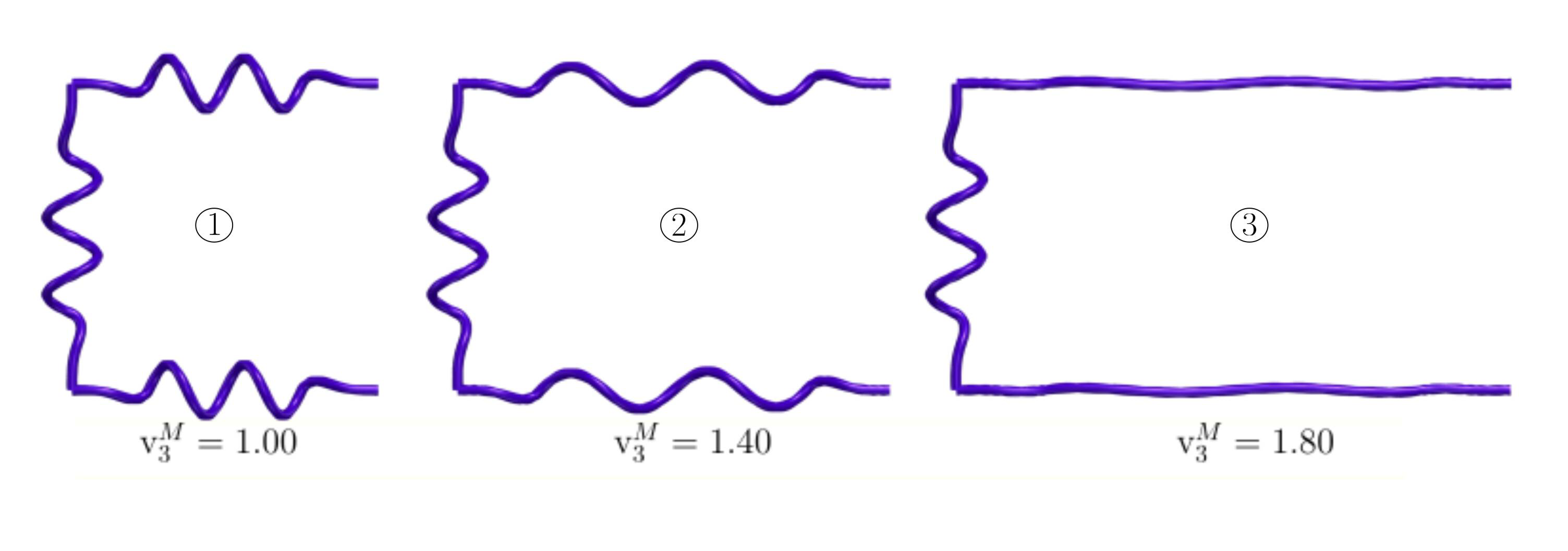}
        \caption{Deformation pattern in a square helical RVE at different levels of stretch}
        \label{subfig:square_helical_deformed_configuration}
    \end{subfigure}
    \caption{Uniform stretching of helical square RVEs. The axial force is normalized as follows: $\bar{\text{n}}_3^M=\text{n}^M_3/(EI/L^2_h)$.}
\label{fig:helical_rve_stretching_config}
\end{figure}
We first obtain the force-stretch response of the single helical rod in Figure \ref{subfig:single_helical_rod_stretching}. A gradual stiffening is observed as is the signature of helical rods. The response is also in good agreement with \cite{chang2022mechanics}. Next, we study the stretching of square helical RVEs. Figure \ref{subfig:square_helical_stretching} and Figure \ref{subfig:square_helical_deformed_configuration} show the force-stretch relationship and the deformed configuration of this RVE, respectively. It is observed that the horizontal helical rods undergo a transition from bending-dominated to stretching-dominated behaviour. This results in low macroscopic extensional stiffness initially which gradually increases as the horizontal beams straighten and enter stretching-dominated mode. Thus, a J-shaped macroscopic response is obtained, as in the case of single helical fibers. This type of response is typical of tissues \citep{vatankhah2017mimicking}. We also see that as the number of turns $N_h$ is increased (keeping $\frac{d}{r_h}$, $\frac{p_h}{r_h}$, $\frac{p_j}{r_h}$ constant), the bending-dominated behaviour is prolonged and thus the transition zone of the J-shaped curve shifts towards higher stretch. Moreover, including stretching in the microscopic rod model predicts significantly lower stiffness.\\\\
Next, the bending deformation for this RVE is studied.
\begin{figure}[h!]
    \centering
    \includegraphics[width=.48\textwidth]{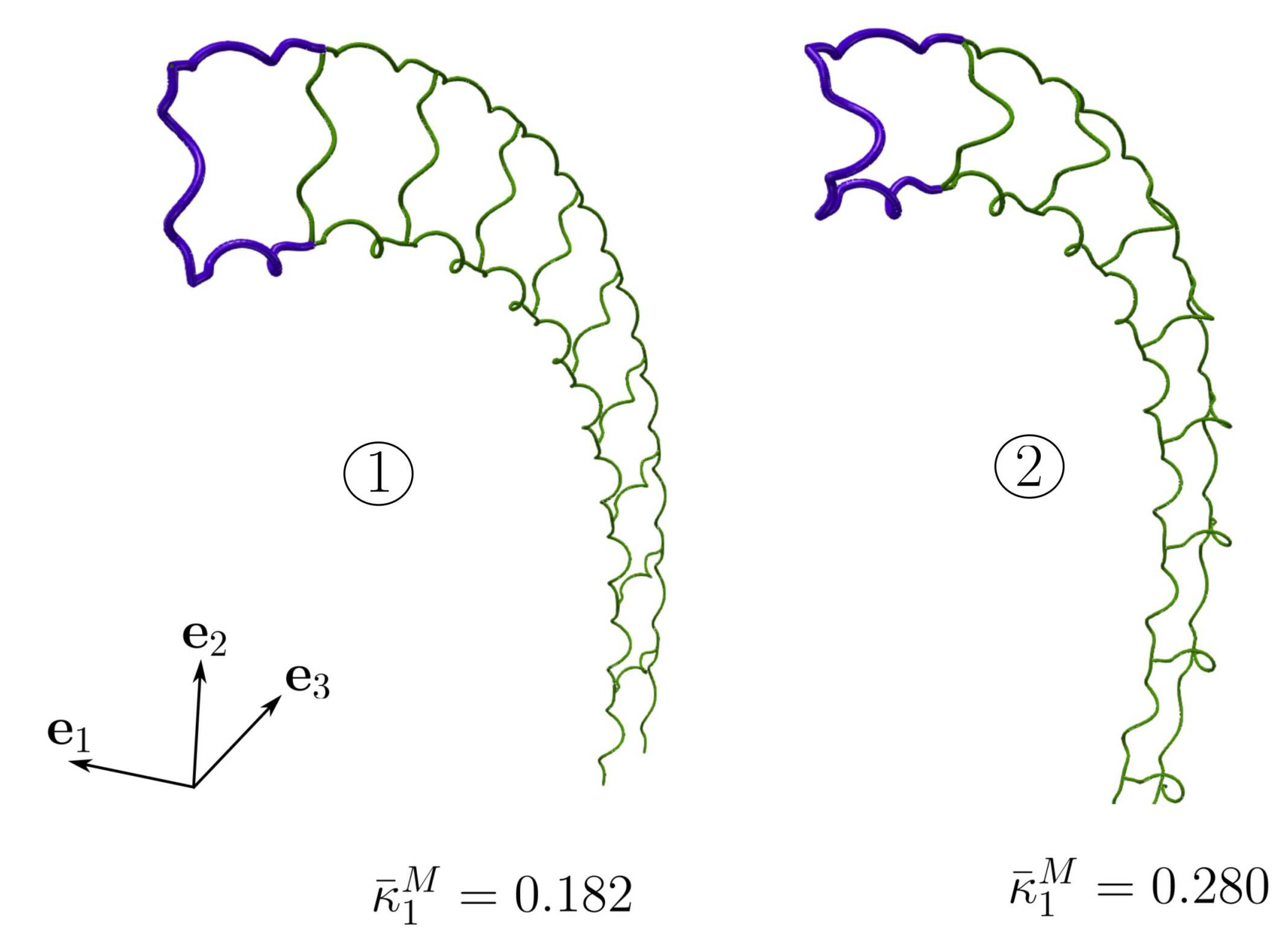}
    \includegraphics[width=0.48\textwidth]{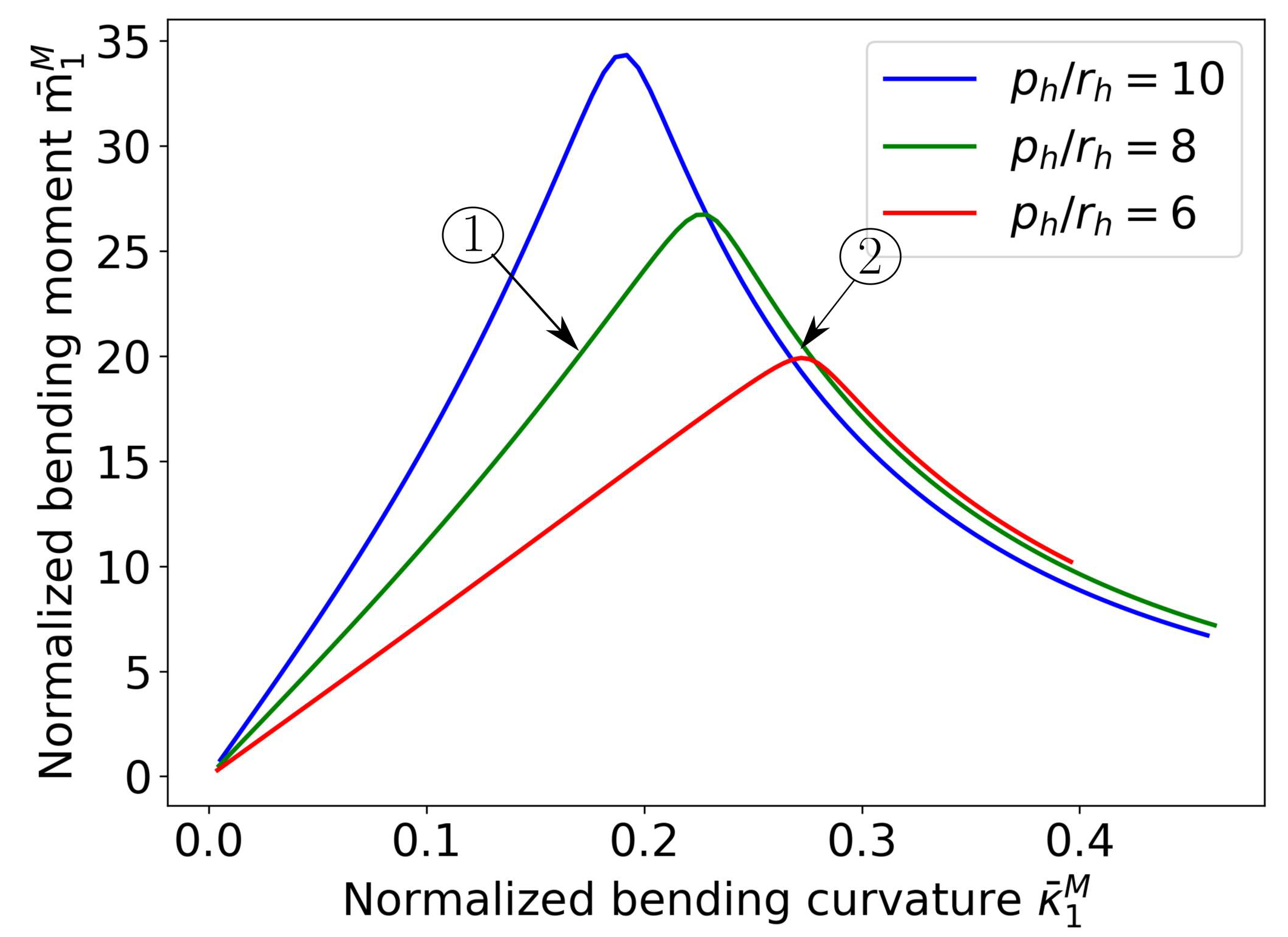}
    \caption{Pure bending of helical square RVEs ($N_h=1$, $d=0.1$). The normalized bending moment $\bar{\text{m}}^M_1=\text{m}^M_1/(EI/L_h)$ and normalized curvature $\bar{\kappa}_1^M = \kappa_1^M L_h$.}
    \label{fig:helical_rve_bending_config}
\end{figure}
Figure \ref{fig:helical_rve_bending_config} shows the normalized macroscopic bending moment $\bar{\text{m}}_1^M$ vs the normalized macroscopic curvature. It shows the response for varying $\frac{p_h}{r_h}$ wherein $r_h$ is kept constant and pitch is varied (the ratios $\frac{d}{r_h}$, $\frac{p_j}{r_h}$ are kept constant). We see that higher helix pitch value leads to higher bending moment at same curvature. This is because, as the pitch increases (with constant helix radius) the helical rod becomes straighter and thus more stretching-dominated, resulting in higher macroscopic bending stiffness. The plot also shows softening behavior at later stage which is attributed to out-of-plane buckling of the interconnecting or cross-linking vertical rod. The direction of this out-of-plane bending is determined by the chirality of the connecting helical rod. Moreover, the softening behavior is achieved earlier in the case of higher pitch value due to larger compressive force developing in the vertical rod. 
\subsection{Auxetic tubular structures}
\begin{figure}[h!]
    \centering
    \includegraphics[width=\textwidth]{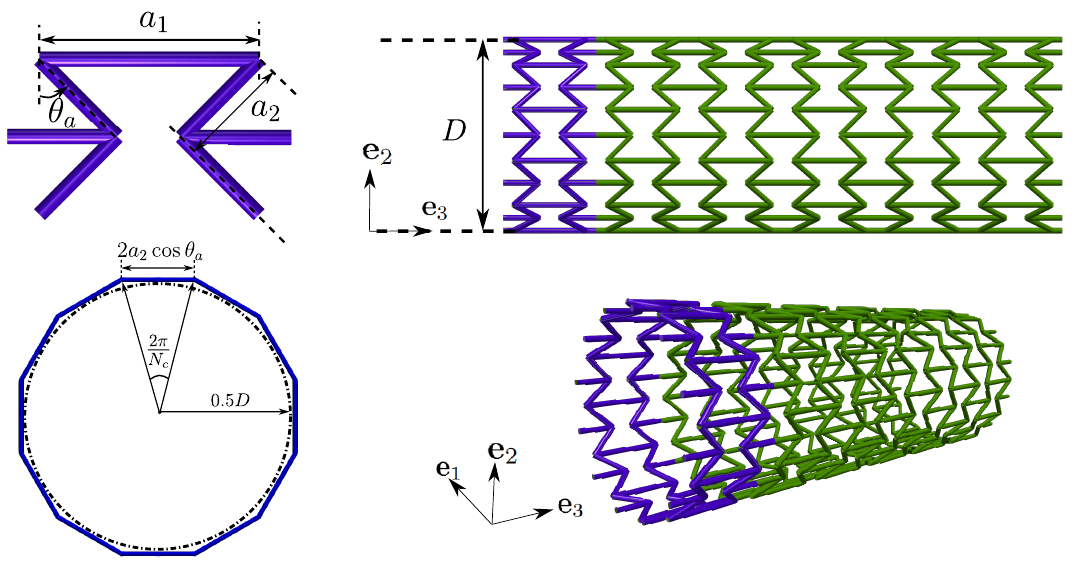}
    \caption{A tubular metamaterial constructed using auxetic unit cells \citep{karnessis2013uniaxial}. To ensure no penetration of rods, the following must hold: $-\frac{\pi}{2}<\theta_a<\arcsin{\left(\frac{a_1}{2a_2}\right)}$.}
    \label{fig:stent_rve_problem}
\end{figure}
Hollow auxetic tubular structures have found use in several applications such as fasteners, stents, energy absorbing devices etc. \citep{tambavca2010mathematical,luo2021design,zhang2021novel}. Some of the important mechanical characteristics of such tubular structures are their radial stiffness, flexural compliance, foreshortening etc. In case of vascular stents, these properties play an important role in dealing with clinical issues such as in-stent restenosis or stent thrombosis \citep{pan2021structural,ebrahimi2023revolutionary}. Moreover, such structures may undergo large three-dimensional deformation, such as in the case of femoral stents \citep{mactaggart2014three}. In this section, we consider a tubular metamaterial whose RVE is constructed by repeating an auxetic unit cell along the circumference of a cylinder of diameter $D$ (see Figure \ref{fig:stent_rve_problem}). Such a geometry has been shown to not only result in favourable mechanical behaviour, but also integrate better with the surrounding auxetic tissues \citep{bhullar2013influence,gupta2025design}. The geometric parameters of the auxetic tubular RVE are: microscopic rod diameter $d$, tube diameter (in an average sense) $D$, number of unit cells along the circumference $N_c$, auxeticity angle $\theta_a$, side lengths $a_1$ and $a_2$. Unless otherwise specified, we take the
following parameter values to simulate the auxetic tubular RVE:
\begin{align}\label{eq:stent_parameters}
    N_c=12,& \quad D=1.0, \quad \theta_a=30^{\circ},\quad d/D = 0.02,\quad \frac{a_1}{a_2}=2.0 \nonumber\\
    &\text{where}~a_2 = \dfrac{0.5D\tan{\frac{\pi}{N_c}}}{\cos{\theta_a}} \approx \frac{0.5 \pi D}{N_c \cos{\theta_a}}\quad(\text{when }\frac{\pi}{N_c} \text{is small}).
\end{align}
 \begin{figure}[h!]
     \centering
     \begin{subfigure}{\textwidth}
         \includegraphics[width=0.49\textwidth]{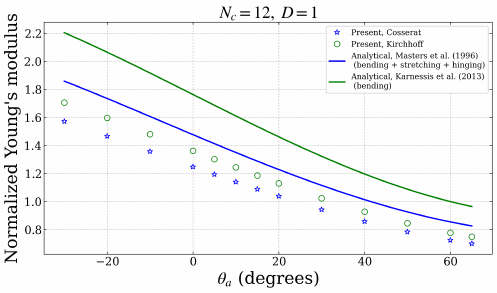}
         \includegraphics[width=0.49\textwidth]{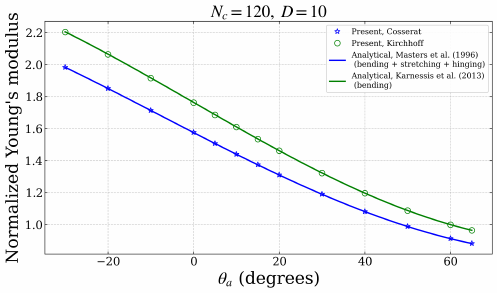}
         \caption{Normalized Young's modulus. It is normalized by its analytical value at $\theta_a=60^{\circ}$ of the bending-only analytical formula.}
         \label{subfig:youngs_modulus}
     \end{subfigure}
     \begin{subfigure}{\textwidth}
         \includegraphics[width=0.49\textwidth]{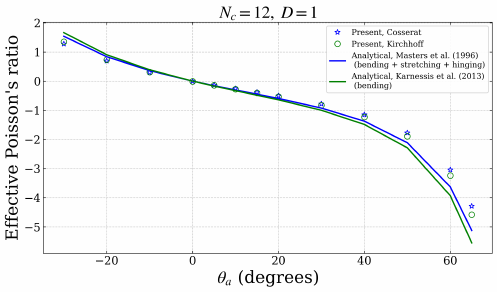}
         \includegraphics[width=0.49\textwidth]{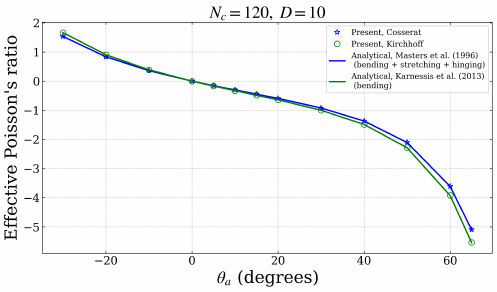}
         \caption{Effective Poisson's ratio}
         \label{subfig:poissons_ratio}
     \end{subfigure}
     \begin{subfigure}{\textwidth}
         \includegraphics[width=0.49\textwidth]{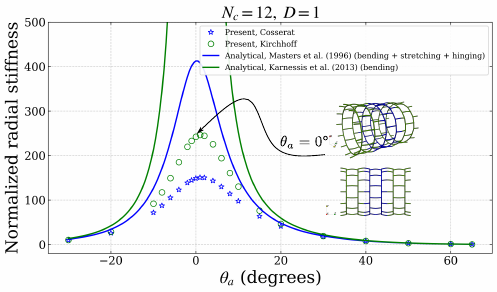}
         \includegraphics[width=0.49\textwidth]{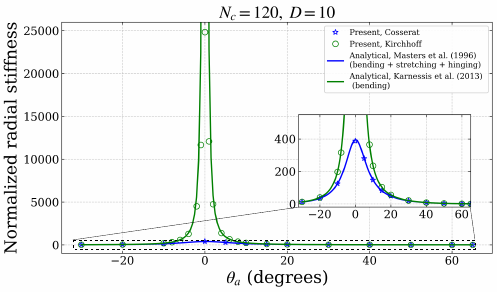}
         \caption{Normalized radial stiffness. It is normalized by its analytical value at $\theta_a=60^{\circ}$ of the bending-only analytical formula. The inset in the left figure shows the scaled deformed configuration of the pressurized tubular metamaterial at $\theta_a=0^{\circ}$. A similar deformation profile is also observed for the case of $D=10$.}
         \label{subfig:radial_stiffness}
     \end{subfigure}
     \caption{Effective properties of the tubular metamaterial in its reference state}
     \label{fig:small_strain_properties_aux_tub}
 \end{figure}
In Figure \ref{fig:small_strain_properties_aux_tub}, the effective macroscopic properties (at zero strain) of the auxetic tubular RVE, namely the longitudinal Young's modulus, effective Poisson's ratio and radial stiffness are presented. It can be seen that the auxeticity angle $\theta_a$ affects all of these properties significantly. The analytical formulas of these properties have been obtained earlier by assuming the discrete tube as an equivalent thin continuum tube \citep{masters1996models,karnessis2013uniaxial}. These formulas are reproduced in \ref{sec:analytical_formulas_aux_tube}. In the present approach, the tube's effective longitudinal Young's modulus is computed as follows:
\begin{align}
    \text{Young's Modulus} = \frac{\mathbb{C}^M_{33}}{A_{tube}} ~\text{or}~ \frac{\mathbb{C}^M_{44}}{I_{tube}}
\end{align}
where $A_{tube}=\pi D d$ and $I_{tube}=\pi D^3 d/8$ are the tube's cross-section's area and second area moment of inertia, respectively. Here, it is assumed that the wall thickness of the tube is the constituent rod's diameter $d$. Figure \ref{subfig:youngs_modulus} shows that the Young's modulus of the tube decreases monotonically as the angle $\theta_a$ increases from $-30^{\circ}$ to $60^{\circ}$. We also note that unlike the case of $D=1$, the obtained Young's modulus matches with the analytical formula very closely for the case when $D=10$. This is because at large diameter of the tube, the discrete metamaterial resembles more as a smooth continuum tube. We then discuss about the tube's effective Poisson's ratio. It is obtained in the present approach by applying a small axial strain to the tubular RVE and measuring the induced radial strain, i.e.,
\begin{align}
    \text{Effective Poisson's ratio} = -\frac{\text{induced radial strain}}{\text{imposed axial strain}} = -\frac{\frac{D^{def}-D}{D}}{\text{v}_3^M-1}
\end{align}
where $D^{def}$ is the average diameter of the tubular RVE in the deformed configuration. Figure \ref{subfig:poissons_ratio} shows that the effective Poisson's ratio is positive when $\theta_a<0^{\circ}$ but becomes negative when $\theta_a>0^{\circ}$. A close agreement with the analytical results is again achieved for the larger diameter case, i.e., when $D=10$. Next, we discuss the tube's radial stiffness. To obtain it, we apply a small outward radial force $F^{\text{radial}}$ on each node of the RVE (except the right boundary nodes). This results in an application of an equivalent internal pressure on the RVE given by
\begin{align}
    P_{eq} = \frac{\text{Total outward radial force}}{\text{Lateral surface area}}=\frac{{N^{RVE}_{\text{nodes}}~F^{\text{radial}}}}{\pi D L}
\end{align}
where $N^{RVE}_{\text{nodes}}$ is the total number of nodes on which the radial load $F^{\text{radial}}$ is applied. The nodes must be uniformly distributed in the RVE to ensure that a uniform pressure is applied throughout the domain. The internal pressure leads to a change in the average tubular diameter. We also relax the RVE such that the axial force $n_3^M$ becomes zero and then compute the radial stiffness using the following formula:
\begin{align} 
    \text{Radial stiffness} = \frac{\text{Equivalent pressure}}{\text{Radial strain (at $n_3^M=0$)}} = \frac{P_{eq}}{\frac{D^{def}-D}{D}}.
\end{align}
where $D^{def}$ is the average diameter of the tubular RVE in the presence of the internal pressure and at $\text{n}^M_3=0$. Figure \ref{subfig:radial_stiffness} shows that the radial stiffness achieves a maximum at $\theta_a = 0^{\circ}$. At this angle, the deformation is mostly stretching dominated as the members corresponding to length $a_2$ are aligned along the circumference of the tube, leading to maximum radial stiffness. It may be noted that the analytical formula (corresponding to bending only mode) predicts infinite radial stiffness at $\theta_a=0$ since no stretching of constituent rods is allowed there and hence the tube gets locked to internal pressure. However, our numerical simulation with Kirchhoff constituent rod shows that the tube is still able to accommodate radial strain but through non-uniform (but periodic) radial deformation along the tube length (shown in the inset of Figure \ref{subfig:radial_stiffness} (left)). Again, a close agreement is observed with the analytical results for the larger diameter case (see Figure \ref{subfig:radial_stiffness} (right)).

\begin{figure}[h!]
    \centering
    \begin{subfigure}{0.49\textwidth}
    \includegraphics[width=\textwidth]{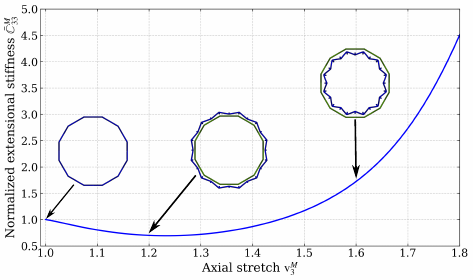}
    \caption{Uniform stretching}
    \label{subfig:stent_stretching}
    \end{subfigure}
\begin{subfigure}{0.49\textwidth}
    \includegraphics[width=\textwidth]{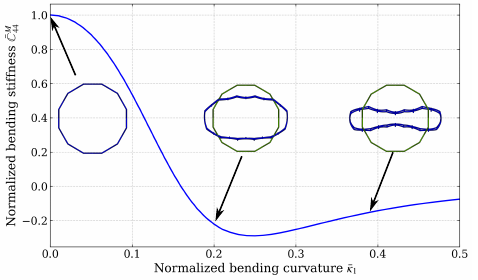}
    \caption{Uniform bending}
    \label{subfig:stent_bending}
\end{subfigure}
    \caption{Uniform stretching and bending of an auxetic tubular RVE ($\theta_a=30^{\circ}$ and $D=1$). The extensional and bending stiffnesses are normalized by their corresponding values at $\text{v}_3^M=1.0$ and $\bar{\kappa_1}^M=0.0$, respectively. The bending curvature is normalized according to $\bar{\kappa}_1^M = \kappa_1^M D$.}.
    \label{fig:stent_stretching_bending}
\end{figure}

Finally, we study the behaviour of the tubular metamaterial at large stretching and bending. Figure \ref{subfig:stent_stretching} shows that, upon stretching, the average radius of the tubular RVE increases initially showing auxetic behavior but at higher axial stretch, the radius starts to decrease eventually. We also see that initially the macroscopic extensional stiffness decreases and later increases. Next, upon uniform bending of the tubular metamaterial, we see in Figure \ref{subfig:stent_bending} that the model captures the well known Bazier's effect, i.e., ovalization of the cross-section \citep{kumar2011geometrically,coman2017bifurcation}.

\pagestyle{plain}

\noindent
\section{Conclusions} \label{sec:conclusions}
We presented a homogenization scheme to obtain the constitutive response of one-dimensional periodic rod networks when modeled as a special Cosserat rod. The HCB rule was used to strain the rod network uniformly (at macroscale) along its length. This, in turn, reduced the elasticity problem of the entire rod network to just its RVE. The RVE problem was solved under joint constraints (to enforce rod connectivity), helical boundary constraint (to enforce helical periodicity) and a set of global constraints. The constituent/microscale rods within the RVE were modeled using geometrically exact Cosserat rod theory. Our macro-to-micro transition map is also geometrically exact and fully nonlinear allowing the scheme to hold for arbitrarily large magnitudes of macroscopic rod's stretch, shear, curvature and twist. We also obtained closed form expressions of internal contact force, moment and stiffnesses of the homogenized rod in terms of the RVE's unknown configuration. Finally, we demonstrated our scheme using a selection of simple and complex RVEs. We began with simpler RVEs such as cross and square ones and showed that our approach results in macroscopic stiffnesses that match with the analytical expressions available in the literature and the ones derived by us. We also showed the effect of including stretch and shear modes in microscopic rods on various macroscopic stiffnesses obtained for such RVEs - shorter microscale rods showed more pronounced effect on the macroscopic response. Next, we considered square RVEs formed of helical microscale rods. We showed that its stretching results in a J-shaped curve which can be tuned by changing the RVE parameters such as the number of turns in the constituent helical rod, its pitch etc. Its bending showed unique softening behaviour in the moment-curvature relation. This was triggered by out-of-plane bending of the cross-linking rod within the RVE. Finally, we did homogenization of auxetic tubular structures and showed how by changing its design parameters, one can obtain desirable radial stiffness, flexural compliance, foreshortening behavior etc.\\\\
It would be interesting to further investigate the stability of rod-network RVEs when subjected to large macroscopic bending, twisting etc. One may have to resort to larger RVEs to obtain stable solutions \citep{dobson2007multilattice, herrnbock2022homogenization}. Introducing contact as a mode of interaction within the RVE to study biological networks, spiral strands, interpenetrating lattices \citep{surjadi2025double}, etc. would also be interesting. The proposed formulation assumes arbitrary constitutive behaviour for microscale rods. This feature can be utilized to see the effect of incorporating elastoplasticity or viscoelasticity in microscale rods on macroscopic behavior. These inelastic effects become an important consideration for applications such as in the case of cardiovascular stent deployment, muscle actuation \citep{wang2023braided} etc. Furthermore, due to increasing interest in deployable structures, 4D printing and robotic actuators/sensors, one could also think of including multiphysics phenomena such as thermoelasticity, chemoelasticity, electroelasticity \citep{barcelo2024computational} etc. at microscale and obtain the coupled multiphysics response at macroscale. We have also assumed the macroscopic strain energy density to depend only on rod's strains. However, in case the strain gradient along the arc-length of the macroscopic rod becomes significant, one would need to resort to higher gradient rod theory at macroscale \citep{yadav2024strain}. We also assumed the RVE to be a rod-network itself. However, it would be interesting to work out the homogenization scheme for RVEs consisting of network of plates or shells as in periodic origami structures \citep{sharma2025programmable} or three-dimensional continuum RVEs as in porous or composite beams. Finally, it would also be interesting to train a machine learning based constitutive model using data obtained from the proposed formulation and readily use such a tool in $\text{FE}^2$ setting \citep{schommartz2025physics,geiger2025multiscale}.

\biboptions{semicolon}
\bibliography{references_final}

\appendix
\pagestyle{plain}
\section{Important rotation related formulas}\label{appendix:rotation_formulas}
The Rodrigues' formula for the rotation tensor corresponding to rotation vector $\boldsymbol{\theta}$ is given by
\begin{align}\label{eq:rotation_formula}
    \textbf{R}(\boldsymbol{\theta})=e^{{\boldsymbol{\Theta}}} &= \textbf{I} + \alpha\boldsymbol{\Theta} + \frac{\beta}{2}\boldsymbol{\Theta}^2
\end{align}
where 
\begin{align}\label{eq:rotation_formula2}
    \boldsymbol{\Theta} = skew(\boldsymbol{\theta}),\quad \alpha(\boldsymbol{\theta}) = \frac{\text{sin}(||\boldsymbol{\theta}||)}{||\boldsymbol{\theta}||},\quad\beta(\boldsymbol{\theta}) = 2\frac{1-\text{cos}(||\boldsymbol{\theta}||)}{||\boldsymbol{\theta}||^2}.
\end{align}
The variation/linearization of the rotation tensor is obtained through a multiplicative perturbation in the following way:
\begin{align}
    \frac{d}{d\epsilon}\textbf{R}^{\epsilon}\bigg|_{\epsilon=0} = \frac{d}{d\epsilon}\left[e^{\epsilon\delta\boldsymbol{\Theta}}\textbf{R}\right]\bigg|_{\epsilon=0} = \delta\boldsymbol{\Theta}\textbf{R}
\end{align}
where $\delta\boldsymbol{\Theta}=skew(\delta\boldsymbol{\theta})$. Thus, we have
\begin{align}\label{eq:multiplicative_rot_variation}
    \delta\boldsymbol{\theta} = \text{axial}\left(\frac{d \textbf{R}^{\epsilon}}{d\epsilon}\bigg|_{\epsilon=0}\textbf{R}^T\right).
\end{align}
Let us denote the variation of rotation vector $\boldsymbol{\theta}$ by $\triangle\boldsymbol{\theta}$. It is related to the multiplicative perturbation $\delta\boldsymbol{\theta}$ using the following identity \citep{simo1988dynamics}:
\begin{align}\label{eq:additive_multiplative_variation_rel}
    \triangle{\boldsymbol{\theta}} =\frac{d\boldsymbol{\theta}^{\epsilon}}{d\epsilon}\bigg |_{\epsilon=0} =\textbf{T}(\boldsymbol{\theta})\delta\boldsymbol{\theta},\quad \delta\boldsymbol{\theta}=\textbf{T}^{-1}(\boldsymbol{\theta})\triangle{\boldsymbol{\theta}}
\end{align}
where
\begin{align}\label{eq:tangent_operator}
    \textbf{T}(\boldsymbol{\theta}) = \textbf{I} - \frac{\boldsymbol{\Theta}}{2} + (1-\frac{\alpha}{\beta})\left(\frac{\boldsymbol{\Theta}}{||\boldsymbol{\theta}||}\right)^2, ~\textbf{T}^{-1}(\boldsymbol{\theta}) = \textbf{I} + \frac{\beta}{2}\boldsymbol{\Theta} + (1-\alpha)\left(\frac{\boldsymbol{\Theta}}{||\boldsymbol{\theta}||}\right)^2.
\end{align}
 Finally, the following identity holds:
\begin{align}\label{eq:tangent_op_identity_1}
    \textbf{T}(\textbf{Q}\boldsymbol{\theta}) = \textbf{Q}\textbf{T}(\boldsymbol{\theta})\textbf{Q}^T \quad \text{and} \quad
     \textbf{T}^{-1}(\textbf{Q}\boldsymbol{\theta}) = \textbf{Q}\textbf{T}^{-1}(\boldsymbol{\theta})\textbf{Q}^T
\end{align}
where $\textbf{Q}$ is an arbitrary rotation tensor. The gradient of the operator $\textbf{T}(\boldsymbol{\theta})$ is written as
\begin{align}\label{eq:derivative_tangent_operator}
    \frac{d}{d\epsilon}\textbf{T}(\boldsymbol{\theta}^{\epsilon})\bigg |_{\epsilon=0}   = -\frac{d\boldsymbol{\Theta}}{d\epsilon}\bigg |_{\epsilon=0}+\frac{1-\gamma}{||\boldsymbol{\theta}||^2}(\boldsymbol{\Theta}\frac{d\boldsymbol{\Theta}}{d\epsilon}\bigg |_{\epsilon=0}+\frac{d\boldsymbol{\Theta}}{d\epsilon}\bigg |_{\epsilon=0}\boldsymbol{\Theta}) + \frac{(1/\beta+\gamma-2)}{||\boldsymbol{\theta}||^4}\boldsymbol{\Theta}^2\left(\boldsymbol{\theta}\cdot\frac{d\boldsymbol{\theta}}{d\epsilon}\bigg |_{\epsilon=0}\right)
\end{align}
\section{Linearization of various constraints}\label{appendix:contraint_linearization}
\subsection{Linearization of global constraints}
Let us first consider the ``mmt"" constraint in \eqref{eq:mixed_moment_gc}, i.e.,
\begin{align}
  \boldsymbol{\mathfrak{m}} =  r_2r_3\textbf{e}_1 +r_1r_3\textbf{e}_2 +\bigg[\text{arctan}\bigg(\frac{r_{2}}{r_{1}}\bigg)-\text{arctan}\bigg(\frac{R_{2}}{R_{1}}\bigg)\bigg]\textbf{e}_3.
\end{align}
Here the indices ($\alpha_k,\beta_k$) have been omitted for simplicity. The linearization of the above constraint is given by
\begin{align}
    \frac{d\boldsymbol{\mathfrak{m}}}{d\epsilon}\bigg\vert_{\epsilon=0}
    = \boldsymbol{\Xi}^{\textbf{r}^T}\delta\textbf{r}
\end{align}
where
\begin{align}
    [\boldsymbol{\Xi}^{\textbf{r}^T}] =\begin{bmatrix}
0 & r_3 & r_2 \\
r_3 & 0 & r_1 \\
-\displaystyle\frac{r_2}{r_1^2 + r_2^2} & \displaystyle \frac{r_1}{r_1^2 + r_2^2} & 0
\end{bmatrix}.
\end{align}
Next we obtain the linearization of ``delta-theta" constraint in equation \eqref{eq:delta_theta_constraint}. Let us consider the change in rotation vector $\Delta\boldsymbol{\theta}$ (omitting the indices ($\alpha_k,\beta_k$)) there and define its pull-back to the reference configuration as
\begin{align}\label{delta_theta0}
    \textbf{R}(\boldsymbol{\theta}) = \textbf{R}(\hat{\boldsymbol{\theta}})\text{exp}^{\Delta\boldsymbol{\Theta}_{0}}
\end{align}
such that
\begin{align}\label{delta_theta_theta0_relation}
    \Delta\boldsymbol{\theta}_{0}=\hat{\textbf{R}}^T\Delta\boldsymbol{\theta}
\end{align}
Linearizing equation \eqref{delta_theta0}, further taking the axial and applying \eqref{eq:additive_multiplative_variation_rel}, we obtain
\begin{align}
    \frac{d}{d\epsilon}\Delta\boldsymbol{\theta}_{0}^{\epsilon}\bigg|_{\epsilon=0} = \textbf{T}(\Delta\boldsymbol{\theta}_{0})\hat{\textbf{R}}^T\delta\boldsymbol{\theta}.
\end{align}
Using \eqref{delta_theta_theta0_relation} and \eqref{eq:tangent_op_identity_1}, we finally have the lineaization of change in rotation vector as follows:
\begin{align}\label{delta_theta_linearization}
    \frac{d}{d\epsilon}\Delta\boldsymbol{\theta}^{\epsilon}\bigg|_{\epsilon=0} &= \hat{\textbf{R}}\textbf{T}(\Delta\boldsymbol{\theta}_{0})\hat{\textbf{R}}^T\delta\boldsymbol{\theta}=\textbf{T}(\Delta\boldsymbol{\theta})\delta\boldsymbol{\theta}= \boldsymbol{\Xi}^{\boldsymbol{\theta}^T}\delta\boldsymbol{\theta}
\end{align}
where 
\begin{align}\label{delta_theta_constraint_lin}
    \boldsymbol{\Xi}^{\boldsymbol{\theta}^T} = \textbf{T}(\Delta\boldsymbol{\theta}).
\end{align}
\subsection{Linearization of internal joint constraints}
In this section, we derive the linearization of internal joint rotational constraint term. In order to do so, we first define the constraint in \eqref{eq:internal_jt_rotation_constraint2} as
\begin{align}
  \boldsymbol{\mathcal{J}}_{i,k}^{rot} \equiv  \boldsymbol{\theta}^{rel}_{i,k} = \textbf{0}
\end{align}
where $\boldsymbol{\theta}^{rel}_{i,k}$ is the relative rotation vector. It is defined using the left multiplicative update as
\begin{align}\label{eq:IJ_theta_rel_left_update}
    (\textbf{R}^{\alpha_k}_{\beta_k})^T\textbf{R}^{\alpha_1}_{\beta_1} = e^{\boldsymbol{\Theta}^{rel}_{i,k}}(\hat{\textbf{R}}^{\alpha_k}_{\beta_k})^T\hat{\textbf{R}}^{\alpha_1}_{\beta_1}
\end{align}
or using the right multiplicative update as
\begin{align}\label{eq:IJ_theta_rel_right_update}
    (\textbf{R}^{\alpha_k}_{\beta_k})^T\textbf{R}^{\alpha_1}_{\beta_1} = (\hat{\textbf{R}}^{\alpha_k}_{\beta_k})^T\hat{\textbf{R}}^{\alpha_1}_{\beta_1}e^{\boldsymbol{\Theta}^{rel}_{0,i,k}}
\end{align}
such that
\begin{align}\label{eq:IJ_theta_rel_left_right_relation}
    \boldsymbol{\theta}^{rel}_{i,k} = (\hat{\textbf{R}}^{\alpha_k}_{\beta_k})^T\hat{\textbf{R}}^{\alpha_1}_{\beta_1}\boldsymbol{\theta}_{0,i,k}^{rel}.
\end{align}
Then the linearization of the rotational internal joint constraint \eqref{eq:internal_jt_rotation_constraint2} is given by
\begin{align}
    \frac{d}{d\epsilon}\boldsymbol{\mathcal{J}}_{i,k}^{rot}\bigg|_{\epsilon=0} = \frac{d\boldsymbol{\theta}^{rel}_{i,k}}{d\epsilon}\Bigg|_{\epsilon=0}.
\end{align}
Now, we linearize equation \eqref{eq:IJ_theta_rel_right_update}, take its $axial$ on both the sides and use \eqref{eq:additive_multiplative_variation_rel} to obtain
\begin{align}
        (\textbf{R}^{\alpha_k}_{\beta_k})^T(\delta\boldsymbol{\theta}^{\alpha_1}_{\beta_1} - \delta\boldsymbol{\theta}^{\alpha_k}_{\beta_k}) =(\hat{\textbf{R}}^{\alpha_k}_{\beta_k})^T\hat{\textbf{R}}^{\alpha_1}_{\beta_1}\textbf{T}^{-1}(\boldsymbol{\theta}_{0,i,k}^{rel})\frac{d\boldsymbol{\theta}_{0,i,k}^{rel}}{d\epsilon}\bigg\vert_{\epsilon=0}.
\end{align}
Finally, using \eqref{eq:IJ_theta_rel_left_right_relation} and \eqref{eq:tangent_op_identity_1}, we get
\begin{align}\label{eq:internal_jt_thete_rel_derivative}
    \frac{d\boldsymbol{\theta}^{rel}_{i,k}}{d\epsilon}\Bigg|_{\epsilon=0} = \textbf{T}(\boldsymbol{\theta}^{rel}_{i,k})(\textbf{R}^{\alpha_k}_{\beta_k})^T(\delta\boldsymbol{\theta}^{\alpha_1}_{\beta_1}-\delta\boldsymbol{\theta}^{\alpha_k}_{\beta_k})
\end{align}
or
\begin{align}\label{eq:internal_jt_const_rot_derivative}
    \frac{d\boldsymbol{\mathcal{J}}^{rot}_{i,k}}{d\epsilon}\Bigg|_{\epsilon=0} = \textbf{T}(\boldsymbol{\mathcal{J}}^{rot}_{i,k})(\textbf{R}^{\alpha_k}_{\beta_k})^T(\delta\boldsymbol{\theta}^{\alpha_1}_{\beta_1}-\delta\boldsymbol{\theta}^{\alpha_k}_{\beta_k}).
\end{align}
\subsection{Linearization of helical constraints}
Now, we derive the linearization of helical constraint terms. The linearization of the translational helical constraint \eqref{eq:position_constraint_hcb} is given by
\begin{align}
    \frac{d}{d\epsilon}\boldsymbol{\mathcal{H}}_k^{trans} = \delta\textbf{r}_{\beta_{k_R}}^{\alpha_{k_R}} - \textbf{R}^M\delta\textbf{r}_{\beta_{k_L}}^{\alpha_{k_L}} - \bigg(\int_0^{L}\delta \textbf{R}^M(l)ds\bigg)\textbf{v}_0^M - \bigg(\int_0^{L} \textbf{R}^M(l)ds\bigg)\delta \textbf{v}_0^M	-\delta \textbf{R}^M(L)\textbf{r}^{\alpha_{k_L}}_{\beta_{k_L}}.
\end{align}
Note that, in the above linearization, we have also included the derivative with respect to macroscopic strains - this won't be needed while linearizing the weak form for prescribed macroscopic strains. The closed form expressions for the terms  $\int \textbf{R}^Mdl$, $\delta \textbf{R}^M$ and $\int \delta\textbf{R}^Mdl$, wherein $\delta=\frac{\partial}{\partial \kappa_i^M}$, can be seen in the Appendix B of \citet{kumar2016helical}. Next, the linearization of rotational helical constraint \eqref{eq:rotation_constraint_hcb} is obtained. First, we define the helical constraint in \eqref{eq:rotation_constraint_hcb} as 
\begin{align}
    \boldsymbol{\mathcal{H}}_k^{rot}\equiv\boldsymbol{\theta}^{rel}_k = \textbf{0}.
\end{align}
where $\boldsymbol{\theta}^{rel}_k$ is in turn defined using the left multiplicative update of \eqref{eq:helical_constraint_rotmat_eq} in the form
\begin{align}\label{eq:left_update_helical_constraint}
    (\textbf{R}^{\alpha_{k_R}}_{\beta_{k_R}})^T\textbf{R}^M\textbf{R}_{\beta_{k_L}}^{\alpha_{k_L}} = e^{\boldsymbol{\Theta}^{rel}_k}(\hat{\textbf{R}}_{\beta_{k_R}}^{\alpha_{k_R}})^T(\hat{\textbf{R}}_{\beta_{k_L}}^{\alpha_{k_L}})
\end{align}
 and using the right multiplicative update as
\begin{align}\label{eq:helical_jt_theta_rel_right}
    (\textbf{R}^{\alpha_{k_R}}_{\beta_{k_R}})^T\textbf{R}^M\textbf{R}_{\beta_{k_L}}^{\alpha_{k_L}} = (\hat{\textbf{R}}_{\beta_{k_R}}^{\alpha_{k_R}})^T(\hat{\textbf{R}}_{\beta_{k_L}}^{\alpha_{k_L}})e^{\boldsymbol{\Theta}^{rel}_{0,k}}
\end{align}
such that
\begin{align}\label{eq:theta_rel_helical_joint_left_right}
    \boldsymbol{\theta}_k^{rel} = (\hat{\textbf{R}}_{\beta_{k_R}}^{\alpha_{k_R}})^T(\hat{\textbf{R}}_{\beta_{k_L}}^{\alpha_{k_L}})\boldsymbol{\theta}_{0,k}^{rel}.
\end{align}
The linearization of the helical constraint is then given by
\begin{align}
    \frac{d}{d\epsilon}\boldsymbol{\mathcal{H}}_k^{rot} = \frac{d}{d\epsilon}\boldsymbol{\theta}^{rel}_k.
\end{align}
In order to obtain the linearization above, we first linearize equation \eqref{eq:helical_jt_theta_rel_right}, take its $axial$ on both the sides and then apply \eqref{eq:additive_multiplative_variation_rel} to obtain
\begin{align}
    (\textbf{R}^{\alpha_{k_R}}_{\beta_{k_R}})^T(\textbf{R}_M\delta\boldsymbol{\theta}^{\alpha_{k_L}}_{\beta_{k_L}}-\delta\boldsymbol{\theta}^{\alpha_{k_R}}_{\beta_{k_R}}) +(\textbf{R}^{\alpha_{k_R}}_{\beta_{k_R}})^Taxial(\delta\textbf{R}^M(\textbf{R}^M)^T) = (\hat{\textbf{R}}^{\alpha_{k_R}}_{\beta_{k_R}})^T\hat{\textbf{R}}^{\alpha_{k_L}}_{\beta_{k_L}}\textbf{T}^{-1}(\boldsymbol{\theta}_{0,k}^{rel})\frac{d\boldsymbol{\theta}_{0,k}^{rel}}{d\epsilon}.
\end{align}
Using \eqref{eq:theta_rel_helical_joint_left_right} and \eqref{eq:tangent_op_identity_1}, we finally obtain
\begin{align}
    \frac{d\boldsymbol{\theta}^{rel}_k}{d\epsilon} = \textbf{T}(\boldsymbol{\theta}^{rel}_k)(\textbf{R}^{\alpha_{k_R}}_{\beta_{k_R}})^T(\textbf{R}^M\delta\boldsymbol{\theta}^{\alpha_{k_L}}_{\beta_{k_L}} - \delta\boldsymbol{\theta}^{\alpha_{k_R}}_{\beta_{k_R}})  +\textbf{T}(\boldsymbol{\theta}^{rel}_k)(\textbf{R}^{\alpha_{k_R}}_{\beta_{k_R}})^Taxial(\delta\textbf{R}^M(\textbf{R}^M)^T)
\end{align}
or
\begin{align}\label{eq:helical_jt_thete_rel_derivative}
    \frac{d\boldsymbol{\mathcal{H}}^{rot}_k}{d\epsilon} = \textbf{T}(\boldsymbol{\mathcal{H}}^{rot}_k)(\textbf{R}^{\alpha_{k_R}}_{\beta_{k_R}})^T(\textbf{R}^M\delta\boldsymbol{\theta}^{\alpha_{k_L}}_{\beta_{k_L}} - \delta\boldsymbol{\theta}^{\alpha_{k_R}}_{\beta_{k_R}})  +\textbf{T}(\boldsymbol{\mathcal{H}}^{rot}_k)(\textbf{R}^{\alpha_{k_R}}_{\beta_{k_R}})^Taxial(\delta\textbf{R}^M(\textbf{R}^M)^T).
\end{align}
Here again, the derivatives of macroscopic strain terms vanish if they are assumed constant.
\section{Governing equations of discrete helical rod}\label{appendix:euler_lagrange_discrete element}
 The governing equations of the discrete helical rod is obtained by evaluating the derivative \eqref{deltaI}. To this end, the following perturbation is applied:
\begin{align}
    \textbf{r}^{\epsilon}_i = \textbf{r}_i + \epsilon\delta\textbf{r}_i\nonumber\\
    \textbf{R}_i^{\epsilon} = \text{exp}(\epsilon\delta\boldsymbol{\Theta}_i)\textbf{R}_i
\end{align}
Then, using \eqref{eq:energy_discrete_sc_rod} and neglecting $W^{ext}$, the derivative \eqref{deltaI} is given by:
\begin{align}\label{energy_derivative}
    \frac{d\mathcal{I}}{d\epsilon}\bigg |_{\epsilon=0} = \sum_{i=1}^{N_{el}}\bigg[\textbf{n}_0^i\cdot\frac{d\textbf{v}_0^{i,\epsilon}}{d\epsilon}\bigg|_{\epsilon=0}+\textbf{m}_0^i\cdot\frac{d\textbf{k}_0^{i,\epsilon}}{d\epsilon}\bigg|_{\epsilon=0}\bigg]h^i = 0
    \end{align}
where $\textbf{n}_0^i$ and $\textbf{m}_0^i$ are the stress resultants in the $i^{th}$ discrete helical element. Linearizing \eqref{eq:k0_discrete} and using \eqref{eq:tangent_operator}, we get
\begin{align}\label{derivative_k0}
    \frac{d \textbf{k}_0^{i,\epsilon}}{d \epsilon}\bigg|_{\epsilon = 0} &= \textbf{T}^i\textbf{R}_1^{i,T}\bigg(\frac{\delta\boldsymbol{\theta}_2^i-\delta\boldsymbol{\theta}_1^i}{h^i}\bigg)\nonumber\\
    &= \boldsymbol{\mathcal{K}}^i\mathbb{R}_1^T\delta\textbf{x}_{12}
\end{align}
where
\begin{align}
\quad\mathbb{R}_1^i &= \text{diag}(\textbf{R}^i_1,\textbf{R}^i_1,\textbf{R}^i_1,\textbf{R}^i_1),\quad
    [\boldsymbol{\mathcal{K}}^i] = \begin{bmatrix}
        \textbf{0} & \boldsymbol{\mathcal{K}}_{\boldsymbol{\theta}1}^i & \textbf{0} & \boldsymbol{\mathcal{K}}^i_{\boldsymbol{\theta}2}
    \end{bmatrix}_{3\times12},\nonumber\\
    \quad
    \boldsymbol{\mathcal{K}}^i_{\boldsymbol{\theta}1} &=  -\frac{\textbf{T}^i}{h^i}\quad \text{and} \quad 
    \boldsymbol{\mathcal{K}}^i_{\boldsymbol{\theta}2} =  \frac{\textbf{T}^i}{h^i}. 
\end{align}
 Using \eqref{v01} and \eqref{eq:derivative_tangent_operator}, the linearization of $\textbf{v}_0^i$ is given by
 \begin{align}\label{eq:derivative_v0_discrete}
     \frac{d\textbf{v}_0^{i,\epsilon}}{d\epsilon}\bigg |_{\epsilon=0} = \boldsymbol{\mathcal{V}}^i(\mathbb{R}_1^{i})^T\delta\textbf{x}^i_{12}
 \end{align}
 where
\begin{align}
\boldsymbol{\mathcal{V}}^i &= \begin{bmatrix}
\boldsymbol{\mathcal{V}}^i_{\textbf{r}1} & \boldsymbol{\mathcal{V}}^i_{\boldsymbol{\theta}1} & \boldsymbol{\mathcal{V}}^i_{\textbf{r}2} & \boldsymbol{\mathcal{V}}^i_{\boldsymbol{\theta}2}
\end{bmatrix}^i_{3\times12}\nonumber\\
\boldsymbol{\mathcal{V}}^i_{\textbf{r}1}
&=-\frac{\textbf{T}^i}{h^i},\quad \boldsymbol{\mathcal{V}}^i_{\boldsymbol{\theta}1} = (\textbf{P}^i\boldsymbol{\mathcal{K}}^i_{\boldsymbol{\theta}_1}+\textbf{T}^i[(\textbf{T}^{-1})^i\textbf{v}^i_0]_{\times}),\quad \boldsymbol{\mathcal{V}}^i_{\textbf{r}2}
=\frac{\textbf{T}^i}{h^i},\quad
\boldsymbol{\mathcal{V}}^i_{\boldsymbol{\theta}2} = \textbf{P}^i\boldsymbol{\mathcal{K}}^i_{\boldsymbol{\theta}_2}\nonumber\\
\textbf{P}^i = \textbf{P}(\textbf{v}_0^i,h^i\textbf{k}_0^i) &= 0.5h^i[(\textbf{T}^{-1})^i\textbf{v}^i_0]_{\times}-\frac{1-\gamma^i}{||\textbf{k}^i_0||^2}\bigg(\textbf{K}^i_0[(\textbf{T}^{-1})^i\textbf{v}^i_0]_{\times}+[\textbf{K}^i_0(\textbf{T}^{-1})^i\textbf{v}^i_0]_{\times}\bigg)\nonumber\\
&+\frac{(1/\beta^i+\gamma^i-2)}{||\textbf{k}^i_0||^4}(\textbf{K}^i_0)^2[((\textbf{T}^{-1})^i\textbf{v}^i_0)\otimes\textbf{k}^i_0]_{\times}
\end{align}
Substituting \eqref{derivative_k0} and \eqref{eq:derivative_v0_discrete} in \eqref{energy_derivative}, we get
\begin{align}
    &\sum_{i=1}^{N_{el}} \bigg\{\textbf{n}^i_0\cdot(\boldsymbol{\mathcal{V}}^i(\mathbb{R}^i_1)^T\delta\textbf{x}^i_{12})+\textbf{m}^i_0\cdot(\boldsymbol{\mathcal{K}}^i(\mathbb{R}^i_1)^T\delta\textbf{x}^i_{12})\bigg\}h^i=0\nonumber\\
    &\sum_{i=1}^{N_{el}}\mathbb{R}^i_1[\boldsymbol{(\mathcal{V}}^i)^T\textbf{n}^i_0 + (\boldsymbol{\mathcal{K}}^i)^T\textbf{m}^i_0]h^i\cdot\delta\textbf{x}^i_{12}=0
\end{align}
Collecting the coefficients of the global degrees of freedoms $\delta\textbf{x}_i$ for internal nodes, we have
\begin{align}
    \sum_{i=2}^{N_{el}}
    &\big\{\textbf{R}_1^{i-1}[(\boldsymbol{\mathcal{V}}_{\textbf{r}2}^{i-1})^T\textbf{n}^{i-1}_0]h^{i-1}+\textbf{R}_1^{i}[(\boldsymbol{\mathcal{V}}_{\textbf{r}1}^{i})^T\textbf{n}^{i}_0]h^{i}\big\}\cdot\delta\textbf{r}_i\nonumber\\
    &+\big\{\textbf{R}_1^{i-1}[(\boldsymbol{\mathcal{V}}_{\boldsymbol{\theta}2}^{i-1})^T\textbf{n}^{i-1}_0+\boldsymbol{(\mathcal{K}}_{\boldsymbol{\theta}2}^{i-1})^T\textbf{m}^{i-1}_0]h^{i-1}+\textbf{R}_1^{i}[\boldsymbol{(\mathcal{V}}_{\boldsymbol{\theta}1}^{i})^T\textbf{n}^{i}_0+\boldsymbol{(\mathcal{K}}_{\boldsymbol{\theta}1}^{i})^T\textbf{m}^{i}_0]h^{i}\big\}\cdot\delta\boldsymbol{\theta}_i=\textbf{0}.
\end{align}
 Note that the rod's internal node $i$ has a contribution from both the adjoining elements $i-1$ and $i$. The boundary nodes $i=1$ and $i=N_{el}+1$ will have contribution from only the boundary elements $1$ and $N_{el}$, respectively. The boundary conditions on the rod are incorporated on these boundary nodes. For arbitrary $\delta \textbf{r}_i$ and $\boldsymbol{\theta}_i$, we get
\begin{align}
    \textbf{R}_1^{i-1}[(\boldsymbol{\mathcal{V}}_{\textbf{r}2}^{i-1})^T\textbf{n}^{i-1}_0]h^{i-1}+\textbf{R}_1^{i}[(\boldsymbol{\mathcal{V}}_{\textbf{r}1}^{i})^T\textbf{n}^{i}_0]h^{i} = \textbf{0} \quad \text{and}\nonumber\\
    \textbf{R}_1^{i-1}[(\boldsymbol{\mathcal{V}}_{\boldsymbol{\theta}2}^{i-1})^T\textbf{n}^{i-1}_0+\boldsymbol{(\mathcal{K}}_{\boldsymbol{\theta}2}^{i-1})^T\textbf{m}^{i-1}_0]h^{i-1}+\textbf{R}_1^{i}[\boldsymbol{(\mathcal{V}}_{\boldsymbol{\theta}1}^{i})^T\textbf{n}^{i}_0+\boldsymbol{(\mathcal{K}}_{\boldsymbol{\theta}1}^{i})^T\textbf{m}^{i}_0]h^{i} = \textbf{0}
\end{align}
that form the discrete linear and angular momentum balance equations of the rod.

Using the similar approach as above, the governing equations for a Kirchhoff rod are obtained. Here, the Lagrange multiplier $\boldsymbol{\lambda}^i_0$ is an additional unknown and is perturbed as follows:
\begin{align}
    \boldsymbol{\lambda}_0^{i,\epsilon} = \boldsymbol{\lambda}_0 + \epsilon\delta\boldsymbol{\lambda}_0
\end{align}
The first derivative of the constrained energy function \eqref{eq:energy_discrete_kirchhoff_rod} when set to zero yields the following discrete equations for a Kirchhoff rod:
\begin{align}
    \textbf{R}_1^{i-1}[(\boldsymbol{\mathcal{V}}_{\textbf{r}2}^{i-1})^T\boldsymbol{\lambda}^{i-1}_0]h^{i-1}+\textbf{R}_1^{i}[(\boldsymbol{\mathcal{V}}_{\textbf{r}1}^{i})^T\boldsymbol{\lambda}^{i}_0]h^{i} &= \textbf{0},\nonumber\\
    \textbf{R}_1^{i-1}[(\boldsymbol{\mathcal{V}}_{\boldsymbol{\theta}2}^{i-1})^T\boldsymbol{\lambda}^{i-1}_0+\boldsymbol{(\mathcal{K}}_{\boldsymbol{\theta}2}^{i-1})^T\textbf{m}^{i-1}_0]h^{i-1}+\textbf{R}_1^{i}[\boldsymbol{(\mathcal{V}}_{\boldsymbol{\theta}1}^{i})^T\boldsymbol{\lambda}^{i}_0+\boldsymbol{(\mathcal{K}}_{\boldsymbol{\theta}1}^{i})^T\textbf{m}^{i}_0]h^{i} &= \textbf{0} \quad \text{and} \nonumber\\
    \textbf{v}_0^i - \textbf{e}_3 &= \textbf{0}.
\end{align}
The Lagrange multiplier $\boldsymbol{\lambda}_0^i$ has the interpretation of the internal contact force in the helical element. The governing equations for the unshearable rod model can be obtained similarly by taking the derivative of \eqref{eq:energy_discrete_unshearable_rod} whose details we skip here.
\section{Configuration of a nearly helical rod}\label{appendix:helical_config}
The initial configuration of the nearly helical rod in \cite{chang2022mechanics} can be written as $\textbf{r}=X(\Phi)\textbf{e}_1+Y(\Phi)\textbf{e}_2+Z(\Phi)\textbf{e}_3$ where
\begingroup
\allowdisplaybreaks
\begin{align*}
X(\Phi) &= 
\begin{cases}
r_h \sin^2 \left( \frac{\pi Z^2(\Phi)}{2p_j^2} \right) \sin \Phi, & 0 \le \Phi < \frac{\pi}{2} \\[8pt]
r_h \sin \Phi, & \frac{\pi}{2} \le \Phi < \left(2N_h + \frac{1}{2}\right)\pi \\[8pt]
r_h \sin^2 \left( \frac{\pi (N_h \rho_h + 2p_j - Z(\Phi))^2}{2p_j^2} \right) \sin \Phi, & \left(2N_h + \frac{1}{2}\right)\pi \le \Phi \le (2N_h + 1)\pi
\end{cases} \\[10pt]
Y(\Phi) &= 
\begin{cases}
- r_h \sin^2 \left( \frac{\pi Z^2(\Phi)}{2p_j^2} \right) \cos \Phi, & 0 \le \Phi < \frac{\pi}{2} \\[8pt]
- r_h \cos \Phi, & \frac{\pi}{2} \le \Phi < \left(2N_h + \frac{1}{2}\right)\pi \\[8pt]
- r_h \sin^2 \left( \frac{\pi (N_h \rho_h + 2p_j - Z(\Phi))^2}{2p_j^2} \right) \cos \Phi, & \left(2N_h + \frac{1}{2}\right)\pi \le \Phi \le (2N_h + 1)\pi
\end{cases} \\[10pt]
Z(\Phi) &= 
\begin{cases}
\dfrac{p_h}{4\pi^3}\left(3 - \dfrac{16p_j}{p_h}\right)\Phi^3 + \dfrac{p_h}{4\pi^2}\Phi^2 + \dfrac{p_h}{16\pi}\left(\dfrac{48p_j}{p_h} - 5\right)\Phi, \qquad 0 \le \Phi < \frac{\pi}{2} \\[10pt]
\dfrac{p_h}{2\pi}\left(\Phi - \dfrac{\pi}{2}\right) + p_j,  \qquad\frac{\pi}{2} \le \Phi < \left(2N_h + \frac{1}{2}\right)\pi \\[10pt]
N_h p_h + 2p_j - \dfrac{p_h}{4\pi^3}\left(3 - \dfrac{16p_j}{p_h}\right)[(2N_h + 1)\pi - \Phi]^3 \\[4pt]
\quad - \dfrac{p_h}{4\pi^2}[(2N_h + 1)\pi - \Phi]^2 - \dfrac{p_h}{16\pi}\left(\dfrac{48p_j}{p_h} - 5\right)[(2N_h + 1)\pi - \Phi],\\
\hspace{80mm} \left(2N_h + \frac{1}{2}\right)\pi \le \Phi
\le (2N_h + 1)\pi
\end{cases}
\end{align*}
\endgroup
\section{Cross RVE analytical derivations}\label{cross_rve_stiffness}
\begin{figure}[h!]
    \centering
    \includegraphics[width=0.9\linewidth]{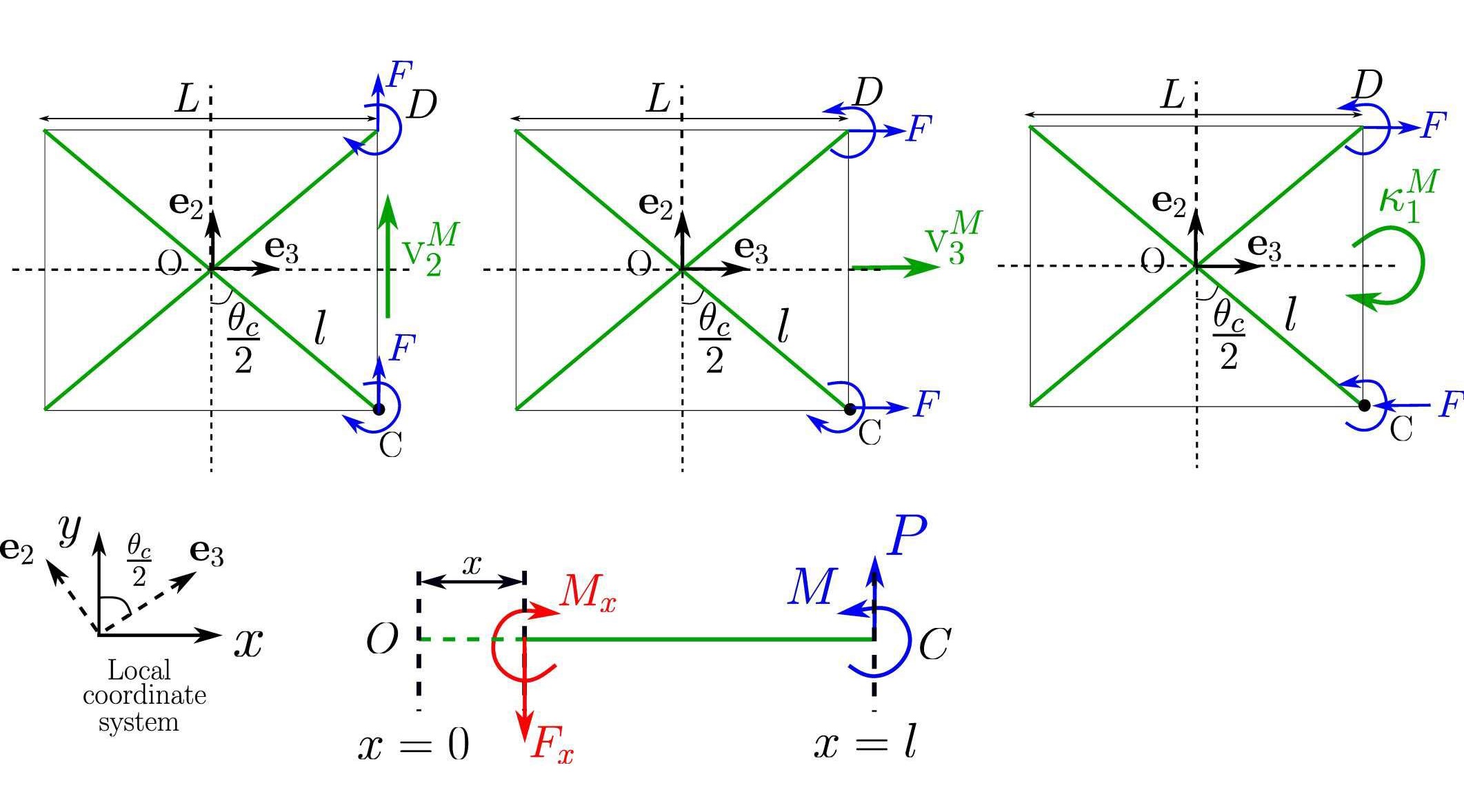}
    \caption{Cross RVE schemetic under uniform in plane shearing ($P = F\sin{\frac{\theta_c}{2}}$), stretching ($P = F\cos{\frac{\theta_c}{2}}$) and bending ($P = -F\cos{\frac{\theta_c}{2}}$). }
    \label{fig:cross_rve_analytic_schematic}
\end{figure}

In this section, we derive the analytical formulas of macroscopic stiffnesses $\mathbb{C}_{22}^M,\mathbb{C}_{33}^M,\mathbb{C}_{44}^M$ given in Table \ref{table:cross_rve_parameters}. The assumptions in the derivations are (i) infinitesimal macroscopic strain and (ii) Euler-Bernouli beam model for microscale rods. Figure \ref{fig:cross_rve_analytic_schematic} shows the three cases of uniform macroscopic in-plane shear (left figure), extension (middle figure) and in-plane bending (right figure). Note that we have centered the RVE at the origin in order to simplify the analysis. The rod $OC$ is isolated from the cross RVE and the bending moment-deflection problem is solved for the three deformation cases.
Firstly, bending moment balance at an arbitrary point $x$ on the microscale rod $OC$ is given by
\begin{align}\label{eq:cross_microscale_moment_balance}
    -M_x +M +P(l-x)=0.
\end{align}
Here the sign convention in the local coordinate system of the beam is - upward deflection and counter-clockwise moment are considered positive. The quantities $P$ and $M$ are unknown reaction force and moment from the joint connections of the rod with the neighbouring image. Assuming small deformation of microscale rod $OC$, one can write
\begin{align}
    M_x = EI\frac{d^2y}{dx^2} 
\end{align}
where $y(x)$ is the transverse deflection. Substituting the above microscale moment-curvature relationship in \eqref{eq:cross_microscale_moment_balance} and integrating, we get the equations for slope and deflection as 
\begin{align}\label{eq:cross_analytic_slope_deflection_general}
    EI\frac{dy}{dx} = Mx + P(lx-\frac{x^2}{2}) + c_1 \quad \quad \text{and}\nonumber\\
    EIy = M\frac{x^2}{2} + P(l\frac{x^2}{2}-\frac{x^3}{6}) + c_1x+c_2,
\end{align}
respectively. Here $c_1$ and $c_2$ are arbitrary integration constants and are obtained for the specific cases of macroscopic strain considering appropriate boundary conditions. Now, let us consider each case of macroscopic strain separately.
\subsection{Extensional stiffness}
In case of uniform extension, the force $P$ on the rod $OC$ is given by
\begin{align}\label{eq:cross_force extension}
    P = F\cos{\frac{\theta_c}{2}}
\end{align}
 and the boundary conditions are
\begin{align}\label{eq:bc extension}
    y(0) = 0,\quad \quad y^{\prime}(0) = 0,\quad\quad y^{\prime}(l) = 0
\end{align}
in the local coordinate system. Using the force expression \eqref{eq:cross_force extension} and the boundary conditions \eqref{eq:bc extension}, we get
\begin{align}
    EIy^{\prime} &= F\cos{\frac{\theta_c}{2}}(lx-\frac{x^2}{2})+Mx,\nonumber\\
    EIy &= F\cos{\frac{\theta_c}{2}}(l\frac{x^2}{2}-\frac{x^3}{6}) + 0.5Mx^2 , \nonumber\\
        M &=-0.5Fl\cos{\frac{\theta_c}{2}}.
\end{align}
From the above equations, the transverse displacement at $x=l$ is given by
\begin{align}
    y(l)=\frac{Fl^3}{12EI}\cos{\frac{\theta_c}{2}}.
\end{align}
The displacement of point $C$ is then given by
\begin{align}
    \delta  = y(l)\cos{\frac{\theta_c}{2}}= \frac{Fl^3}{12EI}\cos^2{\frac{\theta_c}{2}}.
\end{align}
Due to periodicity, the same displacement occurs at the left face of the RVE, resulting in the total deformation of the RVE to be $2\delta$. Thus, the total extensional strain of the RVE is given by
\begin{align}\label{eq:cross_rve_analytic_extension_strain}
    \text{v}_3^M = \frac{2\delta}{L} = \frac{\delta}{l\sin{\frac{\theta_c}{2}}} \frac{Fl^2}{12EI}\frac{\cos^2{\frac{\theta_c}{2}}}{\sin{\frac{\theta_c}{2}}}
\end{align}
and the total axial force on the right face of the RVE is
\begin{align}\label{eq:cross_rve_analytic_extension_force}
    \text{n}_3^M = 2F.
\end{align}
The net moment on the right face is zero which results in the equal and opposite reaction moments on points $C$ and $D$. Finally, using \eqref{eq:cross_rve_analytic_extension_strain} and \eqref{eq:cross_rve_analytic_extension_force}, the extensional stiffness of the RVE is
\begin{align}
    \mathbb{C}_{33}^M = \frac{\text{n}_3^M}{\text{v}_3^M} = \frac{24EI}{l^2}\frac{\tan\frac{\theta_c}{2}}{\cos{\frac{\theta_c}{2}}}.
\end{align}
The above linear relationship between $n_3^M$ and $\text{v}_3^M$ assumes small deformation at the macroscale.
\subsection{In-plane bending stiffness}
In case of uniform in-plane bending curvature, the force $P$ on the rod $OC$ is given by
\begin{align}\label{eq:force_bending}
    P = -F \cos{\frac{\theta_c}{2}}
\end{align}
 and the boundary conditions are
\begin{align}\label{eq:bc_bending}
    y(0) = 0,\quad \quad y^{\prime}(0) = 0,\quad\quad y^{\prime}(l) = -{k_1^Ml\sin{\frac{\theta_c}{2}}}
\end{align}
in the local coordinate system. Using equations \eqref{eq:cross_analytic_slope_deflection_general}, \eqref{eq:force_bending}, and \eqref{eq:bc_bending}, we get
\begin{align}\label{eq:cross_bending_analytic_slope_def_moment}
    EIy^{\prime} &= -F\cos{\frac{\theta_c}{2}}(lx-\frac{x^2}{2})+Mx,\nonumber\\
    EIy &= -F\cos{\frac{\theta_c}{2}}(l\frac{x^2}{2}-\frac{x^3}{6}) + 0.5Mx^2 ,\nonumber\\
        M &= \frac{Fl}{2}\cos{\frac{\theta_c}{2}} - EIk_1^M\sin{\frac{\theta_c}{2}}.
\end{align}
\begin{figure}
    \centering
    \includegraphics[width=0.5\linewidth]{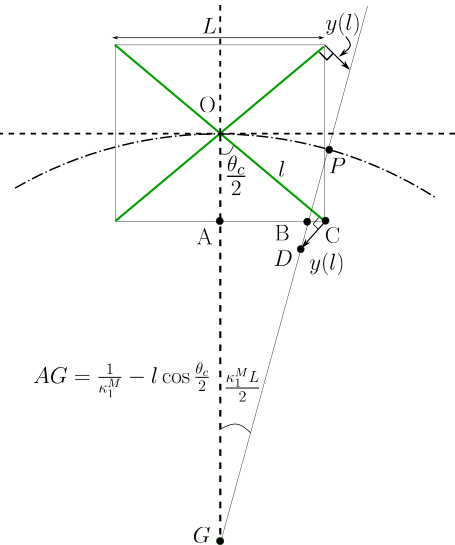}
    \caption{A schematic of cross RVE undergoing small macroscale in-plane bending deformation (scaled here for clarity). The point $G$ is at $\frac{1}{\kappa_1^M}$ distance from RVE origin $O$.}
\label{fig:cross_rve_analytic_bending_deflection_analysis}
\end{figure}
Another equation is needed that is obtained by computing the deflection of the point $C$ through geometry. Considering the figure \ref{fig:cross_rve_analytic_bending_deflection_analysis} wherein the RVE undergoes small bending deformation (scaled for clarity) about the center $G$, we have
\begin{align}
    AG~(\frac{k_1^ML}{2}) = AC - BC.
\end{align}
And since $BD$ is approximately perpendicular to BC in case of small deformation, we can say $BC = y(l)\cos{\frac{\theta_c}{2}}$. This results in
\begin{align}
    \left(\frac{1}{k_1^M}-l\cos{\frac{\theta_c}{2}}\right)\frac{k_1^ML}{2} &= \frac{L}{2}-y(l)\cos{\frac{\theta_c}{2}}\nonumber\\
    \Rightarrow y(l) &= k_1^Ml^2\sin{\frac{\theta_c}{2}}.
\end{align}
 Substituting the above equation in \eqref{eq:cross_bending_analytic_slope_def_moment}(b) with a negative sign because it is a downward deflection in the local coordinate system of the beam, we get
\begin{align}\label{eq:cross-bending-reaction-moment2}
   M &= \frac{2Fl}{3}\cos{\frac{\theta_c}{2}}-2EIk_1^M\sin{\frac{\theta_c}{2}}.
\end{align}
Using \eqref{eq:cross_bending_analytic_slope_def_moment}(c) and \eqref{eq:cross-bending-reaction-moment2}, we get
\begin{align}\label{eq:cross-bending-reaction-force-moment}
    F = \frac{6EIk_1^M}{l}\tan \frac{\theta_c}{2}, ~M = 2EIk_1^M\sin{\frac{\theta_c}{2}}.
\end{align}
Now in the RVE coordinate system, the net moment about the axis passing through the center of the right face of the RVE into the paper is given by
\begin{align}\label{eq:cross-bending-net-moment}
    M^{net} = \text{m}_1^M = 2Fl\cos{\frac{\theta_c}{2}} - 2M.
\end{align}
Finally, substituting \eqref{eq:cross-bending-reaction-force-moment} in \eqref{eq:cross-bending-net-moment}, we get
\begin{align}
    \text{m}_1^M = 8EIk_1^M\sin{\frac{\theta_c}{2}}.
\end{align}
Hence, the bending stiffness is
\begin{align}
    \mathbb{C}_{44}^M = \frac{\text{m}_1^M}{\kappa_1^M} = 8EI\sin{\frac{\theta_c}{2}}.
\end{align}
\subsection{In plane shear stiffness}
In case of uniform in plane shear strain, the force $P$ on the rod $OC$ is given by
\begin{align}\label{eq:cross_analytical_shear_force}
    P = F\sin{\frac{\theta_c}{2}}
\end{align}
and the boundary conditions on the rod $OC$ are
\begin{align}\label{eq:cross_analytical_shear_inplane}
    y(0) = 0,\quad\quad y(l) = \text{v}_2^Ml,\quad\quad y^{\prime}(l) = 0.
\end{align}
in the local coordinate system. Using \eqref{eq:cross_analytic_slope_deflection_general}, \eqref{eq:cross_analytical_shear_force} and \eqref{eq:cross_analytical_shear_inplane}, we get
\begin{align}\label{eq:cross_analytical_shear_slope_deflection}
    EIy^{\prime} &= F\sin{\frac{\theta_c}{2}}(lx-\frac{x^2}{2} - \frac{l^2}{2}) + M(x-l),  \nonumber\\
    EIy &= F\sin{\frac{\theta_c}{2}}(\frac{lx^2}{2}-\frac{x^3}{6}-\frac{l^2x}{2}) + M(0.5x^2 - lx),\quad \text{and} \nonumber\\
    EI\text{v}_2^M l &= -\frac{Fl^3}{6}\sin{\frac{\theta_c}{2}} - \frac{Ml^2}{2}.
\end{align}
Taking the moment balance about the $\textbf{e}_1$ axis passing through the center of the RVE and equating it to be zero (since no external moment is applied on the junction), we get
\begin{align}
    2Fl\sin{\frac{\theta_c}{2}}+2M = 0\nonumber\\
    \Rightarrow M = -Fl\sin{\frac{\theta_c}{2}}.
\end{align}
Substituting the above equation in \eqref{eq:cross_analytical_shear_slope_deflection}(c), we get
\begin{align}
    \frac{F}{\text{v}_2^M} = \frac{3EI}{l^2}\frac{1}{\sin{\frac{\theta_c}{2}}}.
\end{align}
The net macroscale force on the right face of the RVE is given by
\begin{align}
    \text{n}_2^M = 2F.
\end{align}
This results in the macroscopic shear stiffness as
\begin{align}
    \mathbb{C}_{22}^{M} = \frac{\text{n}_2^M}{\text{v}_2^M} = \frac{2F}{\text{v}_2^M} = \frac{6EI}{l^2}\frac{1}{\sin{\frac{\theta_c}{2}}}.
\end{align}
\section{Analytical formulas of effective properties of Auxetic tubular lattice}\label{sec:analytical_formulas_aux_tube}
The analytical formulas of the effective Young's moduli and the Poisson's ratio for a planar re-entrant lattice are derived in \cite{masters1996models}. They are as follows:
\begin{align}
    E_l &= \dfrac{1}{d\left(\dfrac{\cos{\theta_a}}{\dfrac{a_1}{a_2} - \sin{\theta_a}}\right)\left[\dfrac{\cos^2{\theta_a}}{K_f}+\dfrac{\cos^2{\theta_a}}{K_h}+\dfrac{2a_1/a_2+\sin^2{\theta_a}}{K_s}\right]},\nonumber\\
    E_c &=\dfrac{1}{d\left(\dfrac{a_1}{a_2}-\sin\theta_a\right)\left[\dfrac{\sin^2\theta_a}{K_f\cos\theta_a}+\dfrac{\sin^2\theta_a}{K_h\cos\theta_a}+\dfrac{\cos\theta_a}{K_s}\right]},
    \nonumber\\
    \nu_{lc} &=-\sin\theta_a \left(
    \frac{a_1}{a_2}-\sin\theta_a
    \right)
    \left[
    \frac{
    -\dfrac{1}{K_f}
    -\dfrac{1}{K_h}
    +\dfrac{1}{K_s}
    }{
    \dfrac{\cos^2\theta_a}{K_f}
    +\dfrac{\cos^2\theta_a}{K_h}
    +\dfrac{2a_2/a_1+\sin^2\theta_a}{K_s}
    }
    \right],
\end{align}
where the subscripts $l$ and $c$ correspond to the longitudinal and circumferential directions, respectively. Also,
\begin{align}
    K_f = \dfrac{12EI}{a_2^3},\quad K_s = \frac{EA}{a_2},\quad K_h = \dfrac{GA}{a_2}.
\end{align}
The above formulas for planar lattices can be used to derive effective properties of a tubular auxetic structure by computing the side length using \eqref{eq:stent_parameters}. With this approach, \cite{karnessis2013uniaxial} obtained the effective properties analytically for auxetic cellular tubes by considering only bending terms (i.e. $\frac{1}{K_h}=0$ and $\frac{1}{K_s}=0$) in the above expressions. The radial stiffness can be obtained from the circumferential Young's modulus $E_c$. Using the thin-walled shell assumption, the circumferential stress in presence of radial pressure $P_{eq}$ is given by
\begin{align}\label{eq:thin_shell_sigma_c}
    \sigma_c = \frac{P_{eq}D}{2d}.
\end{align}
We also know that the circumferential young's modulus is
\begin{align}\label{eq:youngs_mod_circ}
    E_c = \dfrac{\sigma_c}{\epsilon_c}
\end{align}
where $\epsilon_c$ is the circumferential strain. For a thin-walled tube undergoing small deformations, the radial strain is equal to the circumferential strain. Hence, using \ref{eq:thin_shell_sigma_c} and \ref{eq:youngs_mod_circ}, the radial stiffness is
\begin{align}
    \text{Radial stiffness} = \frac{\text{Radial pressure}}{\text{Radial strain}} = E_c\left(\frac{2d}{D}\right).
\end{align}

\end{document}